\newcommand{\ignore}[1]{}
\documentclass[12pt]{article}
\usepackage[left]{lineno}
\usepackage{amsmath}
\usepackage{amssymb}
\usepackage{cancel}
 \usepackage{graphicx}
\usepackage{braket}
\usepackage{authblk}
\usepackage{caption,subcaption}
\usepackage{comment}
\usepackage{enumitem}
\usepackage[utf8]{inputenc}
\usepackage[pdftex, pdfstartview={FitH}, pdfnewwindow=true, colorlinks=false, pdfpagemode=UseNone]{hyperref}
\numberwithin{equation}{section}
\numberwithin{figure}{section}
\graphicspath{{fig/}}
\renewcommand{\baselinestretch}{1.1}
  \newlength{\abstractwidth}
\usepackage[usenames,dvipsnames,svgnames,table]{xcolor}

  \newcommand{\be}{\begin{equation}}
  \newcommand{\bea}{\begin{eqnarray}}
  \newcommand{\eea}{\end{eqnarray}}
  \newcommand{\beq}{\begin{equation}}
  \newcommand{\ee}{\end{equation}}
  \newcommand{\eeq}{\end{equation}}

\newcommand\nn{\nonumber \\}
\newcommand\mD{{\mathbb D}}

\newcommand\mEye{{\mathbb I}}

\renewcommand\ell{l}
\newcommand\levec[1]{\langle #1 |}
\newcommand\revec[1]{| #1 \rangle}

\newcommand*\Laplace{\mathop{}\!\mathbin\bigtriangleup}
\newcommand\onehalf{\frac{1}{2}}
\newcommand{\eqn}[1]{Eq.~\ref{#1}}

\title{Multigrid for Staggered Lattice Fermions}

\author[*]{Richard C. Brower}
\author[$\dagger$]{M.A. Clark}
\author[+]{Alexei Strelchenko}
\author[*]{ Evan Weinberg}
\affil[*]{Boston University, Boston, MA 02215, USA}
\affil[$\dagger$]{NVIDIA Corporation, Santa Clara, CA 95050, USA} 
\affil[+]{Scientific Computing Division, Fermilab, Batavia, IL 60510-5011, USA}

\date{}

\begin{document}


\maketitle


\begin{abstract}
  {\em Critical slowing down} in Krylov methods for the Dirac operator presents a
  major obstacle to further advances in lattice field theory as it
  approaches the continuum solution. Here we formulate a multi-grid algorithm
  for the Kogut-Susskind (or staggered) fermion discretization which has
  proven  difficult relative to Wilson multigrid due to its first-order anti-Hermitian
  structure. The solution is to introduce a novel spectral
  transformation by the K\"ahler-Dirac spin structure prior to the
  Galerkin projection. We present numerical results for the two-dimensional, two-flavor Schwinger model, however, the general formalism is agnostic to dimension and is directly applicable to four-dimensional lattice QCD. 
\end{abstract}

\setlength{\parskip}{0in}
\thispagestyle{empty}
\setcounter{page}{-1}
\pagebreak
\tableofcontents
\setcounter{page}{1}
\setlength{\parskip}{.2in}

\newpage
\section{\label{sec:intro}Introduction}

Increasingly powerful computers and better theoretical insights
continue to improve the predictive power of lattice quantum field
theories, most spectacularly for lattice quantum chromodynamics
(LQCD)~\cite{Wilson:1974sk}.  However, with larger lattice volumes and
finer lattice spacing, exposing multiple scales, the  lattice Dirac linear system
becomes increasing ill-conditioned threatening further progress.  The cause is well known: as the
fermion mass approaches zero, the Dirac operator becomes singular, due
to the exact {\em chiral } symmetry of the Dirac equation at zero
mass, causing {\em critical slowing down}~\cite{Blum:2013mhx}.  The
algorithmic solution to this problem for lattice QCD was recognized 25
years ago. The fine-grid representation for the linear solver should
be coupled to multiple scales on coarser grids in the spirit of Wilson's real space
renormalization group and implemented as a recursive multigrid (MG)
pre-conditioner~\cite{Wilson:1974mb}. Early investigations in the 1990s
introduced a gauge-invariant projective MG algorithm~\cite{Hulsebos:1990er, Brower:1991xv} with encouraging results for
the Dirac operator in the presence of weak (or smooth) background
gauge fields near the continuum.  However, in practice lattice sizes
at that time were too small and the gauge fields were too rough to
achieve useful improvements.

Not until the development of adaptive geometric MG methods~\cite{Brannick:2007ue, Babich:2010qb}, was a fully recursive
MG algorithm found for the Wilson-Dirac discretization, which was able
to transfer the strong background chromodynamics fields onto 
coarser scales and eliminate the ill-conditioning of the
Dirac kernel in the chiral limit. In spite of this achievement for the
Wilson-Dirac and closely related twisted mass
formulation~\cite{Frezzotti:2000nk,Frommer:2013fsa}, these are not the
only important Dirac discretizations in common use in
lattice field theory. Three other discretizations used extensively in
high energy applications, which more faithfully represent {\em chiral}
symmetry on the lattice, are referred to as the domain wall~\cite{Kaplan:1992bt}, overlap~\cite{Neuberger:1997fp}, and staggered~\cite{PhysRevD.11.395} fermions. The application of
adaptive geometric MG to these discretizations has proven to be more
difficult, perhaps related to the improved lattice chiral symmetry.  A two-level MG solver for domain wall fermions has been
implemented~\cite{Cohen:2012sh, Boyle:2014rwa} which shows some promise, and a non-Galerkin algorithm has been implemented for overlap fermions~\cite{Brannick:2014vda}, but there has been no
success at formulating a MG staggered algorithm.  Moreover, since
staggered lattice ensembles  are now the largest available for LQCD, requiring
$O(10^5)$ iterations for good convergence, improving staggered solvers  is a critical issue.
Here we
introduce a novel solver with the K\"ahler-Dirac spin
structure~\cite{Becher1982,PhysRevD.38.1206} that allows, at last, the construction of an
effective multi-level adaptive geometric MG algorithm for
staggered fermions.

The staggered fermion is a remarkable discretization~\cite{PhysRevD.11.395,KAWAMOTO1981100}
which closely resembles the continuum Dirac linear operator,
\be
D^{ij}\psi_j(x) = [ \gamma^{ij}_ \mu (\partial_\mu - i A_\mu(x))  + m \delta^{ij} ]\psi_j(x) \; .
\label{eq:continuum}
\ee
 The  lattice discretization replaces the derivative by    a gauge-covariant central difference,
\begin{equation}
D_{x,y} =  \frac{1}{2a}\sum^d_{\mu = 1} \eta_\mu(x)\left[
  U(x,x + \mu) \delta_{x+\mu, y} 
- U^\dag(x -\mu,x)\delta_{x-\mu,y}\right]
+ m \delta_{x,y} \; ,
\label{eq:staggeredD}
\end{equation}
resulting in a sparse matrix operator on a hypercubic lattice with the
background gauge fields $U(x,y)$ represented by highly oscillatory
$SU(3)$ matrices on each link $\langle x,y\rangle$ of the lattice. The
$\gamma_\mu$ matrices are replaced by a single staggered $\pm 1$-sign:
$\eta_\mu =(-1)^{\sum_{\nu < \mu} x^\nu}$.  Similar staggered lattice
realizations of Dirac fermions have proven valuable not only for
lattice QCD investigations, but also for a variety of physical systems
such as graphene in condensed matter~\cite{Brower:2012zd}, supersymmetry~\cite{Catterall:2014vka},
and strongly interacting conformal fixed points of possible interest
for beyond the standard model (BSM) physics in the Higgs
sector~\cite{Appelquist:2011dp,Hasenfratz:2011xn,Aoki:2012eq,DeGrand:2011cu,Cheng:2013xha,Cheng:2014jba,Lombardo:2014pda,Aoki:2013zsa,Aoki:2014oha}.

Unlike the Wilson and domain wall methods, the staggered
discretization preserves the exact anti-Hermiticity of the continuum
Dirac operator up to real mass shift.  In this sense it represents the
most primitive (or even fundamental) discretization.  It has no
explicit spin matrices ($\gamma_\mu$), so the Dirac spin structure
only emerges in the continuum limit. Each $2^d$ lattice sub-block in four dimensions
reassembles into four Dirac flavors (or tastes), the content of
a single K\"ahler-Dirac fermion~\cite{Kahler2011}. This is the structure that
our MG algorithm exploits: dividing out the $2^d$
K\"ahler-Dirac spin structure transforms the spectrum into a near
``circle'' in the complex plane as illustrated in
Fig.~\ref{fig:freespec}. The striking similarity of the resultant
spectrum to the Wilson and overlap spectra is, we believe, essential
to the success of our staggered MG algorithm.

In LQCD applications with staggered fermions, the system
$D(U,m)_{ij}\psi_j = b_i$ is typically
solved via Krylov methods on the Schur decomposed
even/odd operator (or, equivalently, the red/black operator). Because the
preconditioned operator is Hermitian positive
definite, the system can be solved by the conjugate gradient (CG)
algorithm. This method has proven robust, and there are some well
established methods to fend off critical slowing
down, such as EigCG~\cite{Stathopoulos:2007zi} eigenvalue deflation or block Krylov
solvers~\cite{Hestenes&Stiefel:1952,Nikishin:1995,Dubrulle_retoolingthe,Clark:2017ekr}. Block solvers do not remove critical slowing down,
 and deflation methods scale poorly with the volume
in terms of the number of eigenvectors need to remove critical slowing down.
As explained in our earlier report~\cite{Weinberg:2017zlv}, an adaptive geometric
MG algorithm for the staggered normal operator can be easily
formulated which removes critical slowing down. However this comes with
a heavy overhead. A Galerkin coarsening of the normal
equation introduces next-to-nearest neighbor (or corner) terms,
resulting in a $2d + 2d(d-1)$ site coarse operator stencil; in four dimensions increasing the
off-diagonal terms from $8$  to $32$ terms. This becomes
prohibitively expensive in terms  communication pressure 
in parallel strong scaling  MG solvers~\cite{benson1973iterative,Chow:2001:PIP:1080623.1080641,Sterck06reducingcomplexity,doi:10.1137/140952570}.

The solution to this problem is to develop an MG algorithm directly on the staggered operator. In the interest of algorithm development, we consider a two-dimensional model system as opposed to the full four-dimensional QCD. The two-dimensional staggered fermion, coupled to an Abelian gauge theory,
$U(x,x+\mu) = \mbox{exp}[ i \theta_\mu(x)]$ is
the two-flavor Schwinger model in the continuum limit~\cite{PhysRev.128.2425,Smilga:1996pi}.
This is a fully non-perturbative quantum field theory which is an
ideal analogue to four-dimensional QCD. Like QCD it exhibits confinement with
a zero mass triplet of ``pion-like'' bound states in the chiral (zero
mass) limit, and instantons that present a topological mechanism which
breaks chiral symmetry dynamically in the flavor singlet
channel~\cite{Adams:2009eb}.  As such, this has proven to be a
reliable test framework~\cite{Brannick:2007ue} prior to a full
implementation for four-dimensional QCD. The reader is referred to an
extensive literature to understand the physical features that guide
our construction in two dimensions and the natural generalization to
four dimensions.

The lattice Schwinger model has the action
\begin{align}
S_{lat} = \bar\chi_x [D(U) +  m_0]_{xy} \chi_y + \beta \sum_{x}
  U^{plaq}(x)  \; .
\end{align}
Introducing the lattice spacing $a$, the bare mass ($m$) and the gauge
coupling ($g$) are given by dimensionless parameters, $m_0 = a m$ and
$\beta = 1/( a^2 g^2)$ respectively.  There are two important physical
length scales determined by these parameters: (1) The fundamental
gauge correlation length (or string length) measured by the Wilson
loop area law is $\ell_\sigma = a \sqrt{2 \beta}$.  (2) The
fundamental fermion length scale measured by the ``pion'' Compton
wave length is
$\ell_{M_\pi} = \frac{1}{M_\pi} \approx 0.5a (a m)^{-2/3}
\beta^{1/6}$~\cite{Smilga:1996pi}.  To approach the continuum both
must be large relative to the lattice spacing.  As an analogue to QCD, we
should also approach the chiral regime with
$\ell_{M_\pi}/\ell_\sigma \gg 1$. To control finite volume $L^d$ and
finite lattice spacing $a$ errors, the four length scales should obey
the constraint: $L \gg \ell_{M_\pi} \gg \ell_\sigma \gg a $.

This two-dimensional theory has been carefully selected because of its remarkable
similarity to four-dimensional QCD both in terms of the underlying physics and
the formal mathematical structure.  Although at present our numerical tests are restricted to two dimensions,
the entire formal structure is applicable to higher dimensions.  The numerical analysis of a
four-dimensional algorithm for lattice QCD is under development in
QUDA~\cite{Clark:2009wm,Babich:2011np,Clark:2016rdz}, an efficient GPU
framework for LQCD applications. Results will be presented in a
subsequent publication.  

The organization of the paper is as follows. In
Sec.~\ref{sec:staggered} we give the mathematical framework of the
staggered Dirac operator essential to our subsequent MG
formulations.  In Sec.~\ref{sec:coarse} we consider a Galerkin
projection of the original operator and explain why it fails as a
MG preconditioner. We then constrast it with the coarse
projection of our new K\"ahler-Dirac preconditioned operator. In
Sec.~\ref{sec:results} we present in detail the construction of the
staggered MG algorithm, followed by detailed numerical tests for the
two-dimensional Schwinger model. In Sec.~\ref{sec:hermpres} we discuss some
alternatives to our current implementation, which may be useful in the
application of our staggered MG algorithm to four-dimensional LQCD and
other staggered lattice simulations.  For example, a method for exactly
preserving complex conjugate eigenpairing and numerical tests thereof is
presented in Sec.~\ref{sec:hermpres} and in
Appendix~\ref{app:cplxconjpairs}, respectively.

\newpage
\section{\label{sec:staggered}Mathematical Preliminaries of Staggered
  Fermions}

The geometric structure of the staggered Dirac operator
\eqn{eq:staggeredD} and its relationship to the low-lying
eigenspectrum is important for our analysis. Many of its features are
inherited directly from the discretization of the continuum action,
\begin{align}
\mathcal{S} = \int d^dx~\bar\psi\left[\gamma_\mu\left(\partial_\mu - i A_\mu\right)+m\right]\psi.
\end{align}
The  na\"ive fermion discretization  uses a central difference approximation for the first
derivative, which causes the so-called ``doubling'' (or aliasing)
problem~\cite{Nielsen:1981hk}. In the continuum, a single na\"ive fermion gives $2^{d}$
Dirac fermions: 16 four-component spinors in four dimensions and 4 two-component
spinors in two dimensions. The staggered construction reduces this multiplicity
by spin diagonalizing the Dirac structure, then dropping all but one of
the $2^{d/2}$ copies. Explicitly, this spin
diagonalization,
\be
\bar \psi_{i,x} \gamma^{ij}_\mu [U(x,x + \mu)\psi_{j,x+\mu}  -
U(x,x - \mu)\psi_{j,x-\mu}] \nn
\rightarrow  \bar \psi_{i,x} \eta_\mu(x) [U(x,x + \mu)\psi_{i,x+\mu}  - 
U(x,x -\mu)]\psi_{i,x-\mu} \; ,
\ee
 is achieved by the unitary  field redefinition,
\be
\psi_x \rightarrow \Omega_x \psi_x \quad , \quad \bar \psi_x
\rightarrow \bar \psi_x \Omega^\dag_x \; ,
\ee
with $\Omega_x = \gamma_1^{x_1} \gamma_2^{x_2}
\cdots \gamma_d^{x_d}$. Dropping all but one copy, with the  replacement
\be
\Omega^\dag_x \gamma^{ij}_\mu \Omega_{x + \mu} = \eta_\mu(x)
\delta_{ij} \rightarrow \eta_\mu(x) = 
\pm 1 \; , 
\ee
results in a partial solution of the {\emph{doubling}} problem by
reducing the fermion content to $2^{d/2}$ staggered Dirac fermions: 4
in four dimensions and 2 in two dimensions.  It is convenient to write the staggered operator succinctly
as
\be
D\psi_x =   \frac{1}{2}\sum_{\mu} \eta_\mu(x) \left[ \mD_\mu -
  \mD^\dag_{\mu} \right] \psi_x + m \psi_x  \; ,
\ee
in terms of the gauge covariant forward difference operator $\mD_\mu\psi (x) \equiv U(x,x+\mu)
\psi(x+\mu) - \psi(x) $
and its  conjugate $\mD^\dag_\mu\psi (x) \equiv  U^\dag(x,x-\mu)
\psi(x-\mu) - \psi(x) $. 

The staggered operator has a few special properties not shared by other
fermion discretizations.  The staggered operator is  anti-Hermitian up
to a mass shift and is normal: $ [D(U,m), D^\dagger(U,m) ] = 0$, just like the continuum operator. This is in contrast to the Wilson  discretization, 
\be
D_{W}(m) =  \frac{1}{2}\sum_{\mu} [\gamma_\mu \left[ \mD_\mu
   - \mD^\dag_{\mu} \right]  + r \mD_\mu^\dagger \mD_\mu] + m,
\label{eq:Wilson}
\ee
with its Hermitian second order Wilson (stabilization) term that
decouples doublers but makes $D_{W}$ non-normal in the interacting
case. The Wilson term also explicitly breaks chiral symmetry. On the
other hand, the staggered operator retains a single exact chiral 
symmetry in the interacting case,
\be
\gamma_5 \rightarrow \epsilon(x) = \Omega^\dag_x \gamma_5 \Omega_x =
(-1)^{x_1 + x_2 + \cdots +x_d}, \label{eq:epsilon}
\ee
with $\epsilon(x)$ being the generator of the chiral symmetry. These good chiral properties give $\epsilon(x) D + D \epsilon(x) = 2 m \epsilon(x) \rightarrow 0$ as $m \rightarrow 0$. The chiral projectors, $\frac{1}{2}(1 \pm \epsilon(x))$, partition the lattice into
even and odd sub-lattices,
\be
D\psi = b \to
  \begin{bmatrix} m & D_{eo} \\ D_{oe} &m \end{bmatrix} 
\begin{bmatrix} \psi_e \\ \psi_o \end{bmatrix}
=
 \begin{bmatrix} b_e \\ b_o \end{bmatrix} \; .
\ee
Furthermore, $D$ features an $\epsilon(x)$ Hermiticity, analogous to $\gamma_5$
Hermiticity of the continuum Dirac operator. 

\begin{figure}[t]
\begin{center}
\includegraphics[width = .5\textwidth]{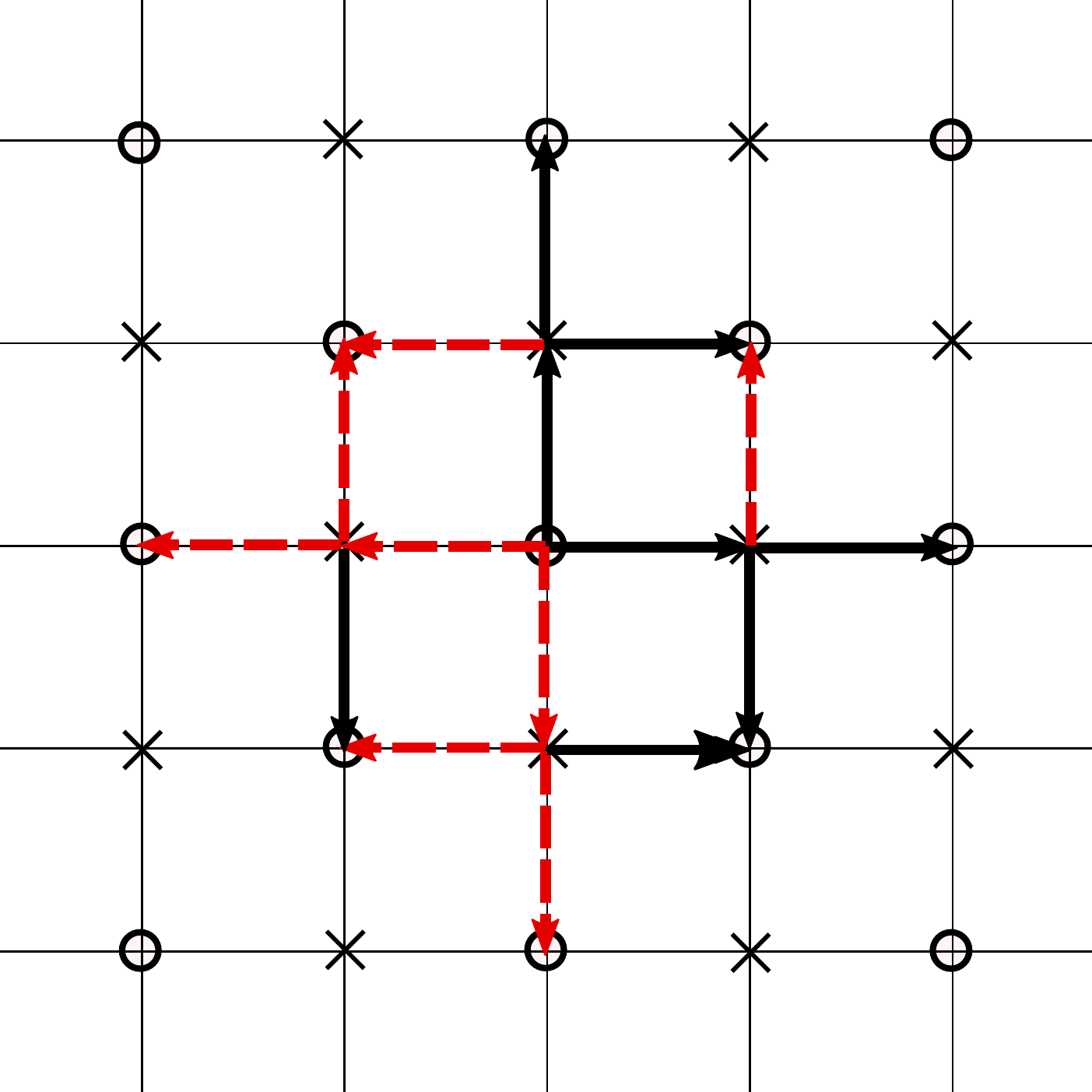}
\end{center}
\caption{\label{fig:2Dlattice} The normal operator applied on an odd site ``o''. All contributions to even sites ``x'' cancel due to $D$ being normal. Links in black (solid) and red (dashed) correspond to $\pm 1$, respectively, due to the contributions of $\eta_\mu$ and the anti-Hermiticity of $D$. In the free field, it's clear that corner terms cancel.}
\end{figure}

The normal equations for the staggered operator are diagonal,
\be
D^\dag D \psi = \  
\begin{bmatrix} 
 m^2 - D_{eo} D_{oe} & 0\\  
0 & m^2 - D_{oe} D_{eo}
\end{bmatrix}
\begin{bmatrix} \psi_e \\ \psi_o \end{bmatrix}\label{eq:normal}
\; .
\ee
The Schur-preconditioned system takes on a similar structure and is
also Hermitian positive definite. For the free problem (i.e., unit
gauge fields, $U(x,x+\mu)  = 1$), there is an exact cancellation of all
next-to-nearest neighbor ``around-the-corner'' terms in the normal operator. This is a result of the $\eta_\mu$ phases preserving a key property of the
Dirac algebra when taking the product of $\eta$s around a plaquette,
\be
\gamma_\mu \gamma_\nu(- \gamma_\mu) (- \gamma_\nu)   = -1 \rightarrow 
\eta_\mu(x)  \eta_\nu(x+\mu) \eta_\mu(x + \nu) \eta_\nu(x)
= -1.
\label{eq:frustration}
\ee
The result is a set of $2^d$ decoupled Laplace operators on a lattice with spacing $2a$ illustrated in Fig.~\ref {fig:2Dlattice}. In this sense, the free staggered
operator is truly the ``square root'' of the Laplace operator, similar
to the continuum Dirac operator. We can immediately write down the eigenvalues of the free staggered operator,
given by
\begin{align}
\lambda(p,m) &= m \pm i \sqrt{\sum_{\mu} \sin^2\left(p_\mu\right)} \; ,
\end{align}
where the $p_\mu= 2n_\mu\pi/L$ for
integers $n_\mu \in [-L/4+1,L/4]$ due to the shift-by-two
translational invariance. The eigenvalues are imaginary (up to a real
mass shift) and come in complex conjugate pairs. When an interacting
gauge field is turned on, the ``around-the-corner'' terms no longer
vanish, leaving the two decoupled components in Eq.~\ref{eq:normal}.
These next-to-nearest neighbor terms are the standard so called clover term,
resulting in an irrelevant, in the Wilsonian sense, spin gauge interaction ($\sigma_{\mu\nu} F_{\mu\nu}$)
in the continuum. The spectrum cannot be found analytically, but
$\epsilon(x)$ Hermiticity symmetry ensures that the eigenvalues still
appear in exact complex conjugate pairs.

\subsection{\label{sec:transform}K\"ahler-Dirac Preconditioning}

We now consider the spectral transformation which is
essential to the staggered MG algorithm  presented in
Sec.~\ref{sec:coarse} and tested numerically in
Sec.~\ref{sec:results}.  Here we will show that when the staggered
operator is right-preconditioned by the $2^d$ K\"ahler-Dirac blocks, the
spectrum on the resultant $2a$ blocked lattice is dramatically
different. In the free case, we prove that this transformation gives
an exactly circular spectrum in the complex plane, similar to the
overlap lattice Dirac discretization~\cite{Neuberger:1997fp}.  The inclusion of gauge fields
and/or the three link Na\"ik term~\cite{Naik:1986bn} are relatively small modifications
of this basic circular structure.

\begin{figure}[ht]
\center
\includegraphics[width=.4\textwidth]{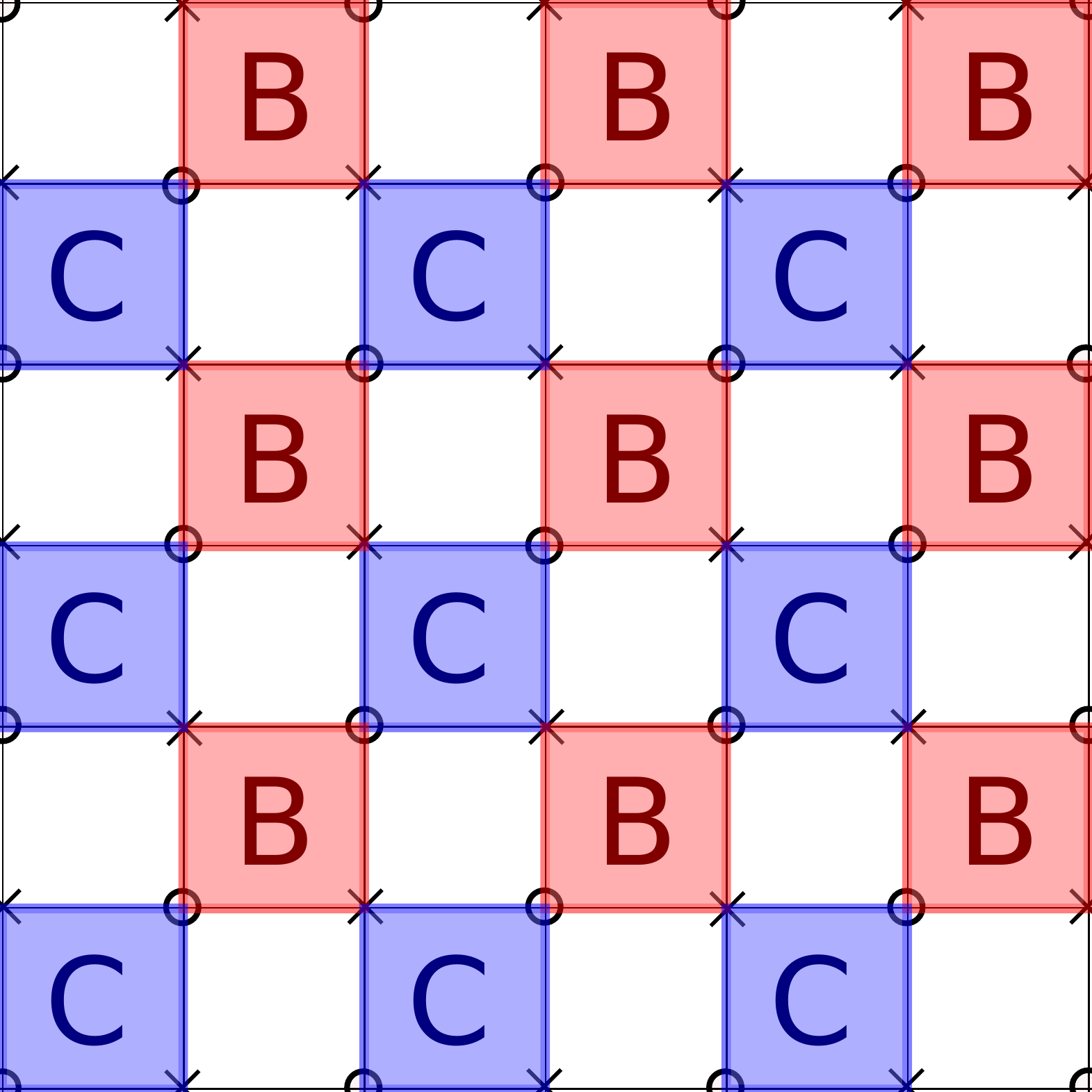}~~~~
\includegraphics[width=.45\textwidth]{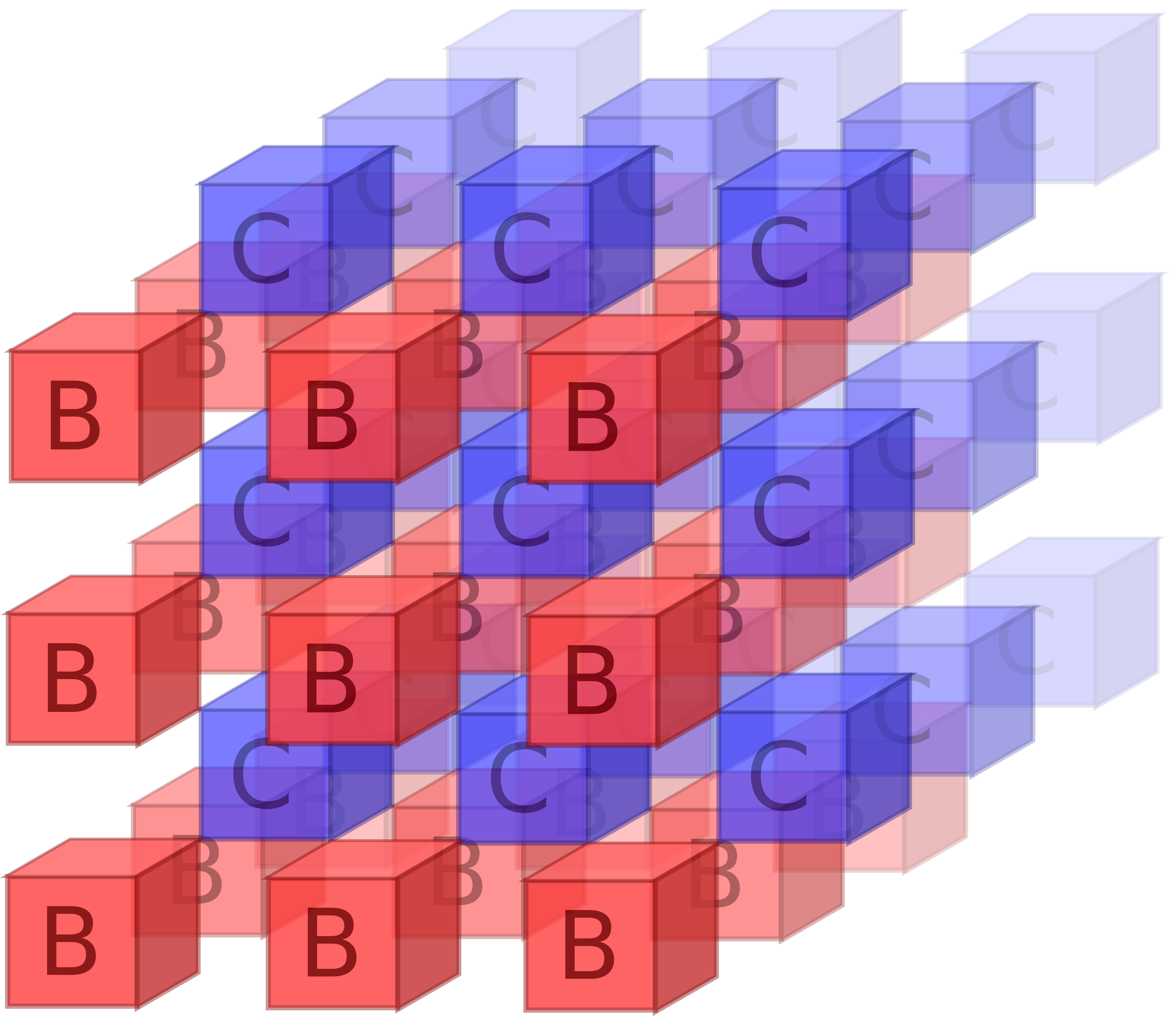}
\caption{\label{fig:BC_partition} On the left is the two-dimensional partition into
  the spin-tasted block B (in red) and the complementary connecting
  block C (in blue). On the right is the three-dimensional partition again into
  spin-tasted block B and the complementary connecting block C. In any
  dimension these partitions into B and C contain all $N$ lattice sites of the
  periodic lattice and each share half of the $d N$ links.}
\end{figure}

The argument proceeds as follows. The staggered operator is composed
of blocks containing $2^d$ sites, corresponding to $2^d$ degrees of freedom that in the continuum
limit are recombined into a multiplet of Dirac
fermions~\cite{Gliozzi:1982ib,KlubergStern:1983dg}. It is straightforward to see that the decomposition of the staggered operator in
$2^d$ blocks of sites partitions the lattice, as illustrated in red in
Fig.~\ref{fig:BC_partition}, into independent $2^d$ blocks $B$
containing a plaquette of links. We will refer to these as
K\"ahler-Dirac blocks. We also include the local mass term into this
$B$ block. The nearest-neighbor terms between the $B$ blocks
contibute to a block hopping term $C$, which is unitarily equivalent, up to
the mass shift, to the block-local contributions in $B$. $B$ and $C$ only share sites at the
corner of squares in two dimensions, cubes in three dimensions, and hypercubes in four dimensions. This
is a dual decomposition: half of the links on the original lattice
contribute to $B$, and half contribute to $C$, as represented in Fig.~\ref{fig:BC_partition}. We denote this
partition between hopping terms within and across blocks as
\be
  D =\frac{1}{2}\eta_\mu [{\mathbb D_\mu} - {\mathbb D^\dag_\mu}] + m = B + C
\ee
We remark that we can interchange this dual description by shifting the
coordinates $\vec{x}_i \to \vec{x}_i + \vec{1}$, where $\vec{1}$ is a
vector of ones. We now construct the right-block-Jacobi or K\"ahler-Dirac preconditioned operator as
\be
\boxed{ A = DB^{-1} =\mEye  + C B^{-1}}\; . \label{eq:defA}
\ee
{\bf This is a remarkably different operator with which we develop our
  MG algorithm.}

To characterize these differences, we will first consider the free
case. After rescaling and multiplying by $\epsilon(x)$, the generator of the exact staggered chiral symmetry, both terms are separately {\bf Hermitian, traceless, and
  unitary}. More concretely, we define
$\widehat B = B \epsilon(x)/\sqrt{d + m^2}$ and
$\widehat C = C \epsilon(x)/\sqrt{d}$ and note:
\be
\widehat B \widehat B^\dag  =\widehat  B^2 = \mEye \quad, \quad
\widehat  C \widehat C^\dag = \widehat C^2 = \mEye \; ,
\ee
as a trivial consequence of the
perfect cancellation of the corner terms for Eq.~\ref
{eq:frustration}. These properties imply that $\widehat B$ and
$\widehat C$ have equal numbers of $\pm 1$ eigenvalues, and further
that the product $\widehat C \widehat B$ is a unitary matrix $U$.
(The addition of a Na\"ik term does not change $\hat B$, but it does
contribute to $\widehat C$.)

\begin{figure}[th]
    \centering
    \includegraphics[width=0.5\linewidth]{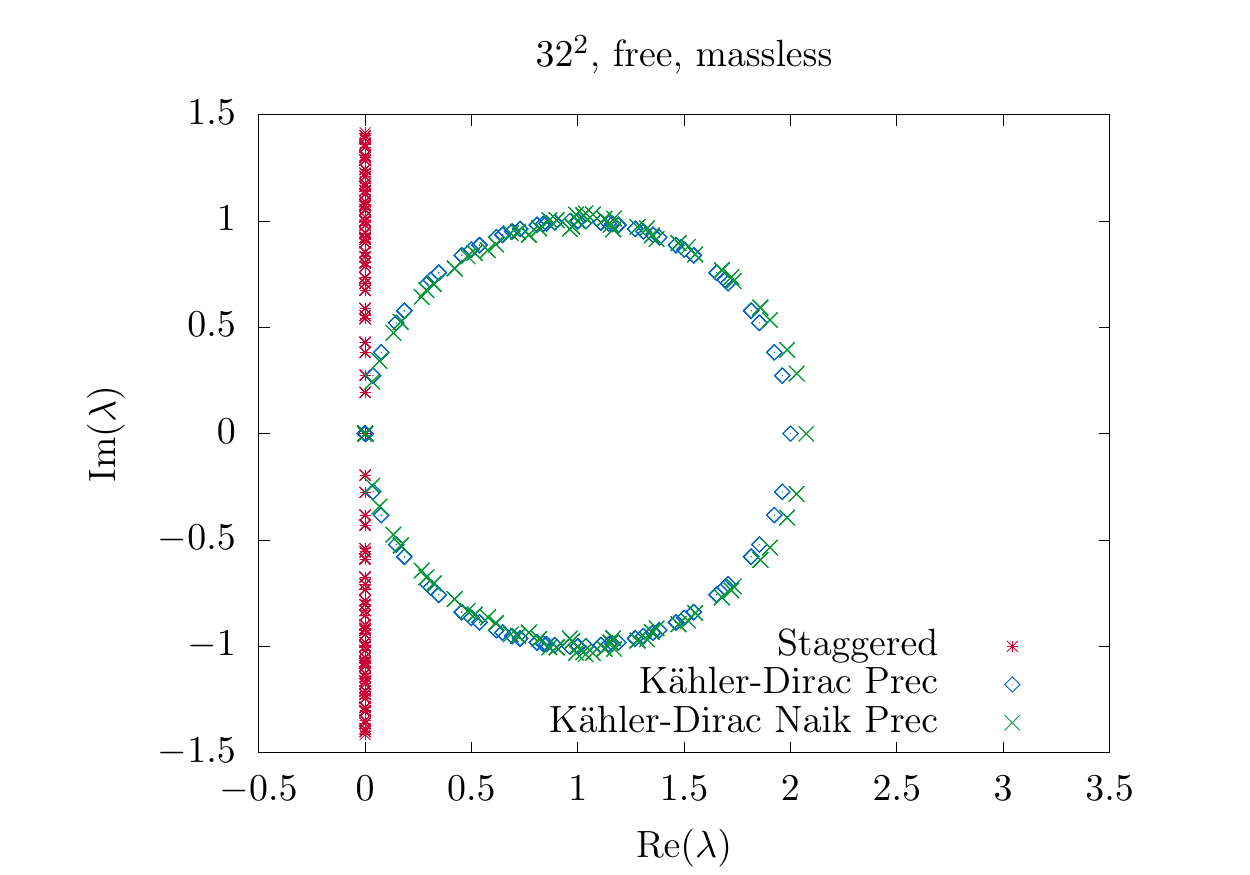}~    \includegraphics[width=0.5\linewidth]{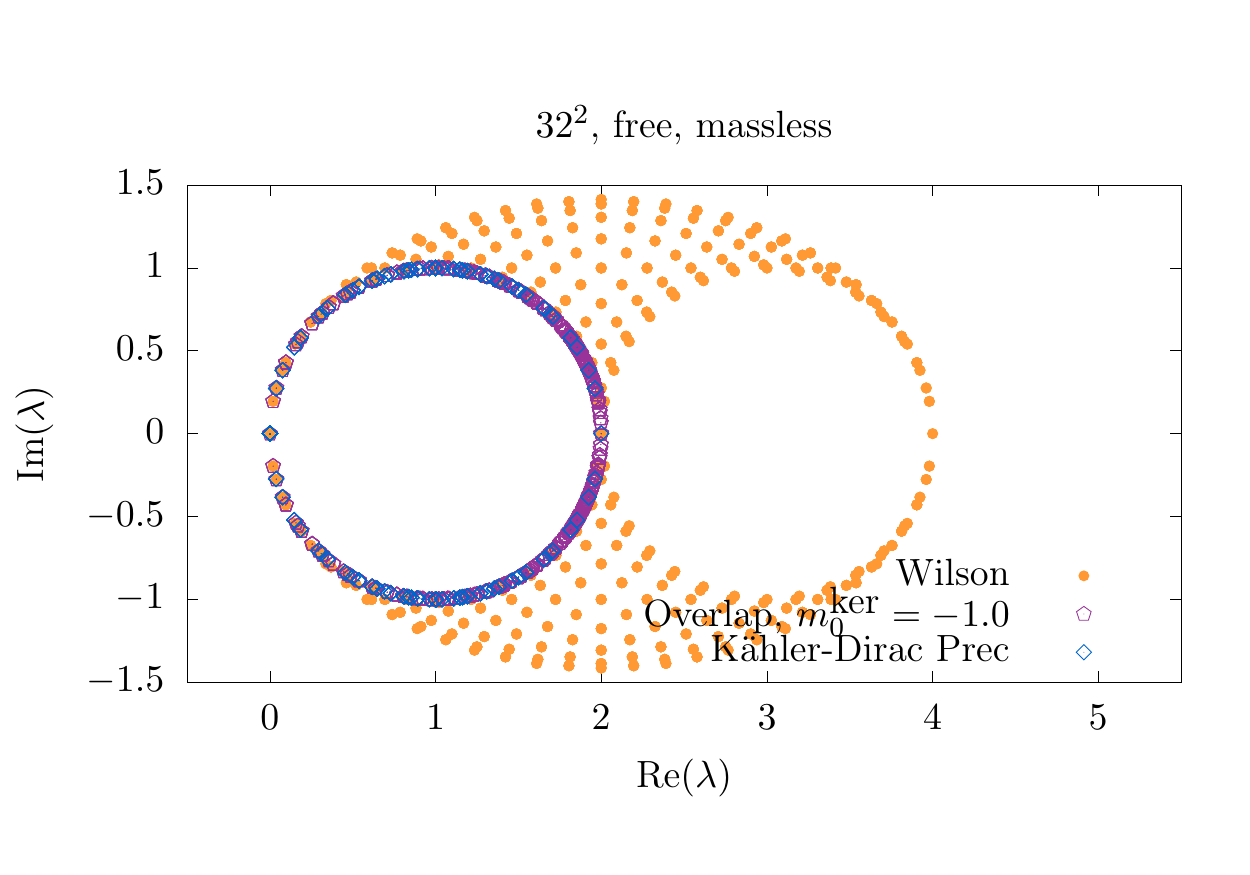}
   \caption{\label{fig:freespec}The spectrum of the free, massless K\"ahler-Dirac preconditioned operator, with and without a Na\"ik term, compared against the Wilson, overlap, and original staggered formulations.}
\end{figure}

With this observation, our free K\"ahler-Dirac staggered operator $A$ 
is given by
\begin{align}
A = DB^{-1} = \mEye + C B^{-1} = \mEye + \sqrt{\frac{d}{d+m^2}} \widehat{C} \widehat{B} = \mEye + \rho U.
\end{align}
The eigenvalues of $D B^{-1}$ lie on a circle centered at 1 as
illustrated in Fig.~\ref{fig:freespec}. The radius of the circle is
$\rho = \sqrt{\frac{d}{d+m^2}}$. In the massless limit, the radius is exactly 1. This leads to an identical structure to the overlap operator,
\begin{align}
 D_{ov} &=  1 + \gamma_5 \widehat\gamma_5 &
 \widehat\gamma_5 &\equiv\mbox{sign}[\gamma_5 D_W (-M)],
\end{align}
under the mapping
$(\gamma_5, \widehat\gamma_5) \rightarrow (\widehat C, \widehat
B)$. Both $\widehat C$ and $\widehat B$ are algebraically similar to
$\gamma_5$ and $\widehat\gamma_5$, being Hermitian and unitary with an equal number of $\pm 1$
eigenvalues. Adding a mass term to the overlap operator similarly
rescales the unitary portion of the spectrum,
\begin{align}
D_{ov} = \mEye + \frac{1-m}{1+m}\gamma_5 \widehat\gamma_5\; ,
\end{align}
introducing a  mass gap.  
For comparison, in the right panel of Fig.~\ref{fig:freespec}, we show
the free spectrum of the massless two-dimensional  Wilson
operator, the two-dimensional overlap operator and our new two-dimensional K\"ahler-Dirac
preconditioned operator. Very similar figures apply to four dimensions, except the
Wilson spectrum now has four arcs in the positive real direction. Finally, we note that if we add a Na\"ik term to the original staggered operator, the right preconditioning perturbs the unitarity of the spectrum but preserves the qualitative geometric features. A comparison of the K\"ahler-Dirac preconditioned operator to the original staggered operator is given in the left panel of Fig.~\ref{fig:freespec}. We compare the massive spectrum against all other fermion discretizations in Fig.~\ref{fig:freespeczoom}.

The K\"ahler-Dirac operator no loner admits a simple ``$\gamma_5$''
Hermiticity condition. {\textbf{However, it does obey a modified
    asymmetric $\gamma_5^{L/R}$ condition}}, which  is essential for our discussions in Sec.~\ref{sec:coarse}. The key observation is to note that we can change the ``convention'' that $A$ is given by a right block preconditioning of $D$ to a left block preconditioning via the transformation $\epsilon(x) A^\dag \epsilon(x) = B^{-1} D$. We can rearrange this identity and note
\be
\epsilon(x) B^{-1}  A = A^\dag  \epsilon(x) B^{-1}  \implies \gamma^L_5 A = A^\dag \gamma_5^L,\label{eq:gamma5L}
\ee
and likewise we can take advantage of the $\epsilon(x)$ Hermiticity of $D$ to note
\be
A B \epsilon = B \epsilon A^\dagger \implies A \gamma^R_5 = \gamma^R_5 A^\dagger.\label{eq:gamma5R}
\ee
Here we have defined $(\gamma^L_5 , \gamma^R_5) = (\epsilon(x) B^{-1},
B \epsilon(x))$. This is a generalization of the idea of $\gamma_5$ Hermiticity: now, $\gamma^L_5  \gamma^R_5 = 1$ and
$A^\dag = \gamma^L_5 A\gamma^R_5$. Also, just
as is the case for the Wilson and staggered operator, these properties are enough to
show that $A$ features complex conjugate eigenpairs. Assume $A \revec{\lambda^R} = \lambda \revec{\lambda^R}$, where the superscript $R$ . We can take
the Hermitian conjugate of each side of the equation. Next, we can
right-multiply by $\gamma_5^L$ and take advantage of $\gamma_5^{L/R}$
Hermiticity. This gives us
$\levec{\lambda^R} \gamma_5^L A = \levec{\lambda^R} \gamma_5^L
\lambda^*$, that is, $A$ also has an eigenvalue $\lambda^*$ with a
left eigenvector
$\levec{\lambda^{*L}} \equiv \levec{\lambda^R} \gamma_5^L$. This same
exercise can be trivially repeated for left eigenvectors using
$\gamma_5^R$ to the same end.

\subsection*{Free Spectrum after K\"ahler-Dirac Preconditioning}

For a detailed analysis of the spectrum, we 
introduce the flavor representation of the staggered operator~\cite{Gliozzi:1982ib,Duncan1982,KlubergStern:1983dg}, which is unitarily equivalent to a lattice K\"ahler-Dirac fermion in the free field~\cite{PhysRevD.38.1206,Bazavov:2009bb}. Here each  submatrix $B$ is expressed  in terms of 
the spin-taste gamma matrices which enumerate the components of a
single continuum K\"ahler-Dirac fermion~\cite{Kahler2011}. Its action is 
\begin{align}\label{eq:kahlerdirac}
\mathcal{S}& = b^d \sum_{X,\mu}
             \bar{q}(X)\left[\nabla_\mu\left(\gamma_\mu\otimes
             1\right) - \frac{b}{2}\Laplace_\mu\left(\gamma_5 \otimes
             \tau_\mu \tau_5\right) + m\right]q(X) \; ,
\end{align}
where $q(X)$ is the K\"ahler-Dirac field containing $2^d$ degrees of
freedom, $X$ is the K\"ahler-Dirac block index, $b = 2a = 2$ is the
lattice spacing between K\"ahler-Dirac blocks, and the finite difference operators are defined as
\begin{align}
\left(\nabla_\mu q\right)(X) &= \frac{q(X+b\hat{\mu})-q(X-b\hat{\mu})}{2b}, \\
\left(\Laplace_\mu q\right)(X) &= \frac{q(X+b\hat{\mu})-2q(X)+q(X-b\hat{\mu})}{b^2}.
\end{align}
In the language of staggered fermions, the $\gamma_\mu$ matrices 
generate the spin algebra, while the matrices $\tau_\mu = \gamma_\mu^\dagger$ generate the so-called taste
algebra.  It should be  noted that
if these lattice fermions  are gauged on the lattice with twice the lattice
spacing $b = 2a$,  the resulting lattice theory of interacting Dirac-K\"ahler
fermions~\cite{Becher1982,RABIN1982315} is no longer  equivalent
to the interacting staggered fermion and, of note, can generate a dynamical mass term. Likewise, on a continuum Riemann manifold, a K\"ahler-Dirac fermion admits a different gravitational gauging than Dirac fermions~\cite{Banks:1982iq}. 

Our decomposition of $D =B + C$ is now partitioning Eq.~\ref{eq:kahlerdirac} into local and nearest neighbor contributions. The local block $B$ is given by
\begin{align}\label{eq:bdefinition}
B= -B^\dagger &\Leftrightarrow \sum_{\mu} \gamma_5 \otimes \tau_\mu \tau_5 .
\end{align}
in the massless case. The inverse is given by
\begin{align}
B^{-1} = -B^{-\dagger} &\Leftrightarrow  -\frac{1}{d}\left(\sum_{\mu} \gamma_5 \otimes \tau_\mu \tau_5\right)\; .
\end{align}
The transformation $D \to A = D B^{-1}$ gives the kernel
\begin{align}
A = D B^{-1} &= -\frac{1}{d}\sum_{\mu,\nu} \left[\nabla_\mu\left(\gamma_\mu \gamma_5 \otimes \tau_\nu \tau_5\right) + \frac{1}{2}\Laplace_\mu\left(1 \otimes \tau_\mu \tau_\nu\right)\right].
\end{align}
This operator can be explicitly diagonalized in arbitrary dimension by noting the Hermitian and anti-Hermitian projections of the operator commute, the Hermitian projection can be trivially diagonalized, and the imaginary part is prescribed by recalling the shifted unitary structure of the spectrum. This gives
\begin{align}
\lambda(p_\mu) &= 1 - \frac{1}{d}\sum_\mu \cos(p_\mu) \pm
                 i\sqrt{1 -( \frac{1}{d}\sum_\mu
                 \cos(p_\mu))^2} \; .
\end{align}
This spectrum is visualized on the left panel of
Fig.~\ref{fig:freespec}. 
The spectrum can be written as $1 - e^{i
  \theta}$ for where $cos(\theta) = d^{-1}\sum_\mu \cos(p_\mu) \in [-1,1]$. We note again that, up to a scaling, the low spectrum is similar to the Wilson, overlap, and staggered spectrum.
\begin{figure}[t]
\centering
\includegraphics[width=.66\textwidth]{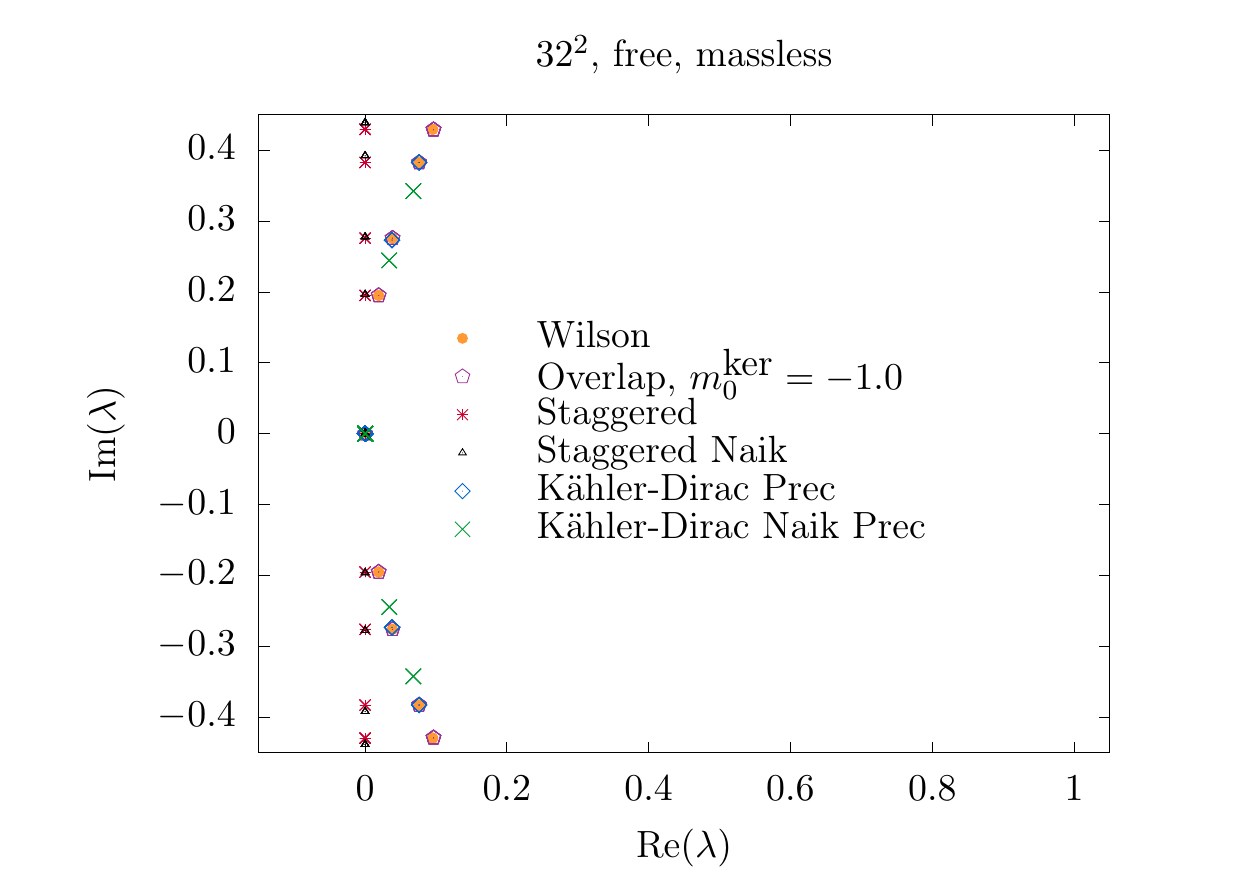}
\caption{\label{fig:freespeczoom} The free field spectrum of the
  staggered, K\"ahler-Dirac  preconditioned (with and without the Na\"ik
  contribution), Wilson, and overlap operators.}
\end{figure}

\paragraph{Non-zero Mass Term}
The spectrum undergoes a minor change when the original staggered
operator is massive. The local block now becomes
$\sum_\mu \gamma_5 \otimes \tau_\mu \tau_5 + m$, and the
preconditioned spectrum becomes
\be
\lambda(p_\mu) = 1 - \rho \left[ \frac{\rho}{d}\sum_\mu 
\cos(p_\mu) \pm i \sqrt{1 - ( \frac{\rho}{d}\sum_\mu \cos(p_\mu))^2} \right] \; .
\ee
which parameterizes the arc of a circle 
$1 - \rho e^{i
  \theta}$centered at $(1,0)$ with
radius $\rho = \sqrt{\frac{d}{d + m^2}}$. The arc is bounded to the range $\cos(\theta) = \rho
d^{-1}\sum_\mu \cos(p_\mu) \in [-\rho,\rho]$, giving a gap $\lambda(0)  = 1 - \rho^2 \pm i \sqrt{1 - \rho^2}$.

\paragraph{Na\"ik Term}
Many modern LQCD simulations add a next-to-next-to-nearest neighbor improvement term known as a Na\"ik~\cite{Naik:1986bn} term. Two common realizations of this improvement, equivalent in the free-field limit, are
AsqTad~\cite{Orginos:1999cr} and HISQ~\cite{Follana:2006rc} fermions. The free operator~\cite{Bazavov:2009bb} is given by
\begin{align}\label{eq:freehisq}
D(m)^{\mbox{\scriptsize Naik}}_{x,y} = -\frac{1}{2}\sum_{\mu}\eta_{\mu}(x) \left[\frac{9}{16}\left(\delta_{x,y-\hat\mu}-\delta_{x,y+\hat\mu}\right) - \frac{1}{48}\left(\delta_{x,y-3\hat\mu}-\delta_{x,y+3\hat\mu}\right)\right]+m\delta_{x,y}.
\end{align}
The improved action admits the spectrum
\begin{align}
\lambda_{\mbox{\scriptsize Naik}}(p,m) &= m \pm i \sqrt{\sum_{\mu} \sin^2\left(p_\mu\right)\left[1+\frac{1}{6}\sin^2\left(p_\mu\right)\right]^2}.
\end{align}
The effect of the Na\"ik term on the K\"ahler-Dirac action is to modify the nearest neighbor term and add a next-to-nearest neighbor term. The $2^d$ K\"ahler-Dirac block $B_{Naik} = \frac{9}{16}B$ is {\emph{unchanged}} up to a trivial rescaling. The new contributions are confined to $C$, which is no longer unitary: $\widehat{C}_{Naik}^\dagger \widehat{C}_{Naik} \neq \mEye$. Likewise, the spectrum is no longer a shifted unitary spectrum. Indeed, in two dimensions, the massless free spectrum is given by
\begin{align}
\lambda(p_\mu) &= 1 - \left[\left(1-x\right)S_1 + x S_2\right] \pm i \sqrt{1 - \left[\left(1-x\right)S_1 + x S_2\right]^2 - 2x\left[S_2 - S_1 + x\left(S_3-1\right)\right]},
\end{align}
where $S_n = \frac{1}{2}\sum_{\mu = x,y} \cos(n p_\mu)$ and $x =
\frac{-1/48}{9/16}$, the ratio of the improvement coefficients in the
Na\"ik-improved action. The improved spectrum is shown on the left panel
of Fig.~\ref{fig:freespec}, with the low modes emphasized in Fig.~\ref{fig:freespeczoom}. We again make the critical observation that the spectrum is {\emph{qualitatively}} similar to the original K\"ahler-Dirac spectrum, Wilson spectrum, and overlap spectrum.

\paragraph{Interacting Staggered Fermons in K\"ahler-Dirac Form:\label{sec:interactingtransform}}

We are ultimately interested in performing this right-block-Jacobi preconditioning on the interacting staggered operator, not the free operator. Procedurally, this is done by first gauging the staggered operator, and then performing the same unitary blocking transformation between the staggered form and the K\"ahler-Dirac form. The local block no longer has a simple structure because of gauge links~\cite{PhysRevD.38.1206}. In two dimensions, the K\"ahler-Dirac block $B$ attached to a unit corner at $2\vec{n}$ on the original staggered lattice is given by
\begin{align}
\label{eq:localblock}\left(\begin{matrix} m & 0 & -\onehalf U_x(2\vec{n}) & -\onehalf U_y(2\vec{n}) \\ 0 & m & -\onehalf U^\dagger_y(2\vec{n}+\hat{x}) & \onehalf U^\dagger_x(2\vec{n}+\hat{y}) \\ \onehalf U^\dagger_x(2\vec{n}) & \onehalf U_y(2\vec{n}+\hat{x}) & m & 0 \\ \onehalf U^\dagger_y(2\vec{n}) & -\onehalf U_x(2\vec{n}+\hat{y}) & 0 & m\end{matrix}\right).
\end{align}
Like the free case, the block $B$ is still anti-Hermitian plus a
massive shift. However, unlike the free case, the interacting
$\widehat{B}$ and $\widehat{C}$ are not unitary, and as such the
product $C B^{-1}$ does not have a unitary spectrum. Nonetheless, the
spectrum is still approximately circular and centered at 1, as can be
seen later in Fig.~\ref{fig:zoomcircle}. This is a desirable property
for matrix preconditioning in general~\cite{Day2001,Benzi2007}, and is
essential for a successful MG algorithm.  Importantly, this
operator still maintains the $\gamma_5^{L/R}$ Hermiticity defined in
Eq.~\ref{eq:gamma5L} and~\ref{eq:gamma5R}. This is true because the
proofs of $\gamma_5^{L/R}$ Hermiticity solely depend on the $\epsilon(x)$ Hermiticity
of $D$, which holds in the interacting case. By extension, the proofs
of complex conjugate eigenpairing still hold. These comments carry over
as appropriate when a N\"aik term is also included.

\newpage
\section{\label{sec:coarse}Multigrid Coarse Operator}

In forming the Galerkin projection of the staggered operator, we
follow the methods of previous successful formulations of MG
for the Wilson-Dirac discretization for LQCD~\cite{Brannick:2007ue}. Near-null vectors, or vectors which predominantly span the low-right eigenspace of the Wilson operator, are constructed by relaxing on the homogeneous equation with random
initial guess as is discussed in detail in
Sec.~\ref{sec:results}.  Later in this section we will also consider {\textit{exact low eigenvectors}} programmatically as near-null vectors. The resulting near-null vectors are chirally
doubled and block-orthonormalized to construct the rows of the restrictor matrix $R$, which aggregates fine degrees of freedom to a single site on the coarse lattice, and the prolongator matrix $P$, which maps coarse degrees of freedom back to the fine lattice. Unless otherwise noted, $R = P^\dagger$. For the staggered operator, this implies that coarsening preserves the anti-Hermitian plus mass-shift structure. Block orthonormalization implies $P^\dagger P = \mEye$. The prologator and restrictor can be used to define the coarse operator, \be
\widehat D = R D P. \ee The {\bf hat} notation refers to an
operator one level coarser than the ``unhatted'' operator.

We will begin by reviewing the Wilson formulation, largely to
establish notation. We will extend this
formulation to the staggered operator and  show why this
method fails to produce an effective recursive algorithm in this
case. We will last repeat this formulation for the K\"ahler-Dirac preconditioned operator and show that, in contrast to the original staggered case, this method succeeds.

\subsection{Review of Wilson Dirac Coarse Operator}

We begin by a basic restatement of the procedure for the adaptive
geometric MG developed for the Wilson operator in
QCD~\cite{Brower:1991xv,Brannick:2007ue}. It is important to first note that the Wilson operator {\emph{does}} obey $\gamma_5$ Hermiticity, that
is, $\gamma_5 D_W \gamma_5 = D_W^\dagger$. $\gamma_5$ Hermiticity is
sufficient to prove that eigenvalues of $D_W$ come in complex
conjugate pairs as the limiting case of $\gamma_5^L = \gamma_5^R$, as
discussed in Sec.~\ref{sec:transform}. Returning to MG, $n_{vec}^1$ near-null vectors are
generated, where the ``1'' refers to coarsening the finest level, as discussed in Sec.~\ref{sec:results}. A key next step is
chiral doubling: every near-null vector $\revec{\psi_i}$ is
``doubled,'' giving
$\frac{1}{2}\left(1 \pm \gamma_5\right)\revec{\psi_i}$. For this
reason, on the coarse operator, each coarse site has $2n_{vec}^1$ internal
degrees of freedom (dof), or, alternatively, a dense structure of $n_{vec}^1$
``coarse color'' dof times two ``chirality'' dof. A successful implementation of Wilson MG critically
depends on the preservation of chirality. 

\ignore{The Wilson operator itself is an indefinite operator, which is
non-normal in the interacting case. This implies that it has separate
left and right eigenvectors,
\be
D_W \revec{\lambda^R} = \lambda \revec{\lambda^R}\quad,\quad\levec{\lambda^L} D_W = \levec{\lambda^L} \lambda\quad,\quad\revec{\lambda^R} \neq \revec{\lambda^L}.
\ee
 This concludes the standard
coarsening of the Wilson operator. }\ignore{ This methodology can be trivially
extended to a recursive coarsening. Next, we are going to
  reformulate the process of chiral doubling in a new language:
  preserving a $\sigma_1$ Hermiticity on the coarse level. This
  formulation is an essential stepping stone in Let us return to a MG coarsening of the Wilson operator. }
After performing a chiral doubling of the near-null vectors, we pack the doubled vectors into the prolongator
\be
\widetilde {P}  = 
\begin{bmatrix}
\frac{1}{2}\left(1+\gamma_5\right)\revec{\psi_i}, & 
\frac{1}{2}\left(1-\gamma_5\right)\revec{\psi_i}
\end{bmatrix}\label{eq:wilsondoubleeigen},
\ee
and again define $\widetilde{R} = \widetilde{P}^\dagger$. The {\em tilde convention} here is an
indication that we have not (yet) block orthonormalized the $2n_{vec}^1$  vectors on
each block. The chiral doubling implies $\gamma_5 \widetilde{P} = \widetilde{P} \sigma_3$, where $\sigma_3 = \mbox{diag}[1, \cdots, 1, -1, \cdots, -1]$ is a block Pauli matrix, or alternatively, the traditional $\sigma_3$ acting on the coarse chirality dof.
It is easy to see that $\widetilde{P}^\dagger D_W \widetilde{P}$ is ``$\sigma_3$'' Hermitian:
\be
\sigma_3 \widetilde{P}^\dagger D_W \widetilde{P} \sigma_3 = \widetilde{P}^\dagger \gamma_5 D_W \gamma_5 \widetilde{P}
 = \widetilde{P}^\dagger D_W^\dagger \widetilde{P} \\
= \left(\widetilde{P}^\dagger D_W \widetilde{P}\right)^\dagger \; .
\ee
The essential property $\gamma_5 \widetilde{P} = \widetilde{P}
\sigma_3$ is unchanged after we perform the last step, block
orthonormalizing $\widetilde{P}$ to get $P$, because we performed our
chiral doubling with a {\emph{bona fide}} projector. The top chiral
components and the bottom chiral components are already trivially
orthonormal. This gives the final essential properties $\gamma_5 P = P \sigma_3$, and $\widehat{D}_W = P^\dagger
D_W P$ is $\sigma_3$ Hermitian. This methodology can be trivially
extended to a recursive coarsening.

\subsection{\label{sec:multigrid}Failure of Galerkin Projection of
  Staggered Operator}

\begin{figure}[t] \centering
\includegraphics[width=0.9\linewidth]{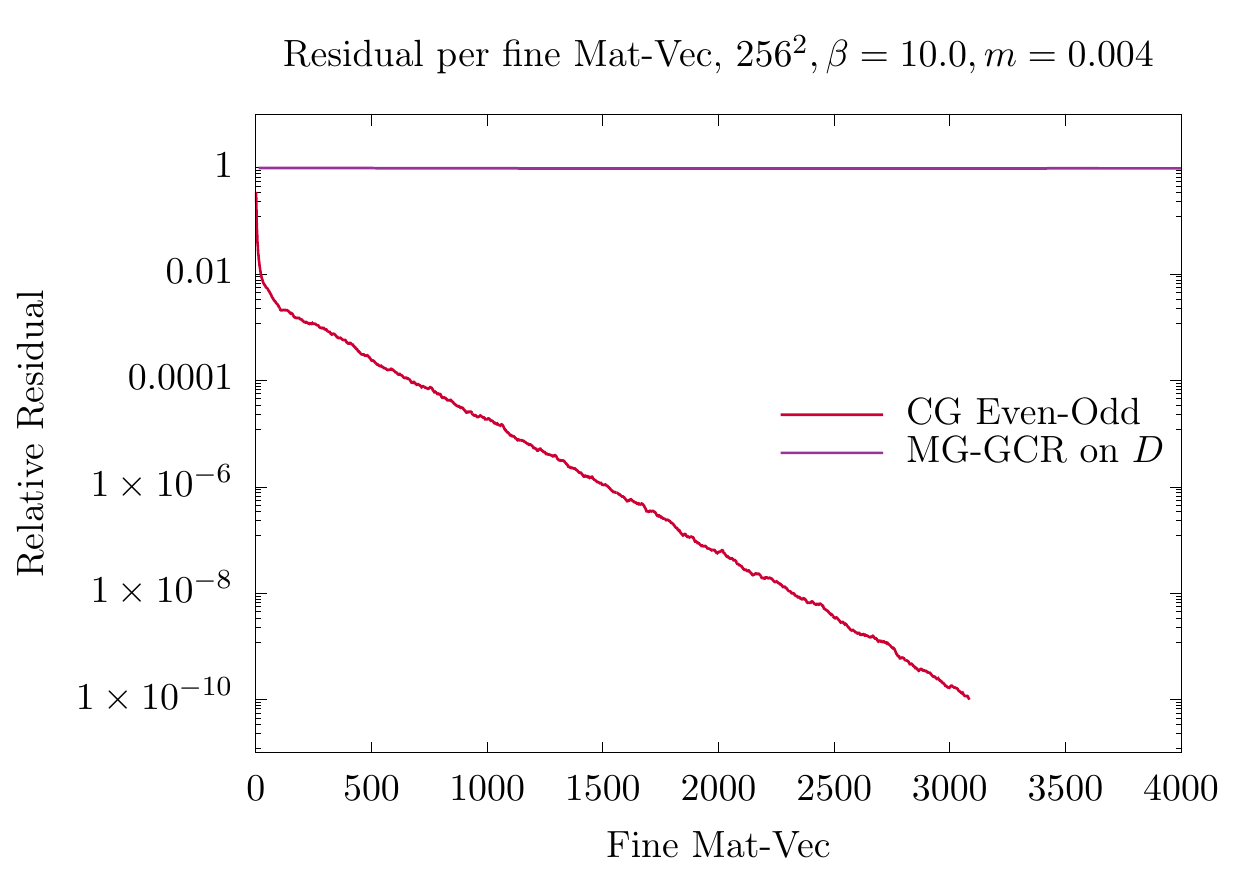}
  \caption{\label{fig:residstag}The relative residual as a function of
    fine $D$ applications for CG on the even/odd system compared with a failing MG solve on the staggered operator. The parameters of the MG solve are given in Table~\ref{tab:kcycle}.}
\end{figure}

The prescription for (recursively) generating a coarse refinement of
the Wilson operator $D_W$ fails when na\"ively translated to the
staggered operator $D$ with the only change being the replacement of $\gamma_5$ with $\epsilon(x)$, as noted by
Eq.~\ref{eq:epsilon}. While the iterative inversion of the
even/odd preconditioned system exhibits critical slowing down, it
does converge. However this attempt at a Galerkin MG on the 
staggered operator $D$ stalls completely at large volumes as  illustrated in
Fig.~\ref{fig:residstag}.  
We need to understand the cause of the failure of the
Galerkin projection $\widehat{D} = P^\dagger D P$  as a preconditioner. 
A  MG algorithm may fail because  the
coarse operator does not accurately reproduce the low eigenspace of the fine operator, or because the coarse error ``correction''
is ineffective.  We will study each of these properties for the
staggered operator to attempt to understand the issue.

\begin{figure}[t] \centering
  \includegraphics[width=.87\linewidth]{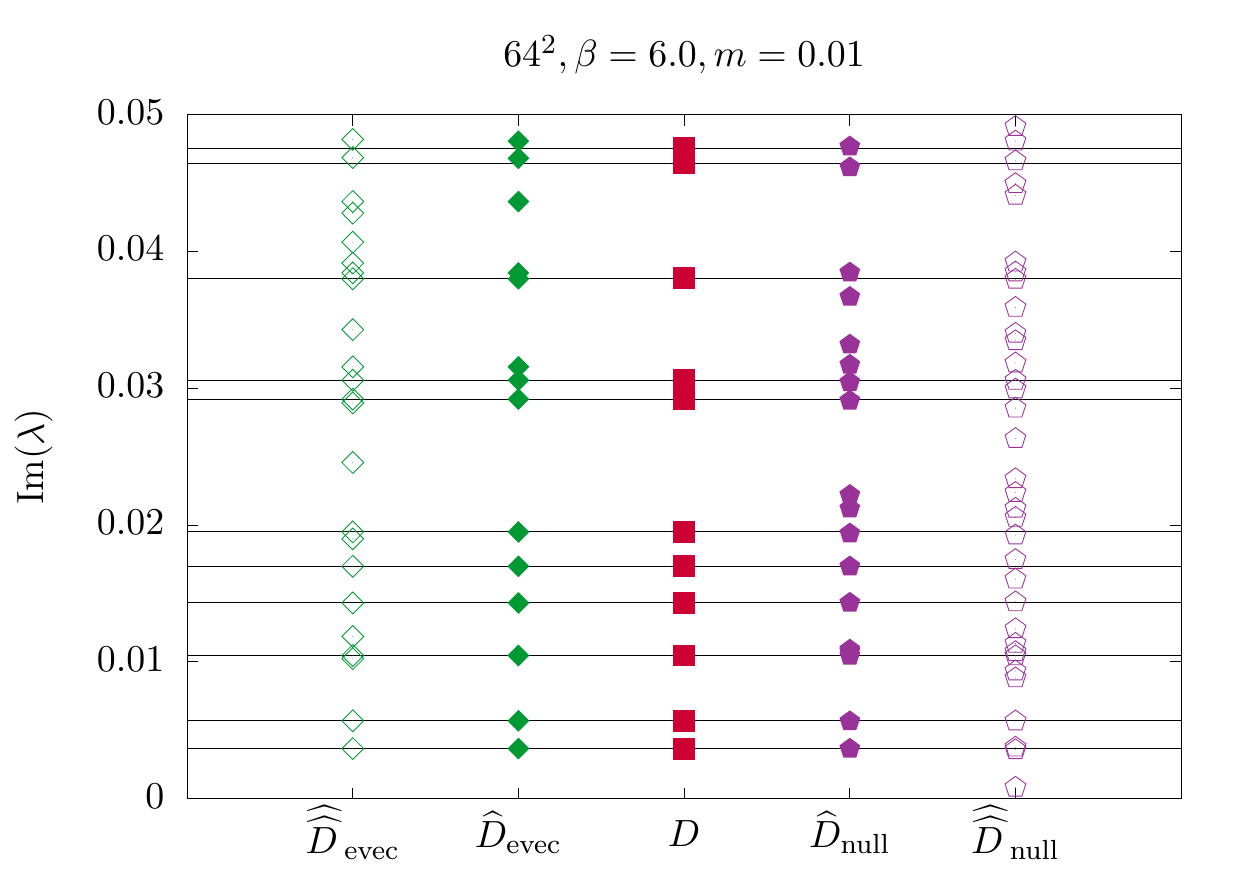}
  \caption{The imaginary part of the spectrum of $D$ for a representative configuration using a recursive symmetric Galerkin projection with both eigenvectors (left) and generated near-null vectors (right). Recursive coarsening introduces ``spurious'' low modes. The parameters of the MG coarsening are given in Table~\ref{tab:kcycle}.}
\label{fig:spectrumStag}
\end{figure}

\begin{figure}[t] \centering
\includegraphics[width=0.87\linewidth]{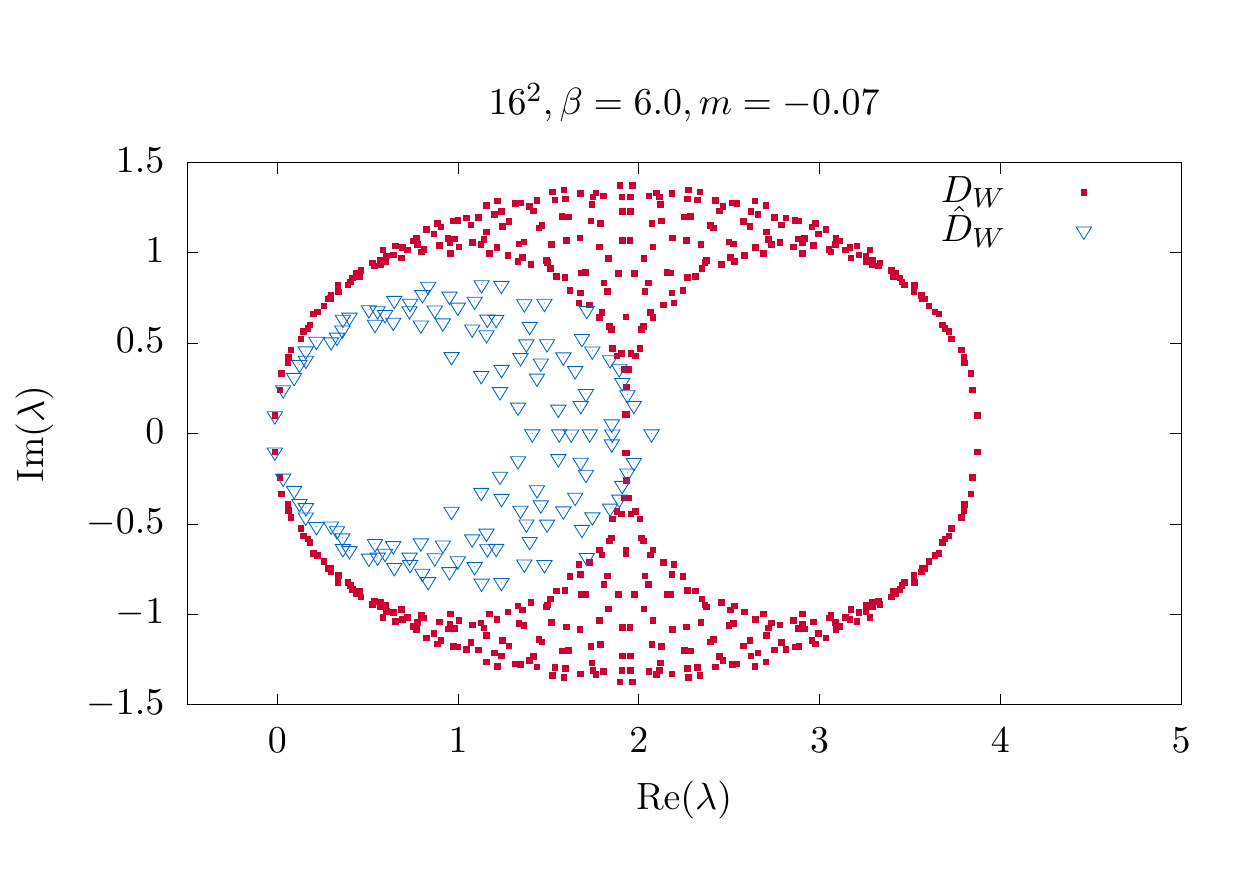}
  \caption{A representative spectrum of an interacting Wilson operator and its Galerkin coarsening. The MG coarsening aggregates fine degrees of freedom over a $4^2$ block into a coarse site using $n_{vec}^1 = 4$ near-null vectors.}
\label{fig:spectrumWilson}
\end{figure}

As a spectral preconditioner, we expect the coarse operator to
approximately preserve the low eigenmodes of the fine operator. In
Fig.~\ref{fig:spectrumStag} we address this issue by comparing our failed
staggered MG spectra with the
  successful Wilson MG  spectra in Fig.~\ref{fig:spectrumWilson}. 

First consider the staggered case. The center  column (in red)  in
Fig.~\ref{fig:spectrumStag} gives the positive
imaginary component of the low-lying spectrum for the fine staggered
operator. The spectrum is exactly paired with complex conjugate
eigenvalues below  the real
  axis due to $\epsilon(x)$ Hermiticity on the fine level
and coarse levels.  The other four columns give the spectrum for a recursively-coarsened operator, constructing the prolongator/restrictor
from exact low eigenvectors (left side) and near-null
vectors (right side), where the near-null vectors are again generated as discussed in Sec.~\ref{sec:results}. Filled shapes correspond to the first coarse
level. Hollow shapes correspond to the operator from a recursive
coarsening. The horizontal black lines trace the low modes of the fine
operator across the coarsened operators. While these physical low modes are well preserved in
all cases, there many additional, {\emph{spurious}} low eigenvalues in
the coarse spectrum.

These spurious eigenvalues have a simple origin. Consider the normalized eigenvector of the coarse operator
$\hat D\revec{\hat\lambda} = \hat\lambda\revec{\hat\lambda}$. We note
that
$\hat\lambda = \levec{\hat\lambda}\hat{D}\revec{\hat\lambda} =
\left(\levec{\hat\lambda}P^\dagger\right)D\left(P\revec{\hat\lambda}\right)$. If
we perform an eigendecomposition of
$P\revec{\hat\lambda} = \sum_{i} c_i \revec{\lambda}$, we find that
$\hat\lambda = \sum_i \left|c_i\right|^2 \lambda$, a consequence of $D$ being
normal. This implies $\hat\lambda$ is some linear combination of
eigenvalues of $D$ in the interval $\left[-\lambda_{\mbox{max}},
  \lambda_{\mbox{max}}\right]$.  A coarse eigenvalue can have an
arbitrarily small imaginary part if, for example,
$P\revec{\hat\lambda}$ is dominantly spanned by a pair of fine
eigenvectors with complex conjugate eigenvalues. This eigenvector may
have nothing to do with the low modes of $D$.

This would be less of an issue if higher modes were gapped along the real axis. This is true of the Wilson operator, as can be seen for a representative case in Fig.~\ref{fig:spectrumWilson}\footnote{In the interacting case, the Wilson operator is no longer normal, and our convex hull proof breaks down. It appears that it is still sufficiently true, perhaps because
  free Wilson operator is exactly normal. In a perturbative sense, the
  interacting Wilson operator is then ``approximately'' normal.}. For the
fine operator, whose eigenvalues are given by red squares, high modes are gapped along the {\emph{real}} axis. For
the coarse operator, whose eigenvalues are blue triangles, low modes are well preserved. Higher modes
``collapse'' towards the complex origin but are still well gapped
along the {\emph{real}} axis. This could be why MG on the
Wilson operator does not break down, and may identically predict
success for the K\"ahler-Dirac preconditioned operator.

\begin{figure}[t] \centering
{\large{$64^2, \beta = 6.0, m = 0.01$}}\\
  \includegraphics[width=.47\linewidth]{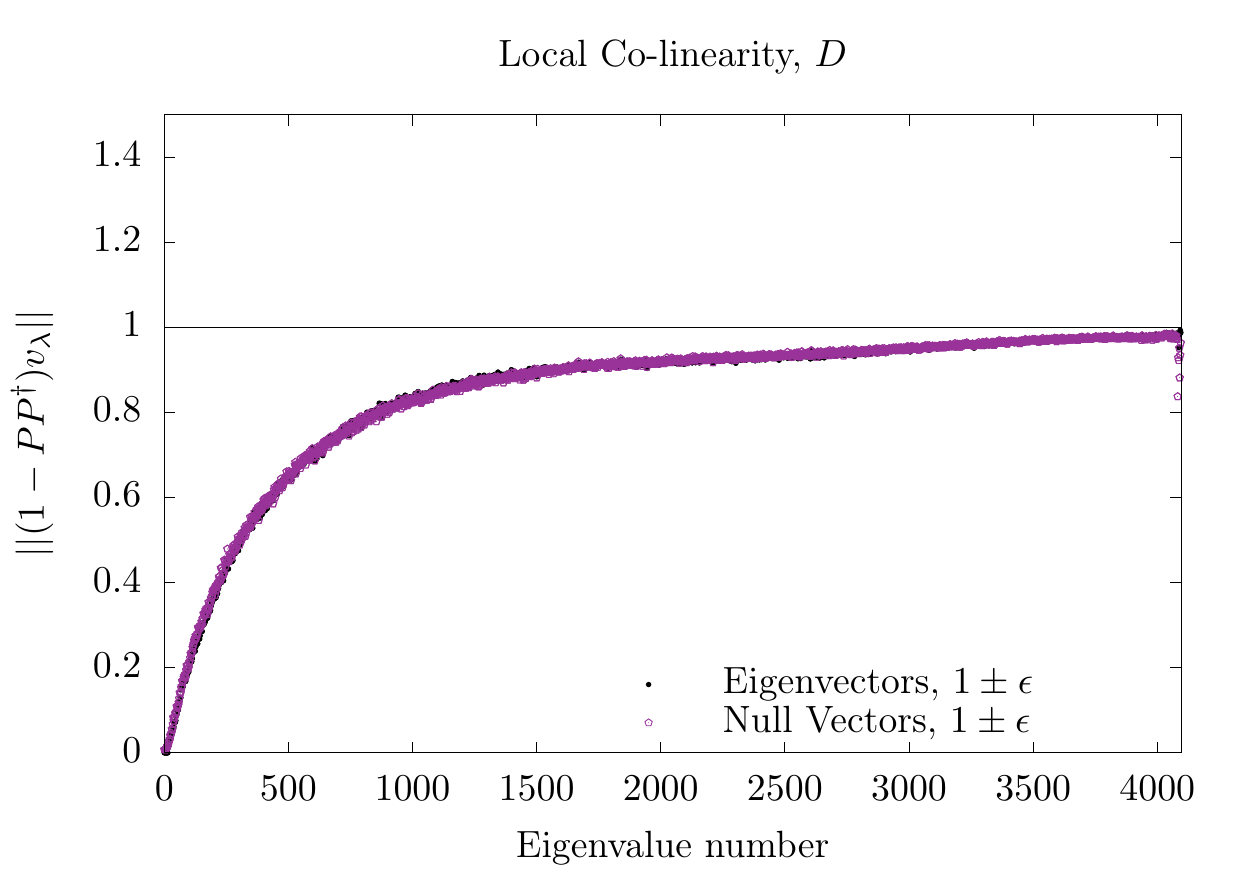}
~\includegraphics[width=0.47\linewidth]{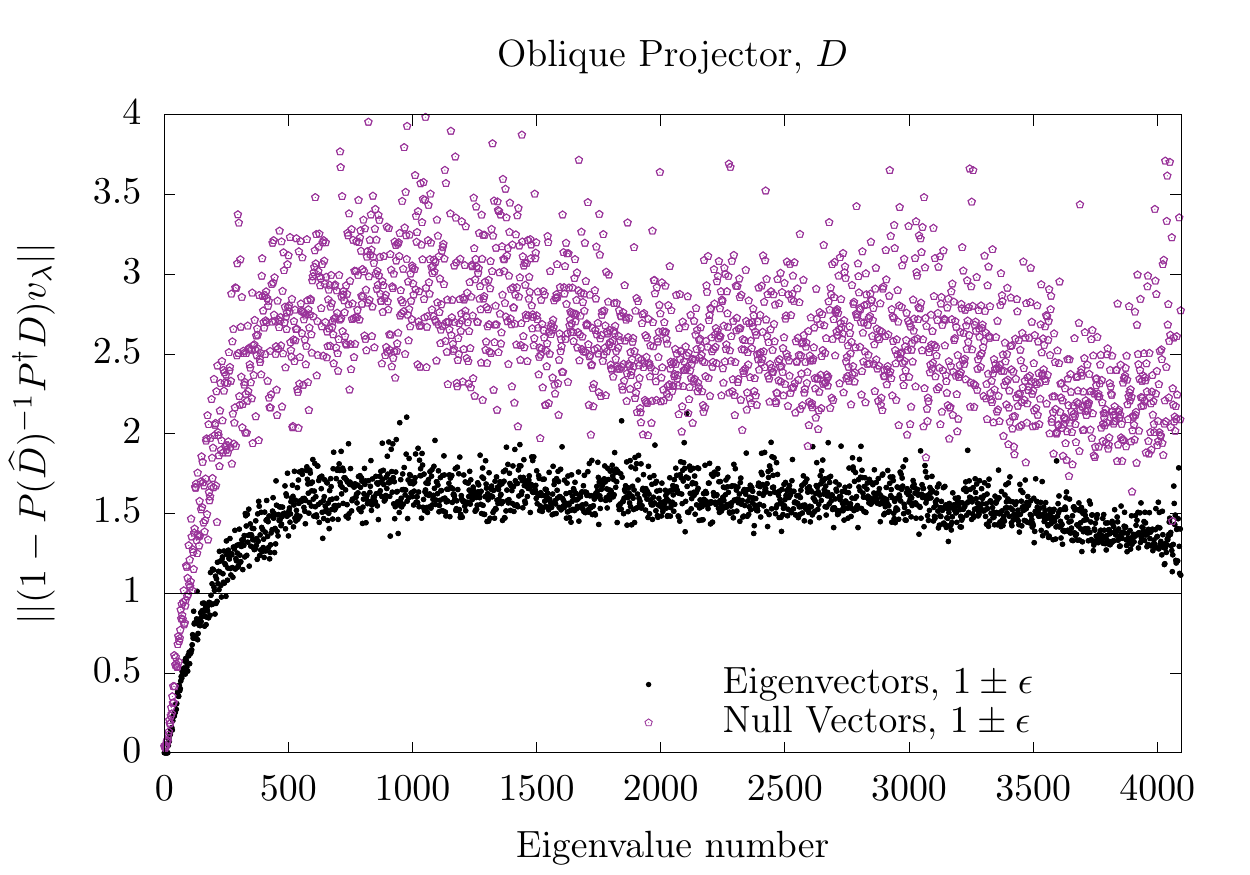}\\\includegraphics[width=.47\linewidth]{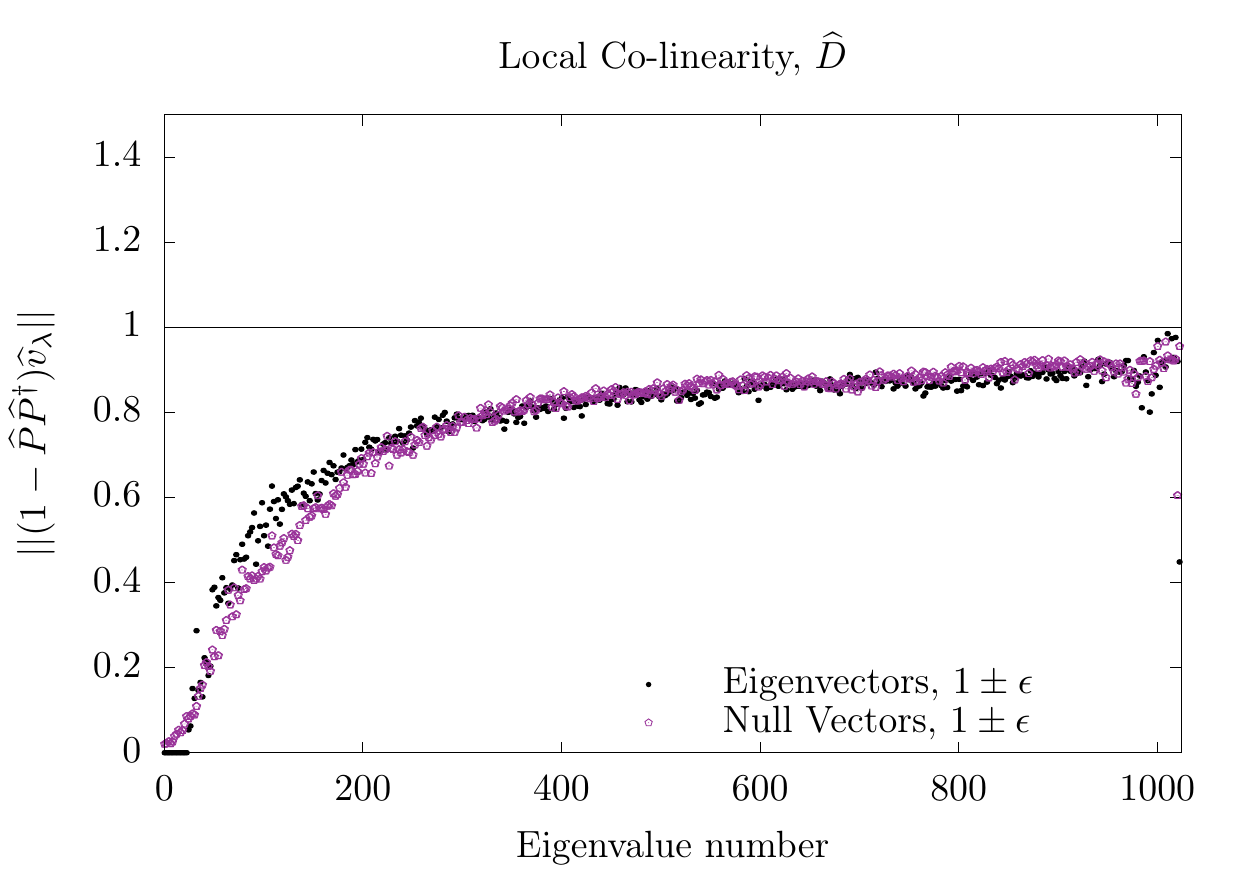}
~\includegraphics[width=0.47\linewidth]{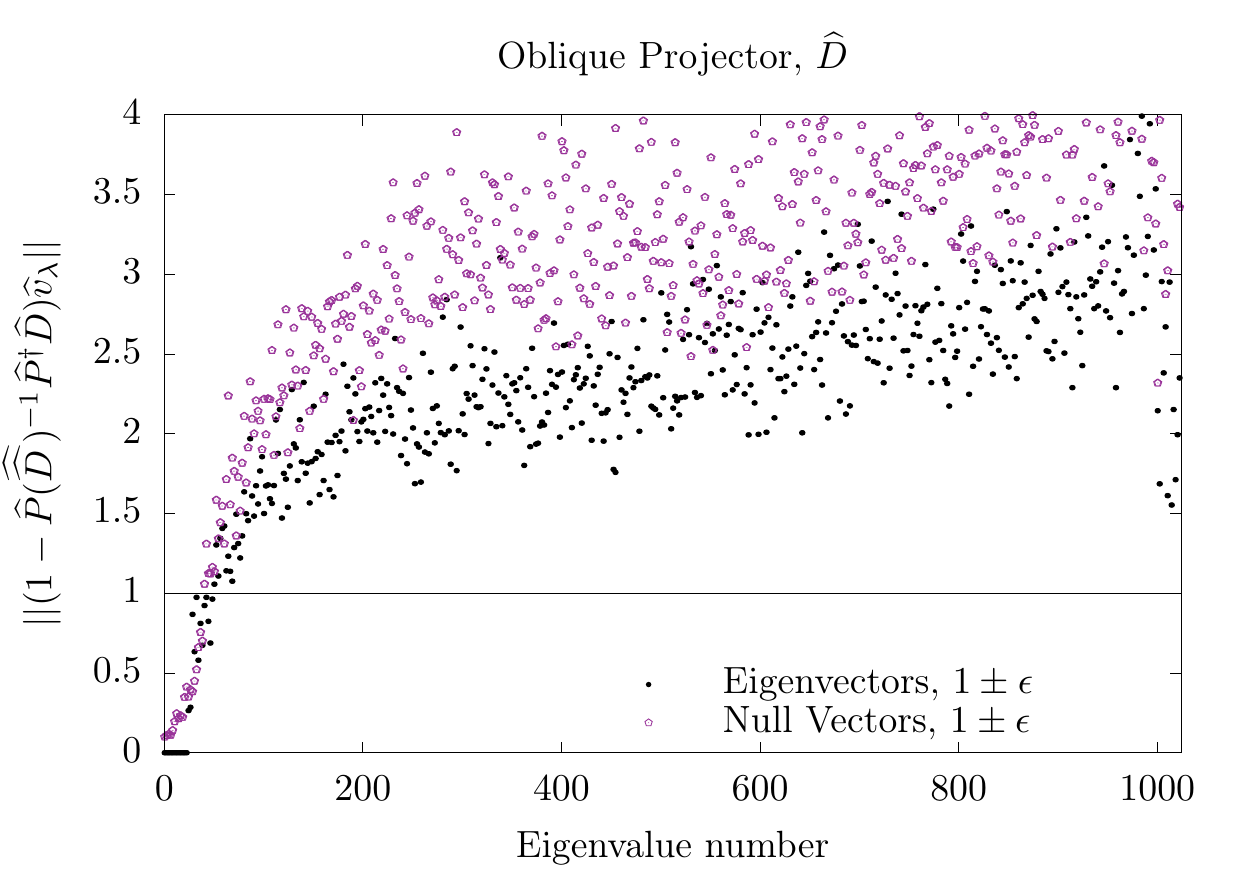}
  \caption{On the left, a measurement of local co-linearity: how well
    low near-null vectors can reconstruct  eigenvectors. On the right,
    the contribution of eigenvector to the error after the coarse
    level solve. The top row of figures is the fine-to-first-coarse operator and the
    bottom row is the next level from the coarse to doubly-coarsened operator. The parameters of the MG coarsening are given in Table~\ref{tab:kcycle}. Both panels are sorted by increasing magnitude of the eigenvalues.}
\label{fig:projectorsspectrum}
\end{figure}

\paragraph{Local Co-linearity versus the Oblique Projector}
The Galerkin MG scheme  involves  two
 different projection operators: 
\begin{itemize}[noitemsep, topsep=0pt]
\itemsep0em 
\item The projection operator, ${\mathbb P}
 =  PR$, from the fine space into a coarse subspace. Using the right eigenvectors as a basis for the fine vector space, $V = \{\revec{v_\lambda}, 0 <
|\lambda| \leq|\lambda_{max}|\}$, our goal is for eigenvectors with small (near-null) eigenvalues, $|\lambda|/|\lambda_{max}| < \varepsilon$, to be approximately represented within the span of the coarse subspace, $\widehat V =
{\mathbb P} V$, in a least-squares sense.
\item The oblique  (or Petrov–Galerkin) projector,
${\mathbb P}_{ob}  = 1 - P\left(R D P\right)^{-1} R
  D$, that defines space of error components that are returned to the
fine level with a complete solve in the coarse subspace. To not
overburden the smoother this should at least not  unduly amplify
large eigenvectors, $|\lambda|/|\lambda_{max}| > \varepsilon$. 
\end{itemize}
 Both are  true projectors dividing the fine vector space $V$ into disjoint subspaces, $ {\mathbb P} (1 -  {\mathbb P} ) =
0$  and ${\mathbb P}_{ob}(1 - {\mathbb P}_{ob}) = 0$, though they do not
define the same subspaces. The orthogonality, 
${\mathbb P}_{ob} {\mathbb P} = 0$ is one-sided  since $ {\mathbb
  P}{\mathbb P}_{ob} \ne 0$.

Let  us see how well the staggered MG handles these two
requirements. In our construction, $R = P^\dag$, so the 
coarse space projector is Hermitian. The statement of preserving the low eigenspace in the least-squares sense can be formulated as sufficiently minimizing
\begin{align}
\left|\left| \left(1 - P R\right)\vec{v}_\lambda\right|\right|_2,
\end{align}
for small eigenvalues of fine operator $D$. Since we
generate our coarse space by geometric aggregation, this can be
thought of as the {\emph{local co-linearity}} of near-null vectors with low eigenmodes.  In the top left panel of
Fig.~\ref{fig:projectorsspectrum}, we see that  starting either with a block-orthonormalized basis of near-vectors or of low eigenvectors results in a good converage of the low spectrum. This is typical of MG methods. At the bottom left this is extended to the 
next coarsest level with similar result. This
has important implications for eigenvector compression methods~\cite{Clark:2017wom}.

However, this is not sufficient for a successful coarse correction in
a MG algorithm. The coarse correction should address the low modes of
the fine operator without introducing large errors in the high mode
subspace.  The error after solving the coarse level is updated as
$e \leftarrow e - P\left(R D P\right)^{-1} R
  D e = {\mathbb P}_{ob} e$. This is quantified by the magnitude of each 
eigenvector acted on by the Petrov–Galerkin or  the so-called {\emph{oblique projector}},
\begin{align}
\left|\left| \left(1 - P\left(R D P\right)^{-1} R
  D\right)\vec{v}_\lambda\right|\right|_2 \; .
\end{align}
The oblique projection of the coarse error (${\mathbb P} e$)  is zero:
${\mathbb P}_{ob} {\mathbb  P}  = [1 - P\left(R D P\right)^{-1} RD] PR
= 0$. However, the oblique projection is not Hermitian so this does
not imply the error in the  orthogonal complement space ($e - PR e$) 
vanishes.  This is illustrated on the right side of Fig.~\ref{fig:projectorsspectrum}.

A magnitude less than or greater than one corresponds to a reduction
or enhancement of the complementary error component, respectively. A
successful coarse operator should strongly reduce the error component
for low eigenmodes. In the context of MG, the enhancement from higher modes is addressed by the smoother. A larger enhancement requires a
more expensive smoother, otherwise the solve stalls. In the top right panel of
Fig.~\ref{fig:projectorsspectrum}, we see that, for high modes, there
is a large error enhancement. This is worse for a prolongator
generated from near-null vectors than one generated from
eigenvectors. In the lower right panel, we see the situation is even worse
for a three-level algorithm. In all cases, an aggressive smoother is
needed, increasingly so at coarser levels. This is why we saw
the MG algorithm fail.  Now we turn to the same analysis for the
Galerkin construction of the K\"ahler-Dirac preconditioned operator,
which has in contrast minimal error enhancement, evident in
Fig. \ref{fig:projectorsspectrumcircle}.

\subsection{Coarse K\"ahler-Dirac  Staggered operator: $\widehat{A}$\label{sec:projkdp}}

We will coarsen the K\"ahler-Dirac preconditioned operator similarly to how we coarsened the staggered operator, still using $\frac{1}{2}\left(1 \pm \epsilon(x)\right)$ (unitarily rotated into the flavor basis) as a chiral projector on the near-null vectors. 
We will denote the method of coarsening the K\"ahler-Dirac
preconditioned operator using $\frac{1}{2}\left(1 \pm \epsilon(x)\right)$
as chiral projectors. Again, we will use $R = P^\dagger$. In Sec.~\ref{sec:hermpres}, we will
discuss an {\emph{asymmetric}} coarsening where $R \neq P^\dagger$. While not being of
merit in two dimensions, it may be an interesting point of investigation
in four-dimensional QCD. In this section we will consider the
spectrum, co-linearity, and oblique projector for a symmetric
coarsening. Looking forward, in Sec.~\ref{sec:results}, we will
demonstrate that symmetric coarsening produces a well behaved and
robust recursive algorithm independent of the volume and the mass for physically relevant values of $\beta$.

\begin{figure}[t] \centering
   \includegraphics[width=0.47\linewidth]{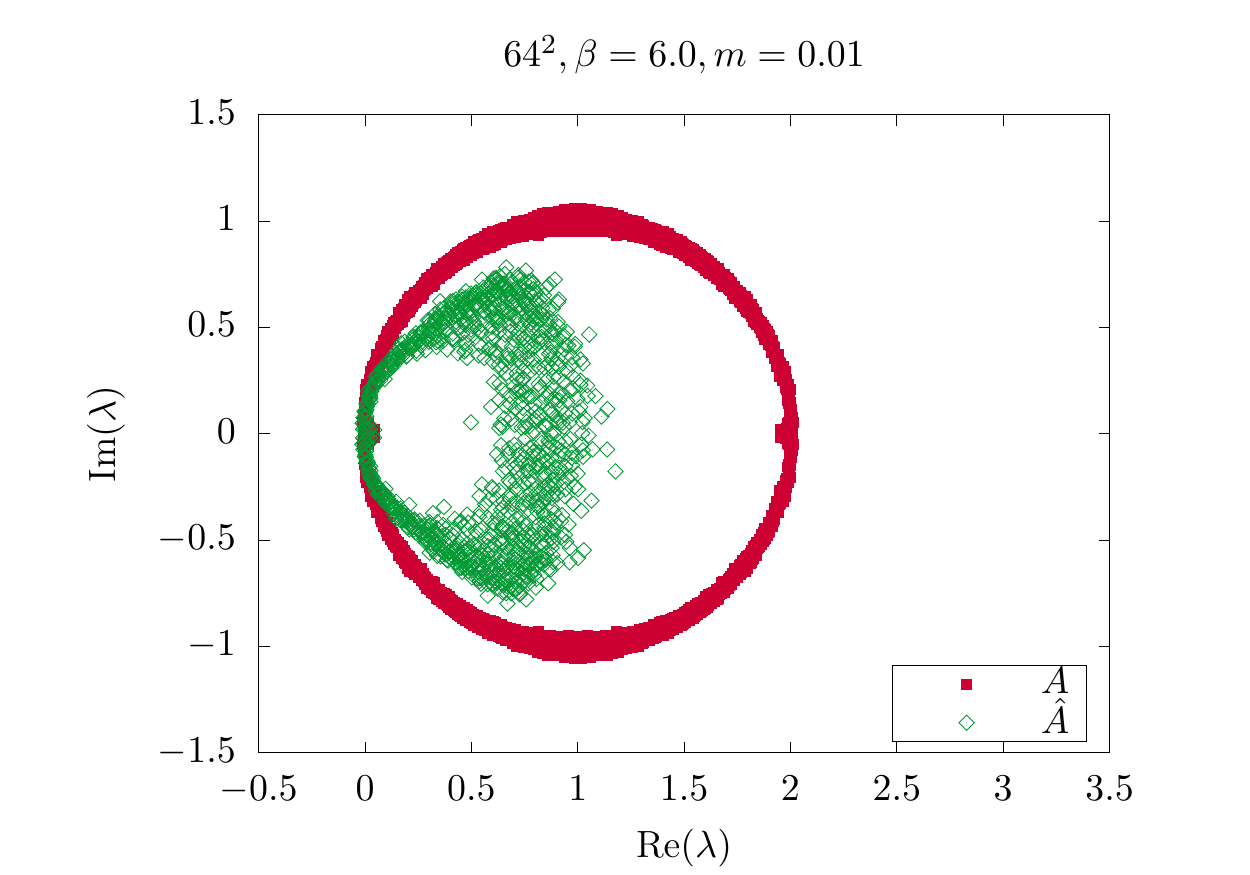}~ \includegraphics[width=0.47\linewidth]{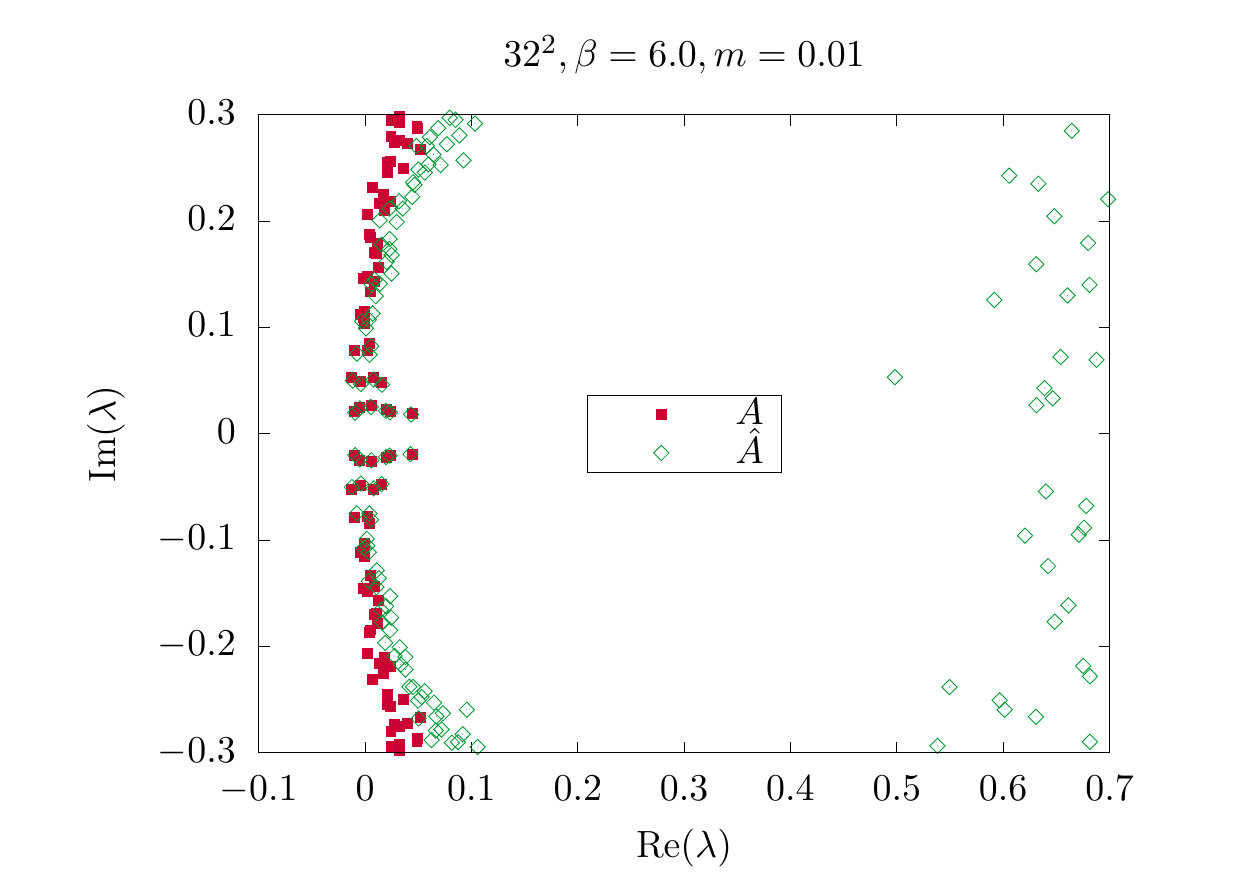}
  \caption{On the left, a representative spectrum of the K\"ahler-Dirac preconditioned operator and its Galerkin projection from an interacting gauge field. On the right, a zoom-in on the low spectrum. The parameters of the MG coarsening are given in Table~\ref{tab:kcycle}.}
\label{fig:zoomcircle}
\end{figure}

\ignore{
\begin{figure}[t] \centering
   \includegraphics[width=0.47\linewidth]{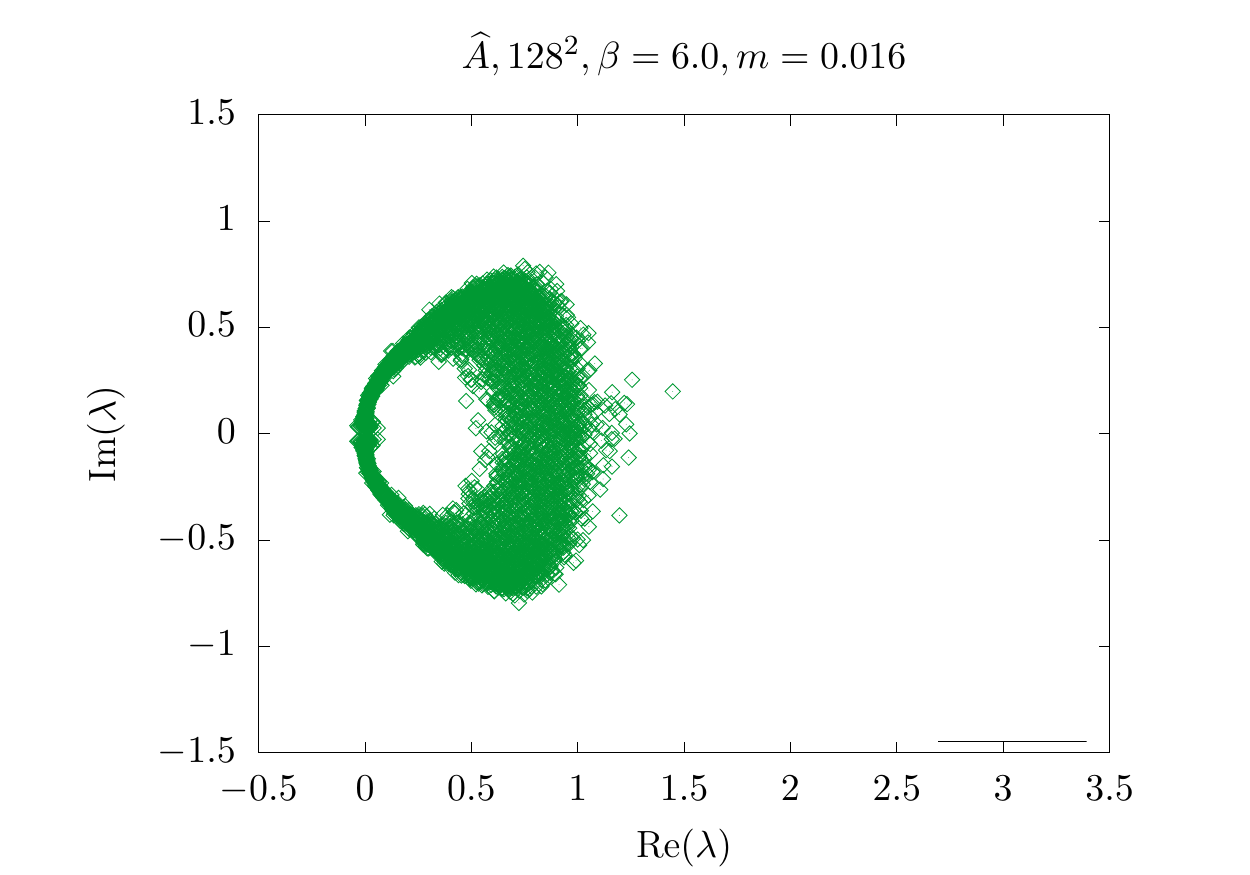}~ \includegraphics[width=0.47\linewidth]{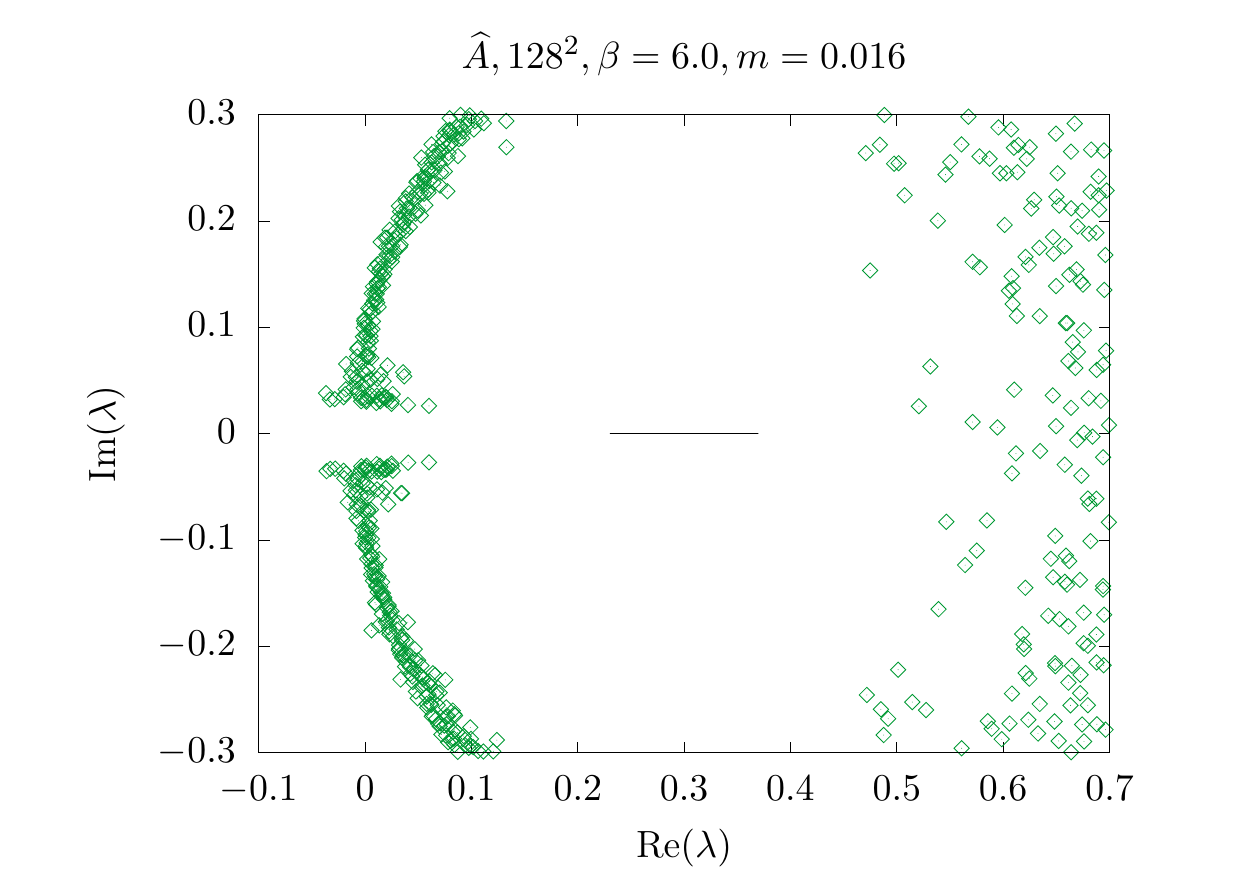}
  \caption{On the left, the spectrum of a coarsened K\"ahler-Dirac preconditioned operator, and on the right, the low spectrum. Computing the fine spectrum was prohibitively expensive.}
\label{fig:zoomcirclebigger}
\end{figure}}

As we described previously, the Wilson operator has a well behaved spectrum for MG as the high modes are well gapped along the real axis. This is also true for the K\"ahler-Dirac
preconditioned operator in Fig.~\ref{fig:zoomcircle}. As we discussed
at the end of Sec.~\ref{sec:transform}, the interacting spectrum is no longer a
perfect circle in the complex plane. This does not undermine the qualitative benefits of the spectrum. Additionally, in the interacting case, a mass term still gaps the spectrum. In the right panel of Fig.~\ref{fig:zoomcircle}, where we zoom in on
the origin of the complex plane, we see that low modes are well
preserved under our coarsening prescription, and there are no spurious modes near the complex origin. Eigenvalues of the coarse operator do not come in exact complex conjugate
pairs, a consequence of using $\epsilon(x)$ as the chiral projector. This is inescapable because, in general, $\gamma_5^{L/R}$ does not define a good projector. The eigenvalues are {\emph{approximately}} paired, which is
consistent with a general preservation of the low spectrum. This may
also be consistent with $\frac{1}{2}\left(1 \pm \epsilon(x)\right)$
becoming equivalent to $\frac{1}{2}\left(1 \pm \gamma_5^{L/R}\right)$,
up to a unitary transformation, in the continuum limit, and as such
preserving complex conjugate eigenpairs.

A careful study of the right panel of Fig.~\ref{fig:zoomcircle} shows
that both the original operator and its coarsening feature eigenvalues
with negative real part, that is, lying in the left-half plane. We
refer to these eigenvalues as {\emph{exceptional}} eigenvalues,
borrowing the language from Wilson-clover fermion
literature~\cite{DeGrand:1998db}. The existence of modes in the left-half plane invalidate proofs which bound the convergence of Krylov
solvers~\cite{doi:10.1137/0907058}. We will see in
Sec.~\ref{sec:results} that, because error components in these exceptional modes are well solved by the coarse error correction, a recursive MG algorithm can successfully address this problem. As we will see in Sec.~\ref{sec:results}, this stabilizes the MG solve, independent of mass and volume, and is
consistent with the success of MG for the Wilson operator beyond the
critical mass.

\ignore{
Indeed, in Fig.~\ref{fig:zoomcirclebigger}, we see an increased density of exceptional eigenvalues from a $128^2$ configuration.\footnote{We only present the spectrum of the coarse operator, $\widehat{A}$. Given the cubic scaling of
  eigenvalue problems, we elected to not spend a month calculating the
  full spectrum of the fine operator.} The scaling has been chosen to
match Fig.~\ref{fig:zoomcircle}. In the right panel, we see a large
number of eigenvalues in the left-hand plane, enclosing the complex
origin more than in the $32^2$ case. Despite this potentially
enhanced concern, }

\begin{figure}[t] \centering
{\large{$64^2, \beta = 6.0, m = 0.01$}}\\
  \includegraphics[width=.47\linewidth]{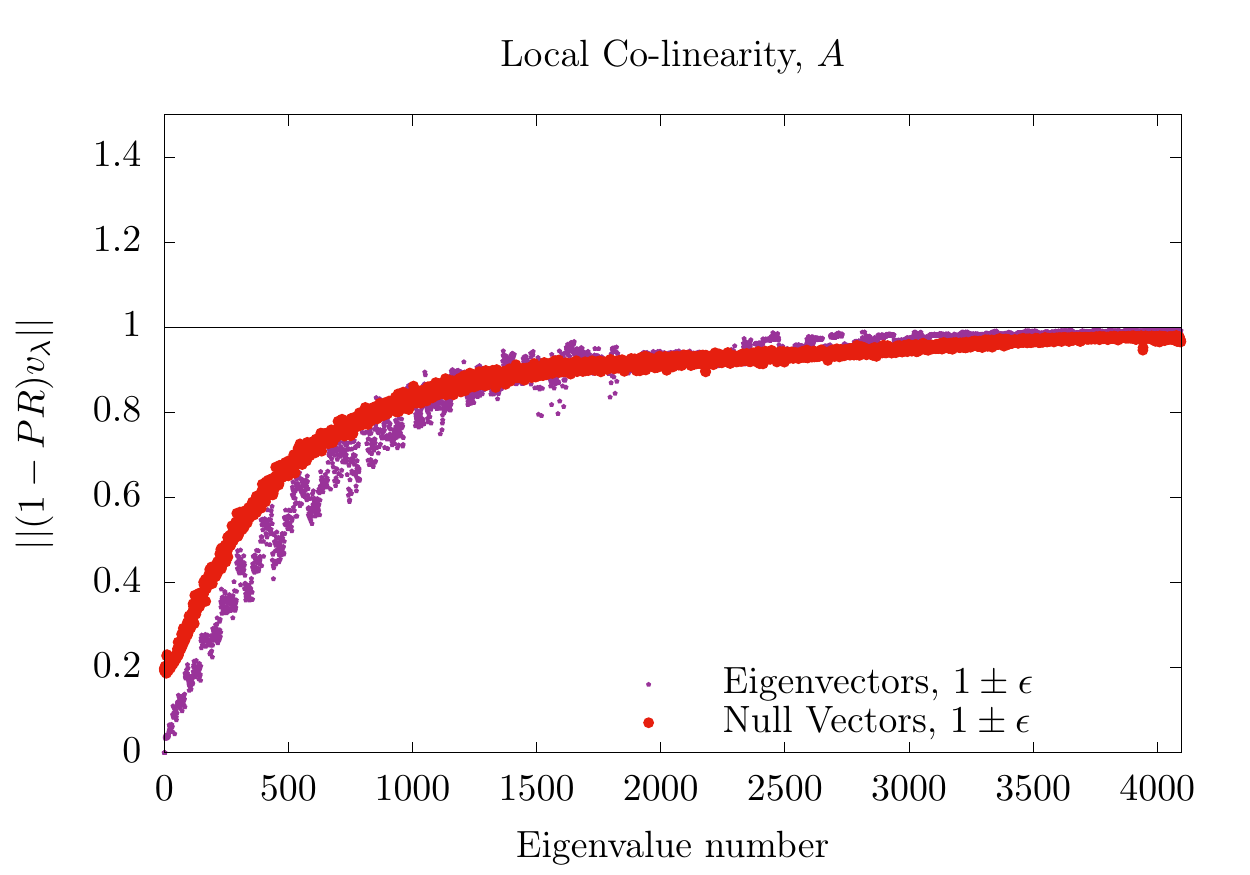}
~\includegraphics[width=0.47\linewidth]{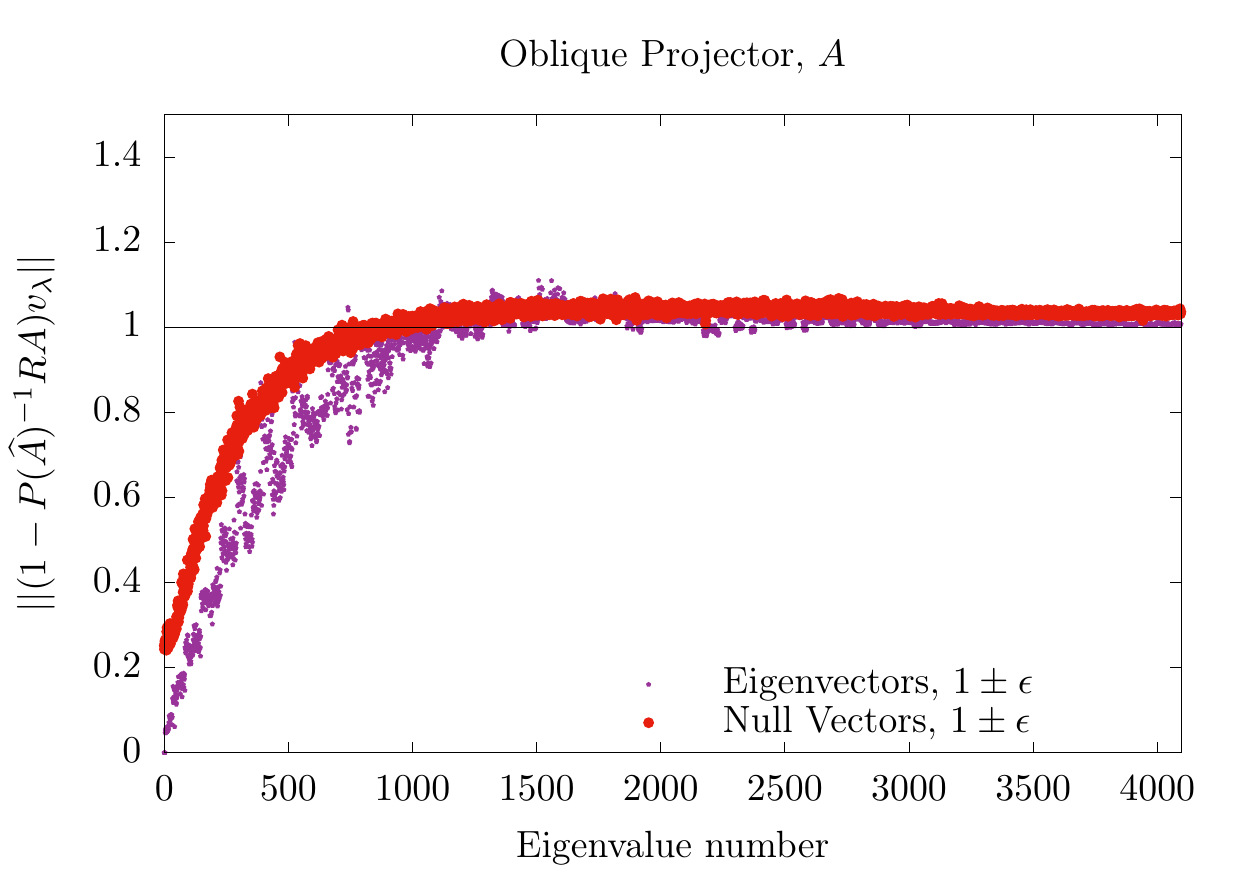}\\\includegraphics[width=.47\linewidth]{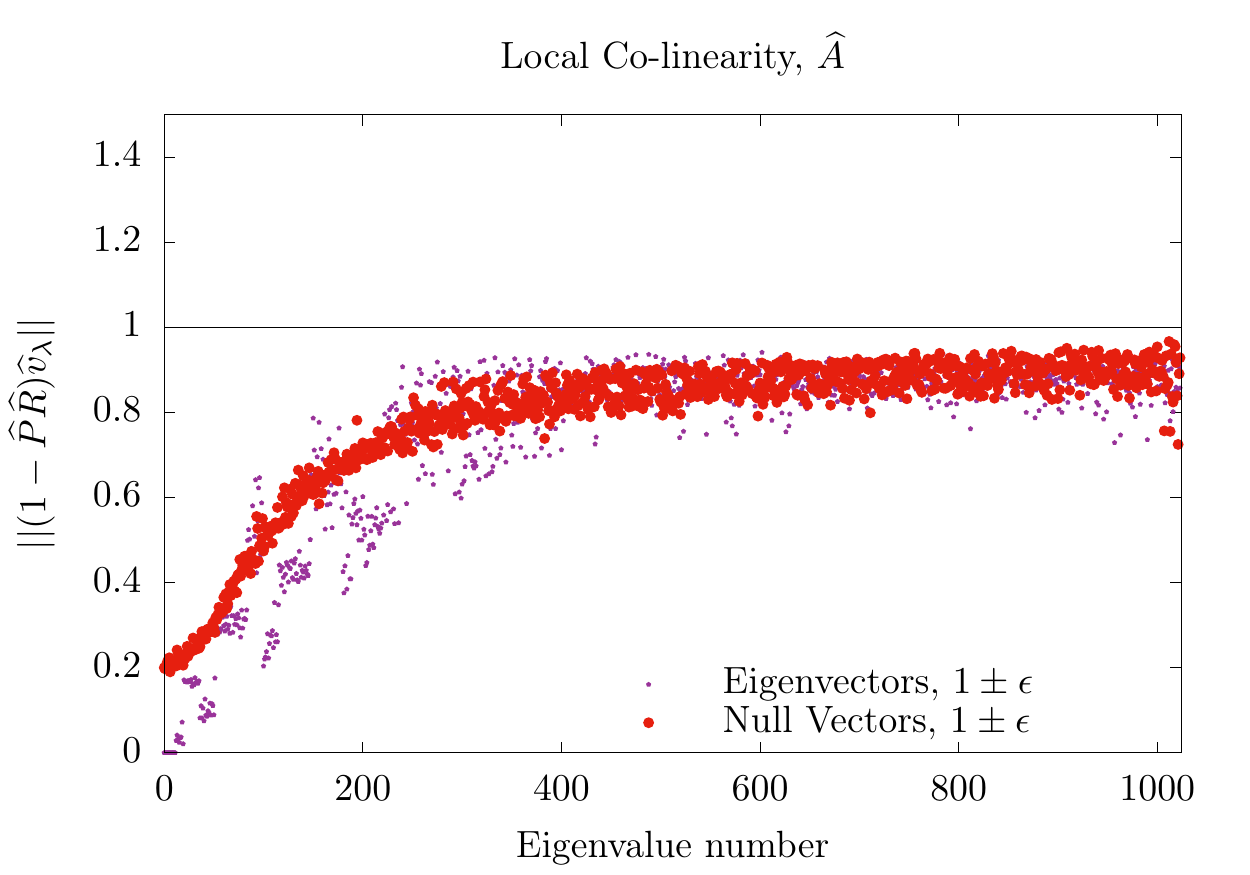}
~\includegraphics[width=0.47\linewidth]{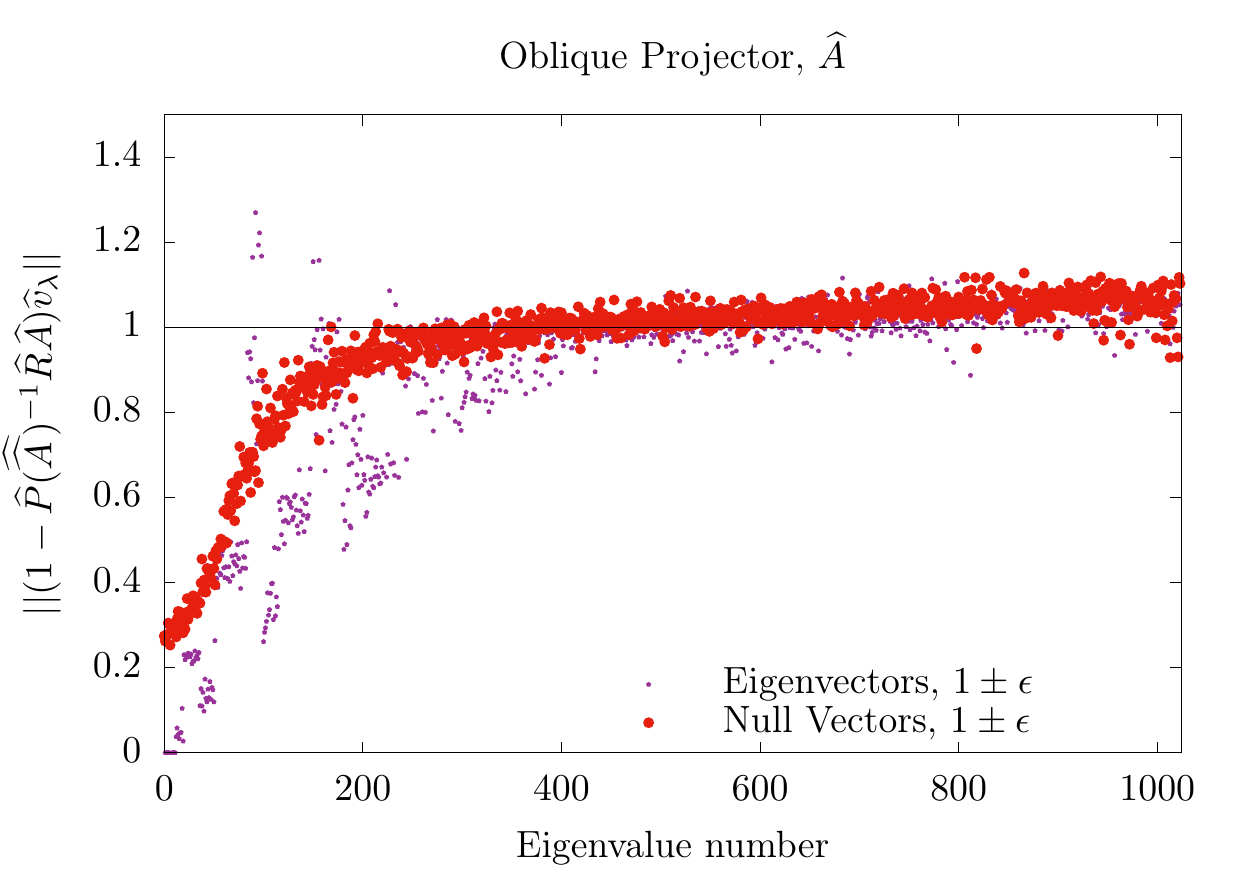}
  \caption{On the left, a measurement of local co-linearity, and on the right, the effect of the oblique projector on eigenvectors. The top row considers a representative K\"ahler-Dirac preconditioned operator; the bottom row considers its coarsening. Both panels are sorted by increasing magnitude of the eigenvalues. The parameters of the MG coarsening are given in Table~\ref{tab:kcycle}.}
\label{fig:projectorsspectrumcircle}
\end{figure}

\paragraph{Local Co-linearity versus the Oblique Projector}
The overall failure of MG for the staggered operator stemmed from the large error enhancement to the high modes from the coarse correction. A predictor of success for MG on the K\"ahler-Dirac preconditioned operator would be a significant reduction of this enhancement. We would also still need to see strong local co-linearity and a significant coarse error correction on low modes.  In the left and right panels of
Fig.~\ref{fig:projectorsspectrumcircle}, we consider the local
co-linearity and oblique projector, respectively, of the K\"ahler-Dirac preconditioned operator on a representative configuration. We explore using both near-null vectors and right eigenvectors to define the prolongator $P$ and
restrictor $R = P^\dagger$.

On the left, we see that local co-linearity of low modes of the K\"ahler-Dirac operator is well maintained, similar to the original staggered operator. The benefit of
coarsening the K\"ahler-Dirac preconditioned operator as opposed to
the original staggered operator is most clearly noted by the action of
the oblique projector as displayed on the right panel of
Fig.~\ref{fig:projectorsspectrumcircle}. The oblique projector reduces the error component on
the fine level for roughly the lowest 15\% of the spectrum. Above this
threshold, the error component is enhanced, but only minimally.

\newpage
\section{\label{sec:results}MG Algorithm Numerical Results}

\begin{figure}[t] \centering
\includegraphics[width=0.9\linewidth]{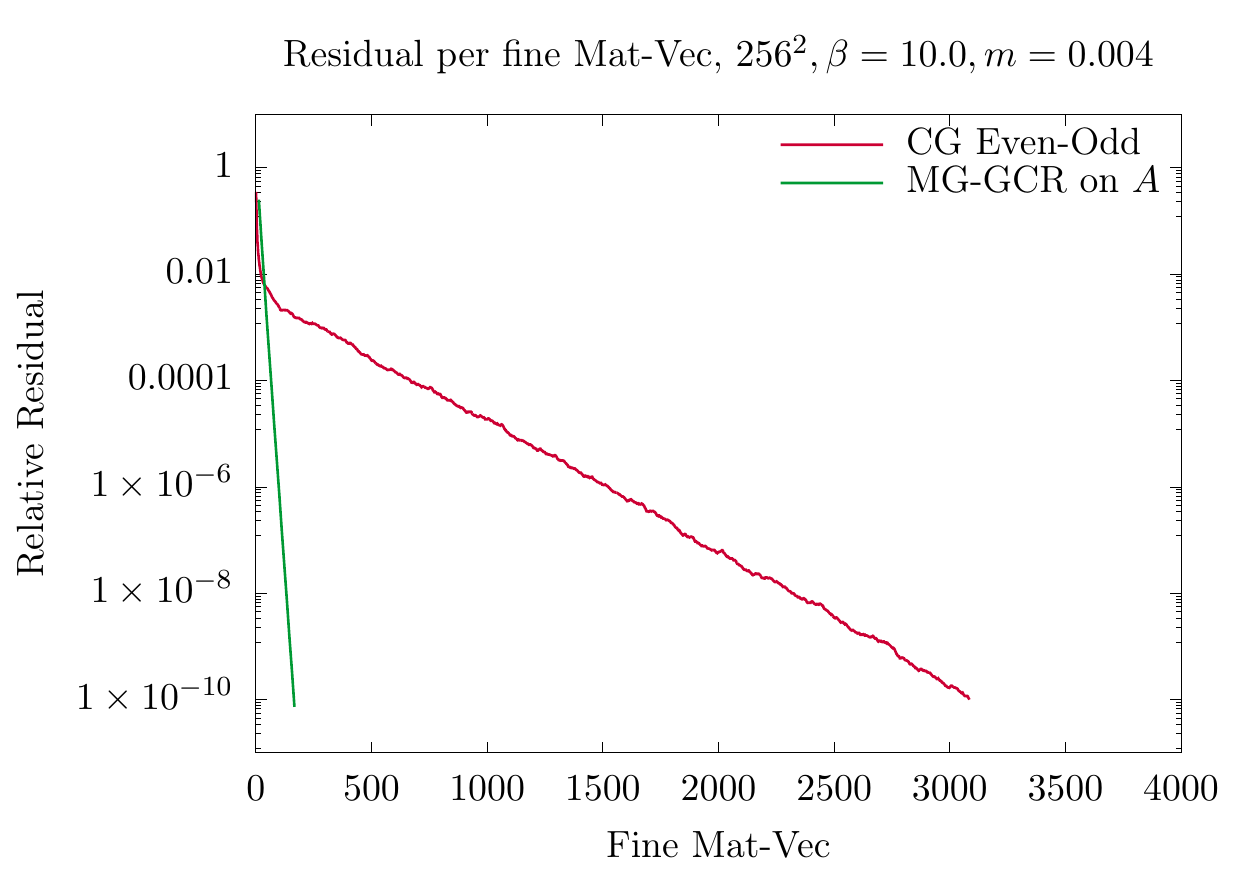}
  \caption{\label{fig:resid}The relative residual as a function of the number of fine operator applications for a representative $\beta = 10.0,~256^2$ configuration. The parameters of the MG solve are given in Table~\ref{tab:kcycle}.}
\end{figure}

\begin{figure}[t] \centering
  \includegraphics[width=0.9\linewidth]{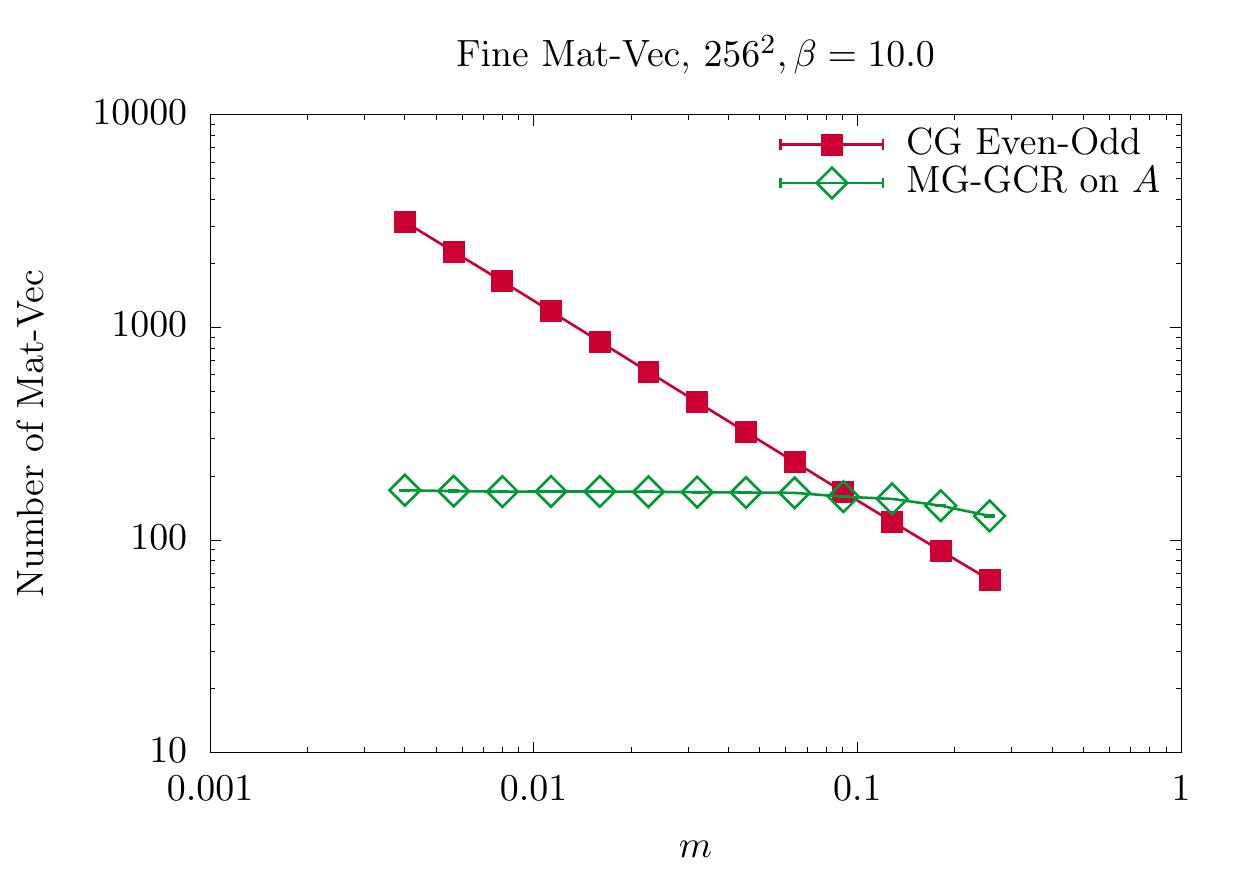}
  \caption{\label{fig:iterworst}The number of fine operations as a function of the bare mass for CG on the even/odd system, which exhibits critical slowing down, and for MG on the K\"ahler-Dirac preconditioned operator, where it is eliminated. Each data point is an average over 100 runs. Error bars are generally too small to be visible on the figures.}
\end{figure}

The convergence rate of our new MG algorithm on the
K\"ahler-Dirac preconditioned operator, illustrated in
Fig.~\ref{fig:resid}, is a dramatic improvement relative to the failed MG
algorithm applied to the original staggered operator in
Fig.~\ref{fig:residstag}. The {\emph{only}} methodological difference
is coarsening the K\"ahler-Dirac preconditioned operator instead of
the original staggered operator. Moreover, as we scan in the quark
mass as shown in Fig.~\ref{fig:iterworst}, we see that our formulation
has eliminated ill-conditioning due to critical slowing down: unlike
using CG on the even/odd preconditioned system, an MG solve takes a
roughly constant number of outer iterations as the chiral limit is
approached.

Let us now describe in detail the new algorithm and the
numerical analysis for MG applied to the  K\"ahler-Dirac
preconditioned staggered operator. The parameters we choose are summarized in
Table~\ref{tab:kcycle}. First, we consider a two-level algorithm. We
construct a right near-null vector $\psi$ by relaxing on
the homogeneous normal system $A A^\dagger \psi = 0$, using
Gaussian distributed random vectors $\psi_0$ as the initial guess\footnote{We remark that $A^\dagger A$ generally
  works just as well. We have also explored relaxing on $A$ directly using
  BiCGstab and BiCGstab($\ell$), $\ell = 6$~\cite{sleijpen1993bicgstab}, which in
  practice works well at small volumes but degrades for larger volumes. The use of the normal operator may be why we can effectively capture exceptional eigenvalues.}. In
practice, this is performed in multiple steps.
\begin{itemize}
\item We convert the homogeneous system to the residual system $A A^\dagger \vec{e} = \vec{r} \equiv -A A^\dagger\psi_0$.
\item We relax on the residual system using CG to a relative tolerance of $10^{-4}$ or a maximum number of 250 iterations.
\item We reconstruct the near-null vector $\psi = \psi_0 + \vec{e}$, where $\vec{e}$ is the result of relaxation.
\end{itemize}

This is performed $n_{vec}^1$ times, and then we {\emph{globally}}
orthonormalize the full set of near-null vectors. We subsequently chirally
double the near-null vectors using
$\frac{1}{2}\left(1 \pm \epsilon(x)\right)$, and form the second-level
operator $\widehat{A} = P^\dagger A P$ from the block-orthonormalized
chirally-doubled null vectors. The coarse correction follows three steps. (1) Relax on the
current residual, a process known as the pre-smoother, (2) approximately solve
the second-level system: 
$\left[R A P\right] Re = R r$ (or, equivalently, approximately solve $\widehat{A} \hat{e} = \hat{r})$,
giving the prolonged error correction $e = P \hat{e}$, and (3)
post-smooth on the error accumulated from steps 1 and 2. In step 2 we
use a Krylov solver, and as such the MG preconditioner is not
stationary. For this reason, we use the restarted generalized
conjugate residual (GCR)~\cite{Axelsson1987} as a flexible outer solver, forming a
K-cycle. We use a global MR for our pre- and post-smoother. The
specific details of these steps are given in Table~\ref{tab:kcycle}. In practice, we iterate on the even/odd preconditioned system on the fine level, with the prescription where we coarsen assuming the odd contributions are all zero, and we also ignore the odd contributions in the prolonged error. This technique proved successful for the Wilson operator~\cite{Osborn:2010mb}.

A two-level algorithm does not fully eliminate critical slowing down, it just shifts it to the second level. We address this by generalizing to a {\emph{recursive}} algorithm, where we perform a still coarser correction to the system in step (2) of the above description. We generate a third level, $\widehat{\widehat{A\, }}$, similar to how we generate the second level: we generate near-null vectors with $\widehat{A} \widehat{A}^\dagger$, chirally double the near-null vectors using $\frac{1}{2}\left(1\pm\sigma_3\right)$, and subsequently form a third level.

This clearly generalizes to still coarser levels. For our numerical experiments in Sec.~\ref{sec:results}, we only study a three-level algorithm. Unlike on the fine level, the Krylov solve we perform on the intermediate level is an iteration directly on $\widehat{A}$, as we found this was more stable in practice. We approximately solve the coarsest level via CG on the normal error. Due to the exceptional eigenvalues which propagate to coarser levels, as noted in Fig.~\ref{fig:zoomcircle}, numerical experiments with Krylov solvers acting on $\widehat{\widehat{A\, }}$ were in general not successful. This was either due to stability reasons (using BiCGstab($\ell$)~\cite{sleijpen1993bicgstab}) or due to cost (using GCR). We believe using the normal operator is of critical importance.

\begin{table}[htp]
\center
\begin{tabular}{rll}
\hline
 & parameter & \\
\hline
setup & setup operator & Normal operator,\footnote{In practice, $A^\dagger A$ works just as well.} $A A^\dagger$ \\
 & setup solver & CG \\
 & max iterations & 250 \\
 & max residual tolerance per null vector & $10^{-4}$ \\
 & number of null vectors, level 1 ($n_{vec}^1$) & 8 \\
 & size of aggregate block, level 1 & $4^2$~($8^2$ in the original lattice) \\
 & number of null vectors, level $\ell > 1$ ($n_{vec}^\ell$) & 12 \\
 & size of aggregate block, level $\ell > 1$ & $2^2$ \\
 & number of levels $\ell_{\mbox{\scriptsize max}}$ & 3 \\
\hline
 solver, level 1 & operator & Schur prec., $\mEye-A_{eo}A_{oe}$ \\
 & restart length of GCR & 32 \\
 & relative residual tolerance & $10^{-10}$ \\
 & MR iterations for pre-,post-smooth & 2 \\
 & MR relaxation parameter & 0.85 \\
\hline
 solver, level 2 & operator & $\widehat{A}$ \\
 & max iterations & 16 \\
 & restart length of GCR & 8 \\
 & relative residual tolerance & 0.2 \\
 & MR iterations for pre-,post-smooth & 2 \\
 & MR relaxation parameter & 0.85 \\
\hline
solver, level 3 & operator & Normal operator, $\widehat{\widehat{A\, }} \widehat{\widehat{A\, }}^\dagger$ \\
 & solver & CGNE \\
 & relative residual tolerance & 0.2\\
 & maximum iterations & 256\\
\hline
\end{tabular}
\caption{\label{tab:kcycle}The parameters we use for our K-cycle. For
  consistency, we use the same setup parameters throughout the procedures described in this paper.}
\end{table}

\subsection{Results\label{sec:elim}}

A successful, recursive MG algorithm will shift critical slowing down to the coarsest level. In the context of the Schwinger model, and four-dimensional QCD, this means we want consistent convergence independent of mass and volume. We are also interested in the MG algorithm being successful in all physically interesting regimes. In the case of our target problems, this means we need to study the behavior with the bare coupling $\beta$. The continuum limit is taking $\beta \to \infty$ at constant physics, where the relevant region is $\ell_{M_\pi} > \ell_\sigma$. When $\beta$ is too small, close to the cutoff scale, we are no longer studying relevant physics. A breakdown of MG for very small $\beta$ is acceptable. The values of $\beta$ studied, 3.0, 6.0, and 10.0, correspond to $\ell_\sigma \approx 2.4, 3.5$, and 4.5, respectively. The lowest value of $\beta$ is becoming rather unphysical.

\paragraph{Elimination of critical slowing down: fine level}

\begin{figure}[t] \centering
  \includegraphics[width=0.45\linewidth]{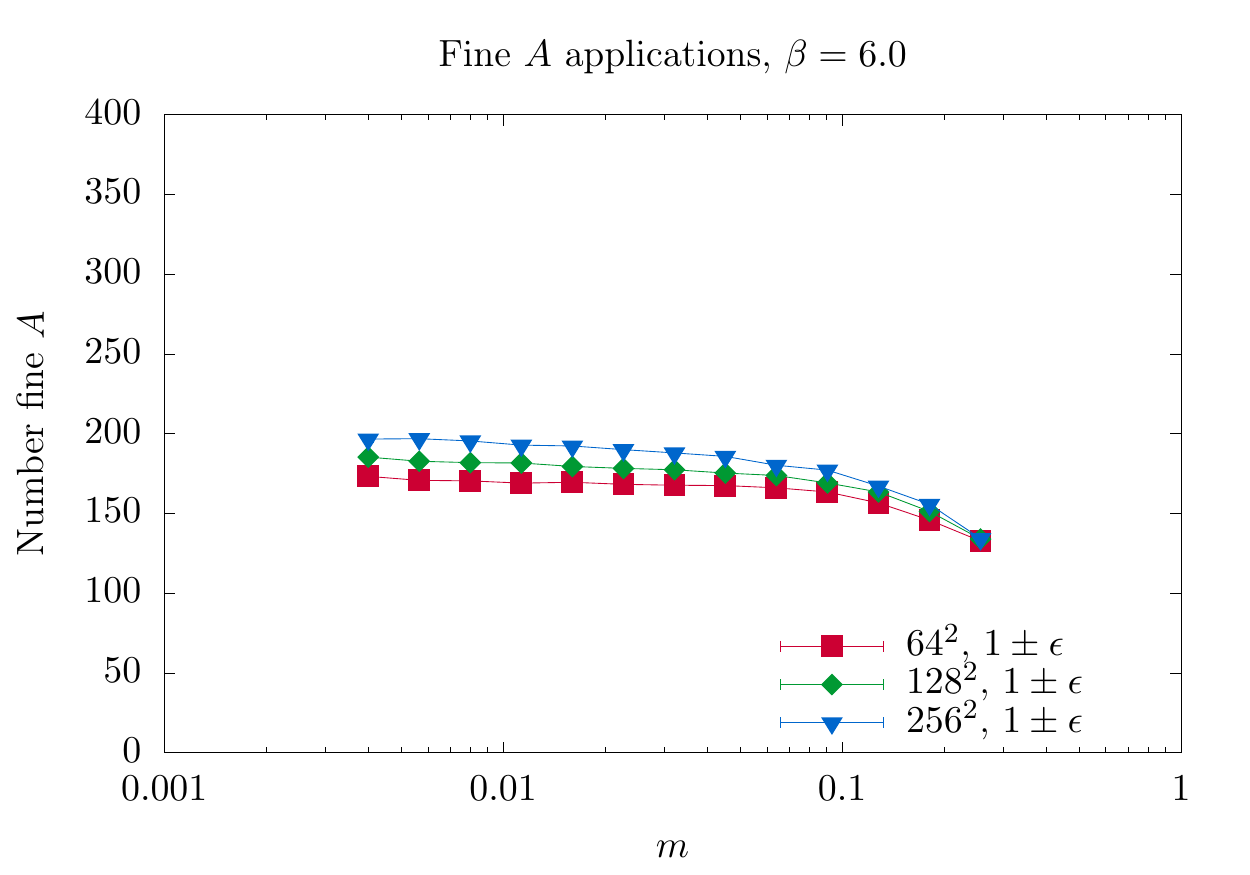}
~\includegraphics[width=0.45\linewidth]{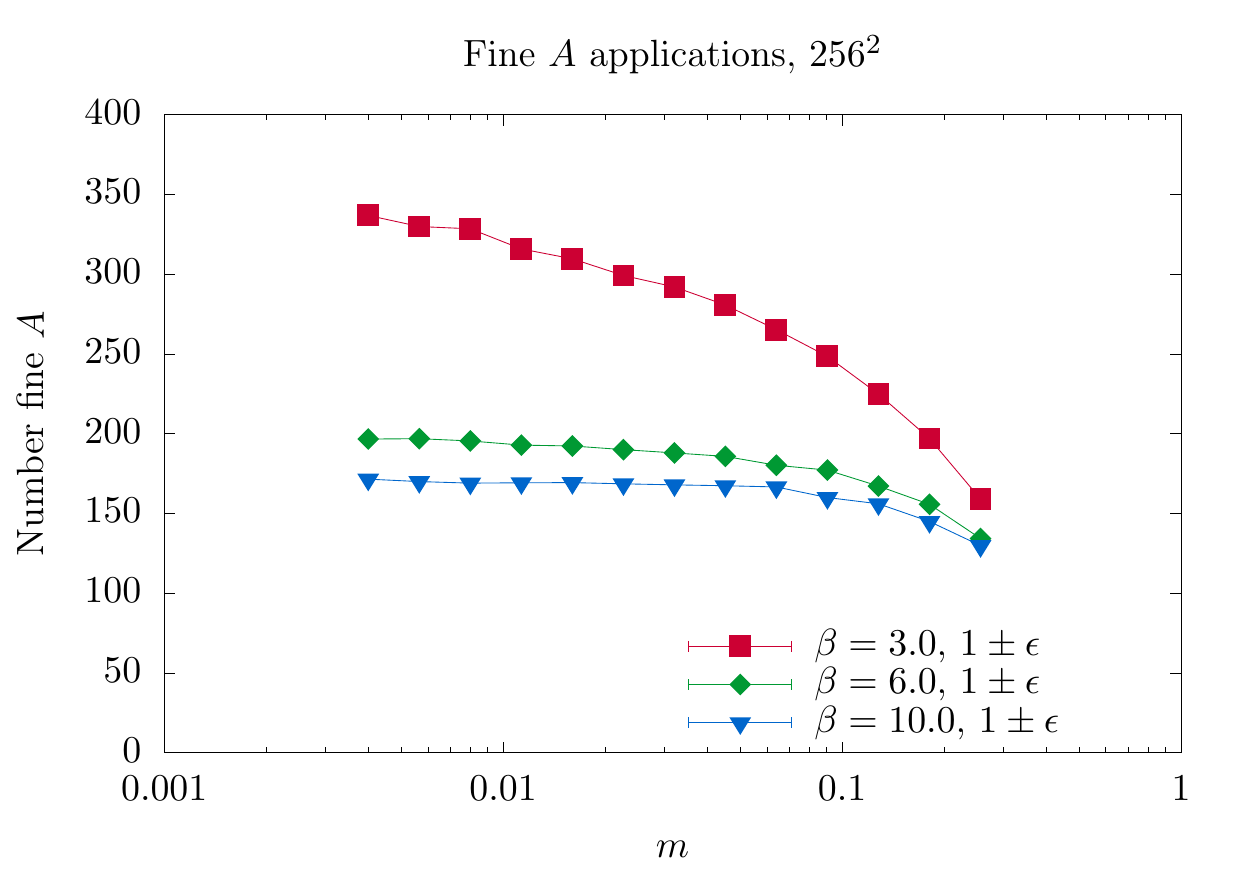}
  \caption{The number of applications of the fine operator $A$ per
    MG-preconditioned solve as a function of mass. On the left, we consider fixed
    $\beta$, and on the right, fixed volume. Each data point is an average over 100 runs. Error bars are generally too small to be visible on the figures.}
\label{fig:iterationsfine}
\end{figure}

The indication of a successful two-level algorithm is the elimination of critical slowing down for the fine operator $A$, that is, constant iterations with respect to the mass and volume per each $\beta$. In Fig.~\ref{fig:iterationsfine}, we present the number of
applications of the fine operator $A$ between the GCR algorithm and the MG preconditioner, which is proportional
to the number of iterations for the outer GCR solve. On the left we
consider the case of fixed physical $\beta = 6.0$ at varying volume. The number
of $A$ applications is roughly constant, independent of the volume and mass in the chiral limit.
In the right panel, we consider our largest volume, $256^2$, fixed for
three different values of $\beta$. We see that at $\beta = 10.0$ and
$6.0$, critical slowing down has been essentially eliminated as a
function of mass. At $\beta = 3.0$, where we are probing somewhat cutoff scale physics, the number of
iterations appears to not be diverging with power law behavior, and as such critical slowing down has still been eliminated.

\paragraph{Elimination of critical slowing down: intermediate level}

\begin{figure}[t] \centering
 \includegraphics[width=0.47\linewidth]{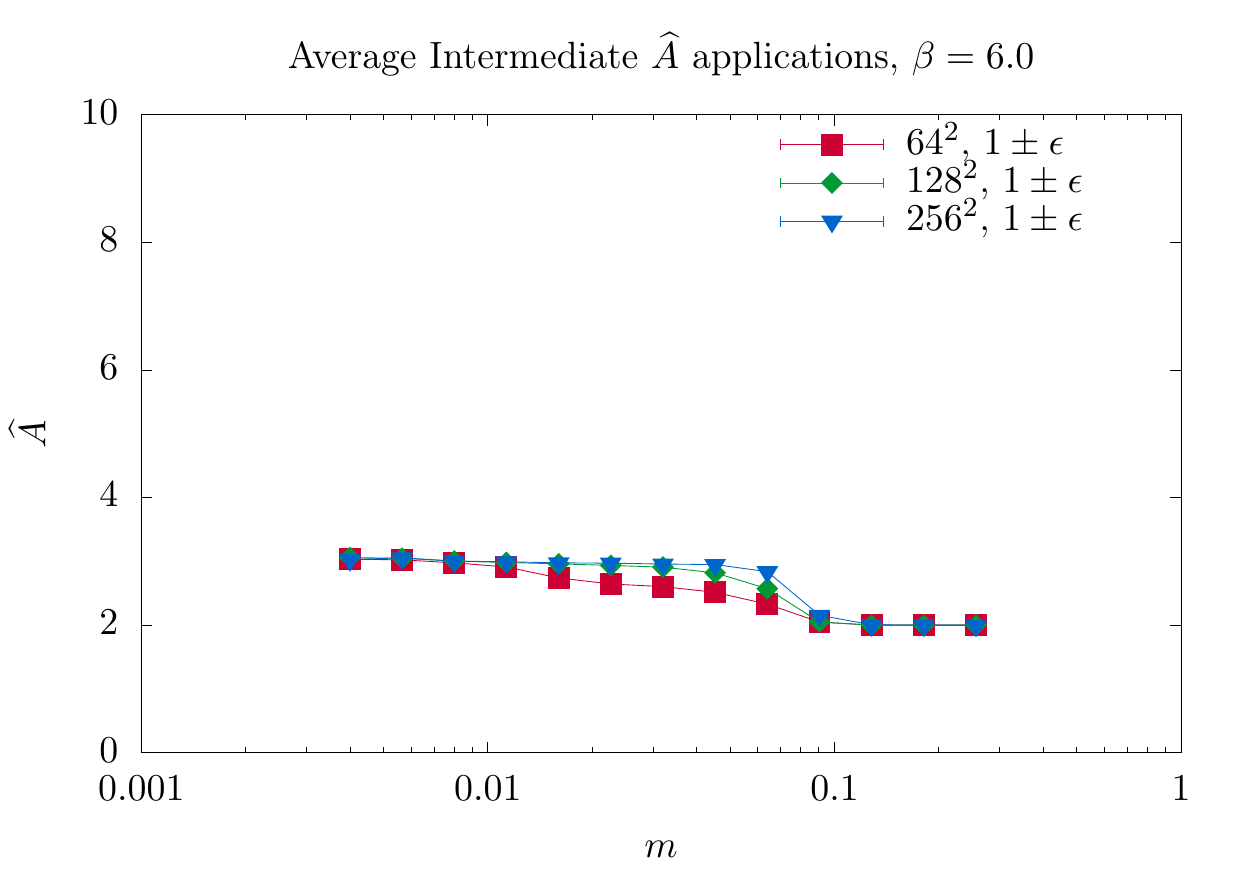}
~\includegraphics[width=0.47\linewidth]{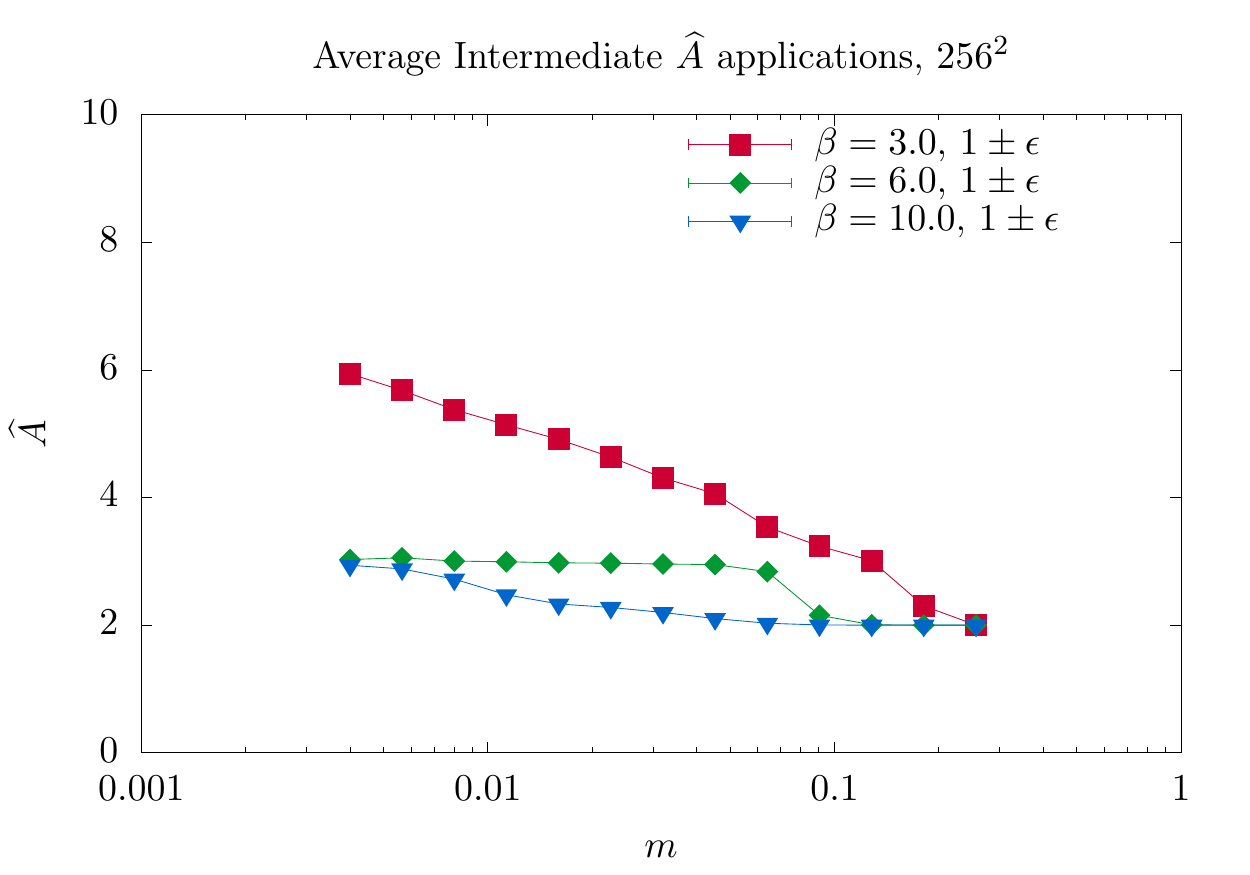}
\caption{The average number of iterations for the inner Krylov solve
  of $\widehat{A}$ as a function of mass. On the left, we consider
  fixed $\beta$, and on the right, fixed volume. Each data point is an average over 100 runs. Error bars are generally too small to be visible on the figures.}
\label{fig:iterationsinter}
\end{figure}

A successful recursive algorithm eliminates critical slowing down at each level. Thus, we consider the average number of iterations for the second level coarse correction, averaged over each outer iteration. In the left panel of Fig.~\ref{fig:iterationsinter}, we consider the
average number of intermediate iterations at fixed $\beta = 6.0$ at
varying volume. In the chiral limit, the average number of
iterations is essentially fixed at 3 independent of volume. In the right
panel we consider a fixed volume of $256^2$ for three different values
of $\beta$. As with the fine level, we see an elimination of critical slowing down for each value of $\beta$.

We remark that in a highly optimized and tuned implementation, it is important that we use a K-cycle at the second level. In such an implementation, the maximum number of iterations on the coarsest level may be capped to some reasonable amount. This would cause the number of iterations on the intermediate level to increase. Since in a K-cycle the second level is solved to a fixed {\emph{residual}} as opposed to a fixed number of iterations, the number of iterations at the {\emph{finest}} level remains stable.

\paragraph{Critical slowing down: coarsest level}

\begin{figure}[t] \centering
  \includegraphics[width=0.47\linewidth]{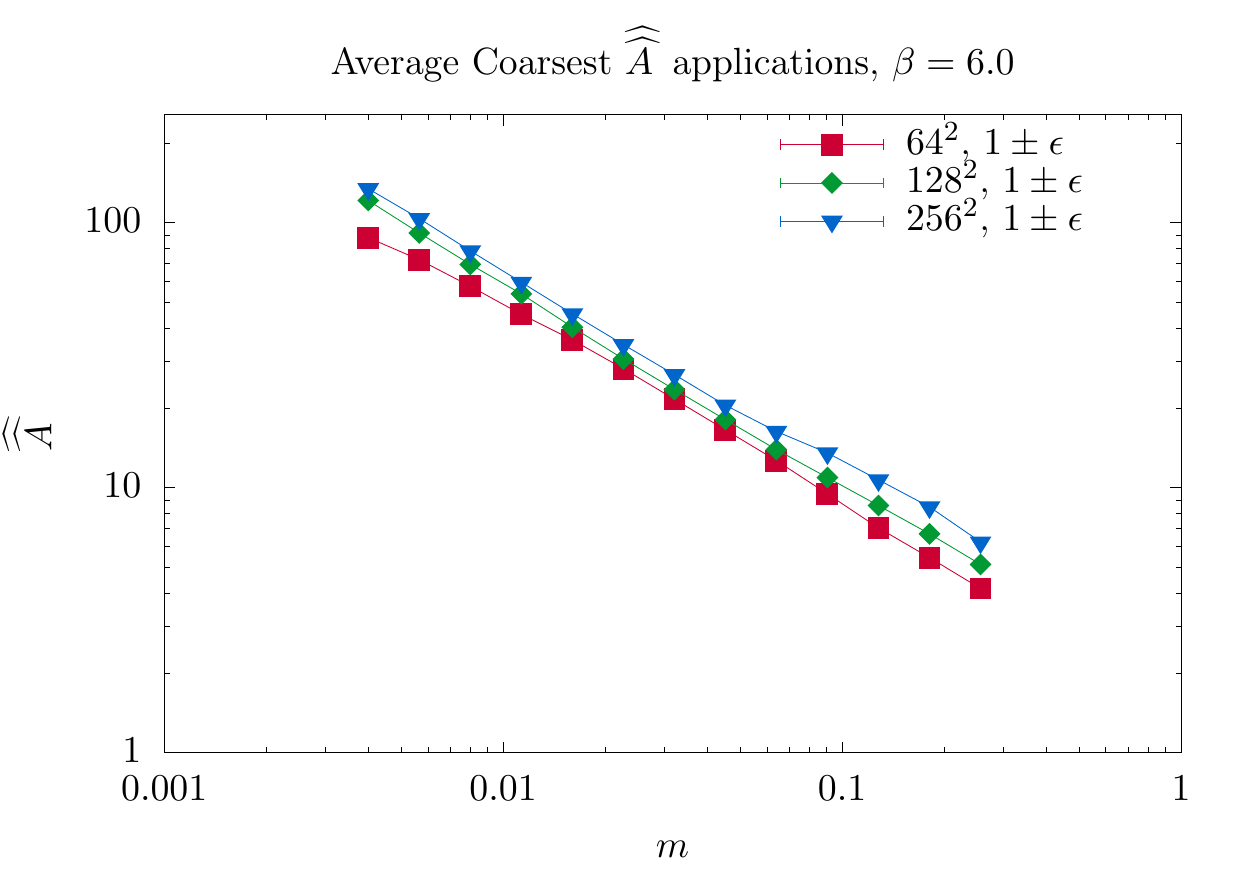}~\includegraphics[width=0.47\linewidth]{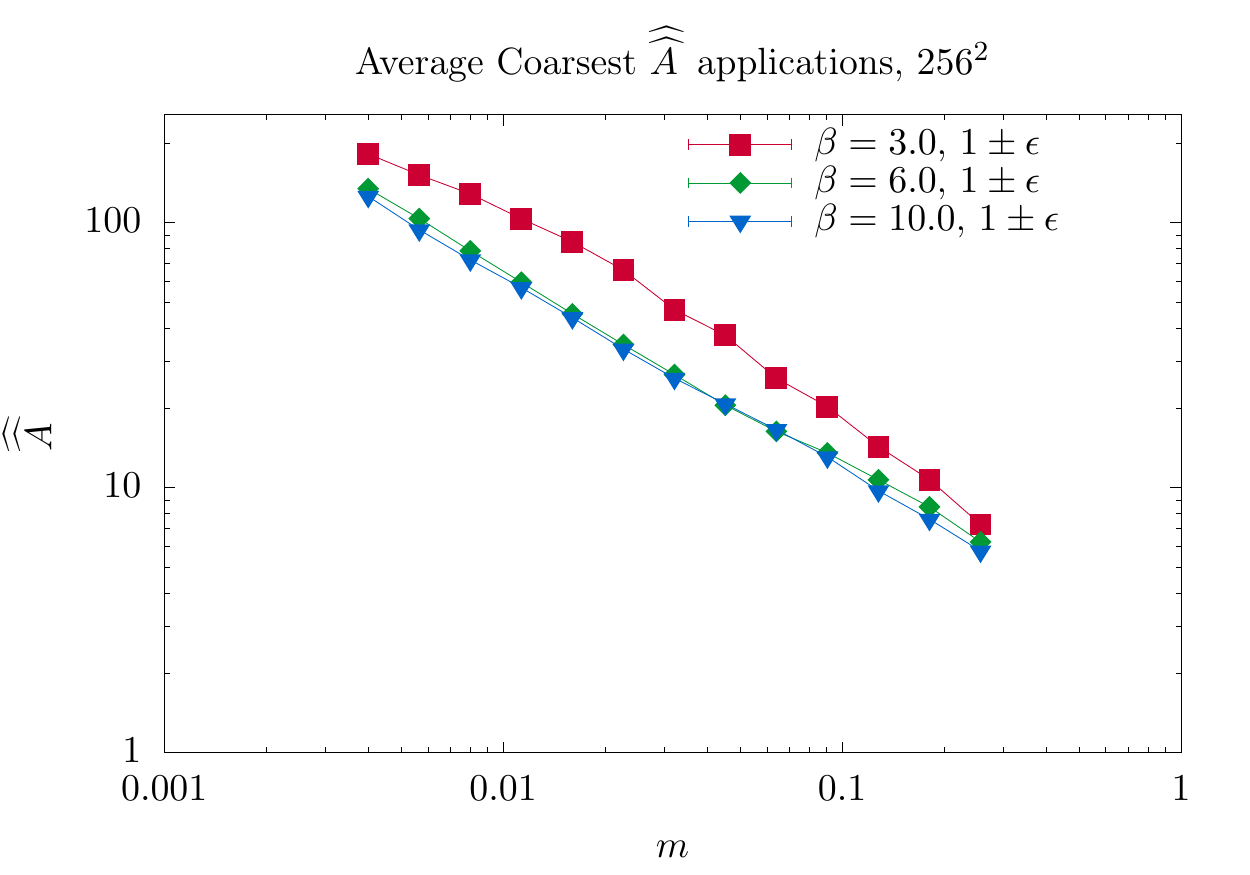}
  \caption{The average number of iterations of CGNE on the coarsest level for, on the left, fixed $\beta$, and on the right, fixed volume. Note that this is a log-log plot. Each data point is an average over 100 runs. Error bars are generally too small to be visible on the figures.}
\label{fig:iterationscoarsest}
\end{figure}

The previous two paragraphs demonstrate an elimination of critical slowing down from finer levels. Thus, there should be critical slowing down on the efficiently solvable coarsest level. In Fig.~\ref{fig:iterationscoarsest} we consider the average number of iterations for the coarsest solve via CGNE. In contrast to the previous two figures, these plots are on a log-log scale instead of a log-linear scale. In the left and right panels, we consider constant $\beta = 6.0$ and a constant volume of $256^2$, respectively. The number of iterations is divergent with power law behavior\footnote{It will not exactly diverge for $m$ extremely small due to finite volume cutoff effects.}. Critical slowing down has been shifted to the coarsest level.

\paragraph{Comparison with a direct solve}

In looking at the outermost level, the intermediate level, and the coarsest level in a three-level solve, we see that we have formulated a MG algorithm which shifts critical slowing down to the coarsest level. Furthermore, the solve is stable: in Fig.~\ref{fig:resid}, we saw that a MG-GCR solve converges smoothly at our most chiral point for $\beta = 10$. There are large reductions in the relative residual on each iteration. On the other hand, the traditional solve with CG on the even/odd operator, despite converging successfully, converges very slowly, an indication of critical slowing down.

This behavior persists independent of mass. In Fig.~\ref{fig:iterworst}, we trace the number of iterations away from the chiral limit, seeing that it is roughly constant. Critical slowing down has been eliminated. On the other hand, the number of iterations for a solve with the even/odd operator diverges with mass with power-law behavior. This is exactly the critical behavior that's been shifted to the coarsest level in Fig.~\ref{fig:iterationscoarsest}. The benefit of our MG algorithm is drastic.

\begin{table}[tp]
\center
\begin{tabular}{ccc|rrr}
\hline
 $L$ & $m$ & $\beta$ & Fine mat-vec & Intermediate avg. iter. & Coarsest avg. iter. \\
\hline
\hline
 64 & 0.01 & 3.0 & 228.6(1.2) & 3.45(4) & 62.0(0.5) \\
\hline
 128 & 0.005 & 12.0 & 159.4(0.4) & 2.60(2) & 95.7(0.6) \\
\hline
 256 & 0.0025 & 48.0 & 147.4(0.5) & 2.09(1) & 205.0(0.6) \\
\hline\hline
 64 & 0.004 & 0.75 & --- & --- & max \\
\hline
 128 & 0.002 & 3.0 & 290.5(1.8) & 5.08(8) & 206.2(1.5)${}^\dagger$ \\
\hline
 256 & 0.001 & 12.0 & 189.9(0.4) & 4.61(2) & 249.6(0.2)${}^\dagger$\\
\hline
\end{tabular}
\caption{\label{tab:zerospacing}The effect of taking the lattice spacing to zero at constant physical box size and mass gap for two sequences of successive refinement. Cases where the maximum number of iterations is sometimes hit on the coarsest level are denoted with a dagger. All quantities are averaged over 100 runs.}
\end{table}
\subsection{Continuum Limit}

{\textbf{It should be emphasized that our fixed prescription is effective in the most relevant regime: towards the continuum, where the lattice spacing vanishes relative
 to fixed physics, and in the chiral limit, where $\ell_{M_\pi}$ diverges relative to $\ell_\sigma$.}}

For the two-dimensional Schwinger model, taking the
continuum limit at constant physics corresponds to simultaneously
doubling the length scale of the fine volume, halving the mass, and quadrupling
$\beta$. In Table~\ref{tab:zerospacing}, we consider the use of MG while taking the
continuum limit from two base configurations.
First, we consider a base configuration of $64^2$ at $m = 0.01$ and
$\beta = 3.0$, where we have discussed earlier that a MG
algorithm is successful. On two successive refinements towards the
continuum limit, we see that there is a reduction in the number of
outer applications of $A$ and in the average number of iterations on
the intermediate level. In tandem, the average number of iterations on
the coarsest level increases: there is more critical slowing down to
shift to the coarsest level, which is to be expected, towards the continuum limit. Our MG
algorithm performs better as the continuum limit is taken.
Next, we consider a base configuration of $64^2$ at $m = 0.004$ and $\beta = 0.75$, an unphysically coarse configuration. In this case, a MG algorithm fails to converge. Again, on progressive refinements, the MG algorithm becomes convergent and becomes better behaved as the continuum limit is taken.

\ignore{
\subsection{Coarse Gauge Fields}

We now return our attention to the exceptional configuration at $256^2, \beta = 6.0$. This configuration is at the largest volume and at the smallest $\beta$ investigated. The source of the issue here is the exceptional eigenvalues noted on the right hand side of Fig.~\ref{fig:zoomcircle}. There are three effects that contribute to an increase in the density of exceptional eigenvalues:
\begin{itemize}
\item As the volume increases, the density of eigenvalues in the complex plane increases. This can be argued from the quantization of eigenvalues in the free case.
\item As $\beta$ decreases, there is a larger spread of eigenvalues from the free field circular spectrum. This causes more eigenvalues to drift to the left half plane. 
\item As the mass decreases, the radius of the spectrum increases. This also causes more eigenvalues to drift to the left half plane.
\end{itemize}
The combination of these three effects leads to an eventual breakdown of the recursive MG algorithm.

However, this is an overly pessimistic view of this situation. It is
more important to frame the problem of increasing lattice sizes and
shrinking masses in the context of taking the continuum limit. As
noted in Sec.~\ref{sec:schwinger}, taking the lattice spacing to zero
at constant physics and box size is asymptotically at large $\beta$
holding $M_\pi \ell_\sigma$ (the mass gap) and $L / \ell_\sigma$
fixed. In terms of bare lattice quantities, this corresponds to
simultaneously doubling $L$, quadrupling $\beta$, and halving $m$.

In Tab.~\ref{tab:zerospacing}, we consider two sequences of
approaching the zero lattice spacing limit. In the first case, we see
that while a coarse configuration at a smaller volume may not
completely eliminate critical slowing down, the effect can completely
disappear at larger volumes. We see an even more drastic example in
the second case: with a $64^2$ lattice at an extremely small
$\beta = 0.75$, the solve completely stalls out. Despite this, at
constant physics for a $256^2$ volume, the MG preconditioner works
without issue.

It's well known, in practice, that there's a crossover point where the original operator demonstrates enough that an MG solve becomes more effective in terms of time-to-solution than a direct solve of the original operator. In light of these exceptional eigenvalues and the detrimental effect they have on a successful MG preconditioner, it becomes increasingly important to determine where an MG solve becomes more cost effective. 
}

\ignore{
\begin{figure}
    \centering
    \includegraphics[width=0.48\linewidth]{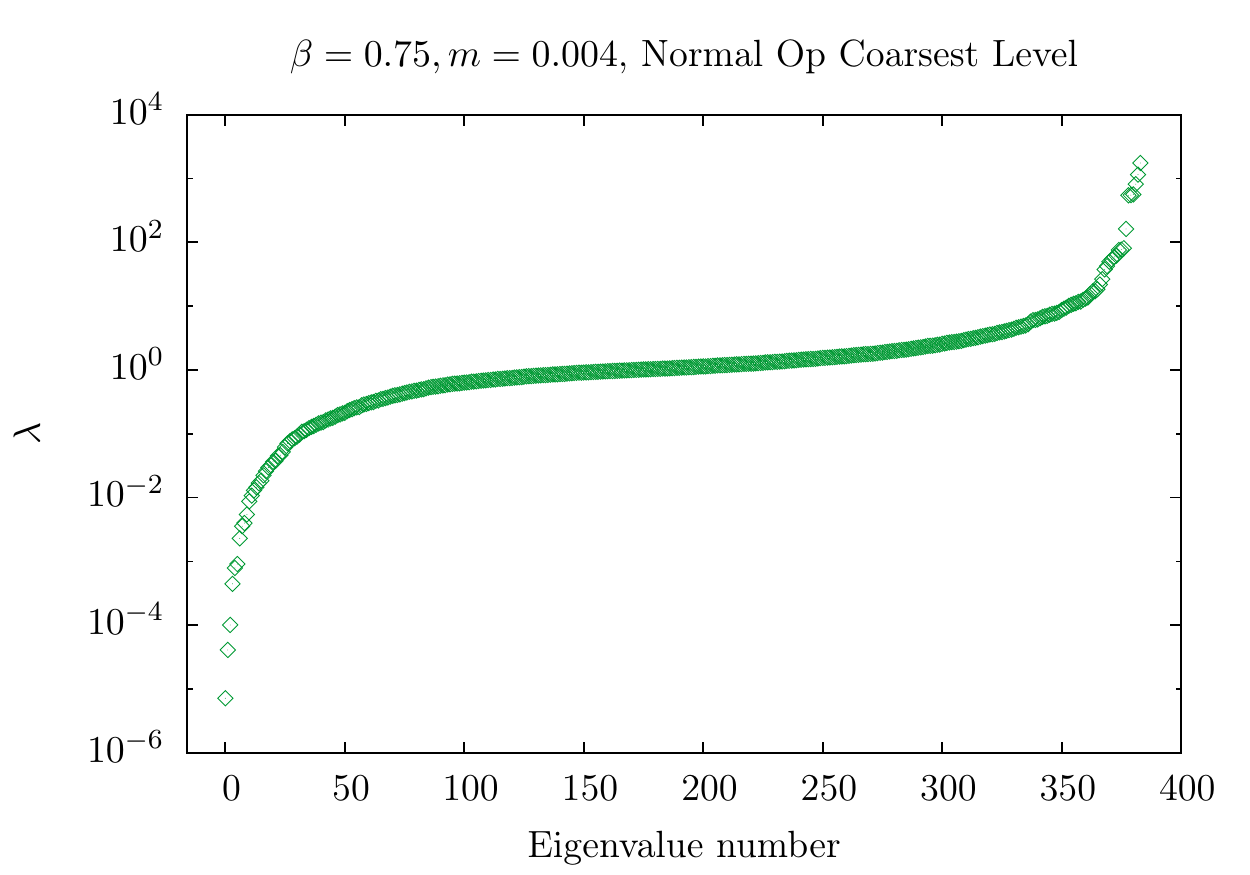}
   \caption{\label{fig:badeigenvalues}The spectrum of the normal operator from the right block Jacobi preconditioned coarsest operator for the $64^2, m = 0.004, \beta = 0.75$ configuration.}
\end{figure}
There is an interesting remark to make about the $64^2, \beta = 0.75$ example: while the coarsest normal operator is a $384 \times 384$ matrix, a CG solve takes around 1400 iterations, indicating an extremely ill-conditioned operator. We inspect the spectrum of this operator in Fig.~\ref{fig:badeigenvalues}: the condition number is approximately $2.44 \times 10^8$! While less drastic, this issue may spread to other configurations. It would be interesting to explore techniques such as eigenvalue deflation on the coarsest level, which could significantly stabilize the solve.
}

\newpage

\section{Preserving Complex Conjugate Pairs\label{sec:hermpres}}

A possible, if not necessary, generalization for four-dimensional QCD or other
staggered fermion problems could be the exact preservation of complex
conjugate pairs upon coarsening.   Indeed it  is possible to develop a prolongator $P$ and a restrictor $R \neq P^\dagger$, abandoning chiral doubling with projectors, which preserves complex conjugate eigenpairs after coarsening the K\"ahler-Dirac preconditioned operator, or any operator satisfying $\gamma_5^{L/R}$ Hermiticity with $\gamma_5^L \gamma_5^R = \mEye$. The resulting formalism gives what we will call an {\emph{asymmetric}} coarsening with $\sigma_1^{L/R}$ Hermiticity on the coarse level.

We consider a set of left and right vectors, $\levec{\bar\psi_i}$ and $\revec{\psi_i}$, respectively, which can generally be arbitrary and unequal. We perform a chiral doubling which gives
\be
 \widetilde R = 
\begin{bmatrix}
\levec{\overline\psi_i}\\
\levec{\psi_i}  \gamma^L_5
\end{bmatrix}
\quad, \quad 
 \widetilde{P} = 
\begin{bmatrix}
\revec{\psi_i} & 
\gamma^R_5\revec{\overline\psi_i}
\end{bmatrix}.\label{eq:prcplx}
\ee
These prolongators and restrictors obey $\widetilde{P} \sigma_1 = \gamma_5^R \widetilde{R}^\dagger$ and $\sigma_1 \widetilde{R} = \widetilde{P}^\dagger \gamma_5^L$. This is sufficient to prove $\widetilde{R} A \widetilde{P}$ is $\sigma_1$ Hermitian. The next step is to {\emph{block bi-orthonormalize}} $R$ and $P$, enforcing $RP = \mEye$, by-products of which give us $\sigma_1^{L/R}$.

As a clarifying tangent, we will consider the case $\gamma_5^L = \gamma_5^R$ and $\revec{\psi_i} = \revec{\bar\psi_i}$, that is, $\widetilde{R} = \widetilde{P}^\dagger$. This is true, for example, for the K\"ahler-Dirac preconditioned operator in the free field limit, or when considering the Wilson operator in general. The critical observation in this case is to recall that the process of (block) orthonormalization via a Gram-Schmidt is equivalent to a thin-QR decomposition. We define the block-dense matrix $\mathbb{M}$ of block dimension (coarse dof) $\times$ (coarse dof) as
\begin{align}
\widetilde{P}^\dagger \widetilde{P} = \mathbb{M} = \Sigma^\dagger \Sigma,\label{eq:thinqr}
\end{align}
where in the last step we have performed a Cholesky decomposition. We can rearrange Eq.~\ref{eq:thinqr} as
\begin{align}
\left(\widetilde{P}\Sigma^{-1}\right)^\dagger\left(\widetilde{P}\Sigma^{-1}\right) \equiv P^\dagger P = \mEye.
\end{align}
By definition, $P \equiv \widetilde{P} \Sigma^{-1}$ is block orthonormal. With the definition $\sigma_1^{\Sigma} \equiv \Sigma \sigma_1 \Sigma^{-1}$, we have $\gamma_5 P = P \sigma_1^\Sigma$, and $P^\dagger A P$ is $\sigma_1^{\Sigma}$ Hermitian.

We return to the (block) bi-orthonormalization of $R$ and $P$. The above procedure generalizes to a ``thin-LU'' decomposition. Eq.~\ref{eq:thinqr} generalizes to
\begin{align}
\widetilde{R} \widetilde{P} = \mathbb{M} = LU,\label{eq:thinlu}
\end{align}
where in the last step we have performed an LU decomposition. We can rearrange Eq.~\ref{eq:thinlu} as 
\begin{align}
\left(L^{-1} \widetilde{R}\right)\left(\widetilde{P}U^{-1}\right) \equiv RP = \mEye.
\end{align}
$R$ and $P$ are block bi-orthonormal. We can show $\widehat{A} \equiv RAP$ admits a $\sigma_1^{L/R}$ Hermiticity condition via defining
\begin{align}
\sigma_1^L &= U^{-\dagger} \sigma_1 L, & \sigma_1^R &= U \sigma_1 L^{-\dagger},
\end{align}
and noting
\begin{align}
\sigma_1^L R &= P^\dagger \gamma_5^L, & P \sigma_1^R &= \gamma_5^R R^\dagger.
\end{align}
The pair $\sigma_1^{L/R}$ obeys $\sigma_1^L \sigma_1^R = \mEye$, as can be verified by explicit calculation, requiring the critical and subtle observation that $\tilde{R}\tilde{P}$ is $\sigma_1$ Hermitian itself.







We emphasize that this construction is fully generic, whether or not $\revec{\psi_i} = \revec{\overline\psi_i}$. We defer a discussion of numerical experiments with preserving complex conjugate eigenpairs to appendix~\ref{app:cplxconjpairs}. Our deference to an appendix reflects our observations that, in two dimensions, (recursively) preserving eigenpairing actually leads to a less effective, and sometimes unstable algorithm. This method, or a further development thereof, may bear some fruit in four dimensions.

We make the additional remark that we can now make the algorithmic choice to right-block-Jacobi precondition $\widehat{A}$, analogous to the transformation we made to the staggered operator in the K\"ahler-Dirac form in the first place, and continue to preserve complex conjugate eigenpairs if we coarsen again. Let us denote $\widehat{A} = \mathcal{B} + \mathcal{C}$, where $\mathcal{B}$ is the block-local contribution. The resulting right-block-preconditioned operator $\widehat{A} \mathcal{B}^{-1}$ obeys a $\sigma_1^{rbj,L/R}$ Hermiticity condition with $\sigma_1^{rbj,L} = \sigma_1^L \mathcal{B}^{-1}$ and $\sigma_1^{rbj,R} = \mathcal{B} \sigma_1^R$. This recursive right-block-Jacobi preconditioning did not lead to an effective algorithm in two dimensions.


\paragraph{Exact Preservation of Eigenvectors\label{sec:eigens}}

In the case of, for example, the Wilson operator, chiral doubling with $\frac{1}{2}(1 \pm \gamma_5)$ preserves complex conjugate eigenpairs. We can choose the vectors $\revec{\psi_i} \equiv \revec{\bar\psi_i}$ to be right eigenvectors $\revec{\lambda_i^{+,R}}$ with eigenvalues $\lambda_i^{+}$, where the ${}^{+}$ denotes that the eigenvalue has positive real part.

The coarse operator $P^\dagger D_{W} P$ exactly preserves the eigenvalue $\lambda_i^{+}$, and $\revec{\lambda_i^{+,R}}$ is exactly preserved on the coarse subspace, that is, $P P^\dagger \revec{\lambda_i^{+,R}} = \revec{\lambda_i^{+,R}}$. However, even though chiral doubling guarantees the eigenvalue $\lambda_i^{-}$ is also preserved by the coarse operator, it is {\textbf{not}} because $\revec{\lambda_i^{-,R}}$ is exactly preserved by the coarse subspace, that is, $P P^\dagger \revec{\lambda_i^{-,R}} \neq \revec{\lambda_i^{-,R}}$. 

We can use {\emph{asymmetric}} coarsening to preserve $\revec{\lambda_i^{-,R}}$. We can choose $\revec{\psi_i} = \revec{\lambda_i^{+,R}}$ and $\levec{\bar\psi_i} = \levec{\lambda_i^{+,L}}$, then chirally double using Eq.~\ref{eq:prcplx} and subsequently block bi-orthonormalize the $P$ and $R$. This operator preserves the eigenvalues $\lambda_i^{\pm}$, {\emph{and additionally}} $PR \revec{\lambda_i^{\pm,R}} = \revec{\lambda_i^{\pm,R}}$ and $\levec{\lambda_i^{\pm,L}} PR = \levec{\lambda_i^{\pm,L}}$. 



\newpage
\section{\label{sec:Conclusion}Conclusion}

The first successful MG algorithm in LQCD was constructed for the Wilson discretization of the Dirac operator nearly a decade ago~\cite{Brannick:2007ue,Babich:2010qb}. This advance relied on, at the time, the novel approach in LQCD to adaptively discover the near-null space and geometrically project onto coarse lattices. Remarkably, with the exception of the similar twisted-mass discretization, the basic method has not been easily generalized to two important methods: staggered and domain wall fermions, each of which feature improved chiral symmetry. A more fundamental understanding of MG methods in LQCD is clearly lacking. Here, we have taken a step towards this. For the staggered operator, we identified the spectral feature that was responsible for the failure of a straightforward generalization of Wilson MG and have overcome this problem by preconditioning by the K\"ahler-Dirac (spin-flavor) block structure. We demonstrate that this has a dramatic effect on the spectrum: in the singular, zero mass limit, the pure imaginary spectrum of the anti-Hermitian operator maps to a unitary circle of the form seen in the overlap operator.

The success of the resultant MG algorithm for this K\"ahler-Dirac preconditioned operator has been demonstrated numerically for the two-dimensional Schwinger model. Both the theoretical framework and the phenomenological features naturally generalize to the case of four-dimensional QCD. On this basis, we are optimistic that our staggered multgrid algorithm will have similar success in this application. Numerical tests for this conjecture are underway by extending the high performance MG framework of the QUDA library to coarsen staggered-like operators. These tests will be made on the largest available lattices to explore the scaling of the algorithm over a range similar to the two-dimensional tests presented here.

We have also made an effort to explore a range of projection methods that are capable of exactly preserving the complex conjugate pairs of eigenvalues present in the K\"ahler-Dirac preconditioned operator. We hope our emphasis on spectral analysis and transformations will provide some flexibility in adapting our algorithm not only to four-dimensional QCD but also to similar Dirac discretizations found in BSM theories~\cite{Blum:2013mhx}, supersymmetric Yang-Mills theory~\cite{Catterall:2014vka}, and quantum critical behavior in condensed matter~\cite{Brower:2012zd}.

\section*{Acknowledgments}

We acknowledge fruitful conversations with Peter Boyle, Carleton DeTar, Arjun Gambhir, Martin L\"uscher, James Osborn, Eloy Romero, Andreas Stathopoulous, and especially with Matthias Rottman and Andreas Frommer who have also explored the staggered problem and gave the insight that it is possible to exactly preserve complex conjugate eigenpairs for the coarsened operator. We also thank the Higgs Centre for Theoretical Physics for its hospitality at the 2016 QCDNA Workshop. R.C.B. and E.S.W. acknowledge support by DOE grant DE-SC0015845 and by the Exasale Computing Project (17-SC-20-SC), a collaborative effort of the U.S. Department of Energy Office of Science and the National Nuclear Security Administration. The work by A.S. was done for Fermi Research Alliance, LLC under Contract No. DE-AC02-07CH11359 with the U.S. Department of Energy, Office of Science, Office of High Energy Physics.

\bibliographystyle{unsrt}
\bibliography{bib/MG}

\newpage
\appendix

\section{Studies of Preserving Complex Conjugate Eigenpairs\label{app:cplxconjpairs}}

\begin{figure}[t] \centering
   \includegraphics[width=0.47\linewidth]{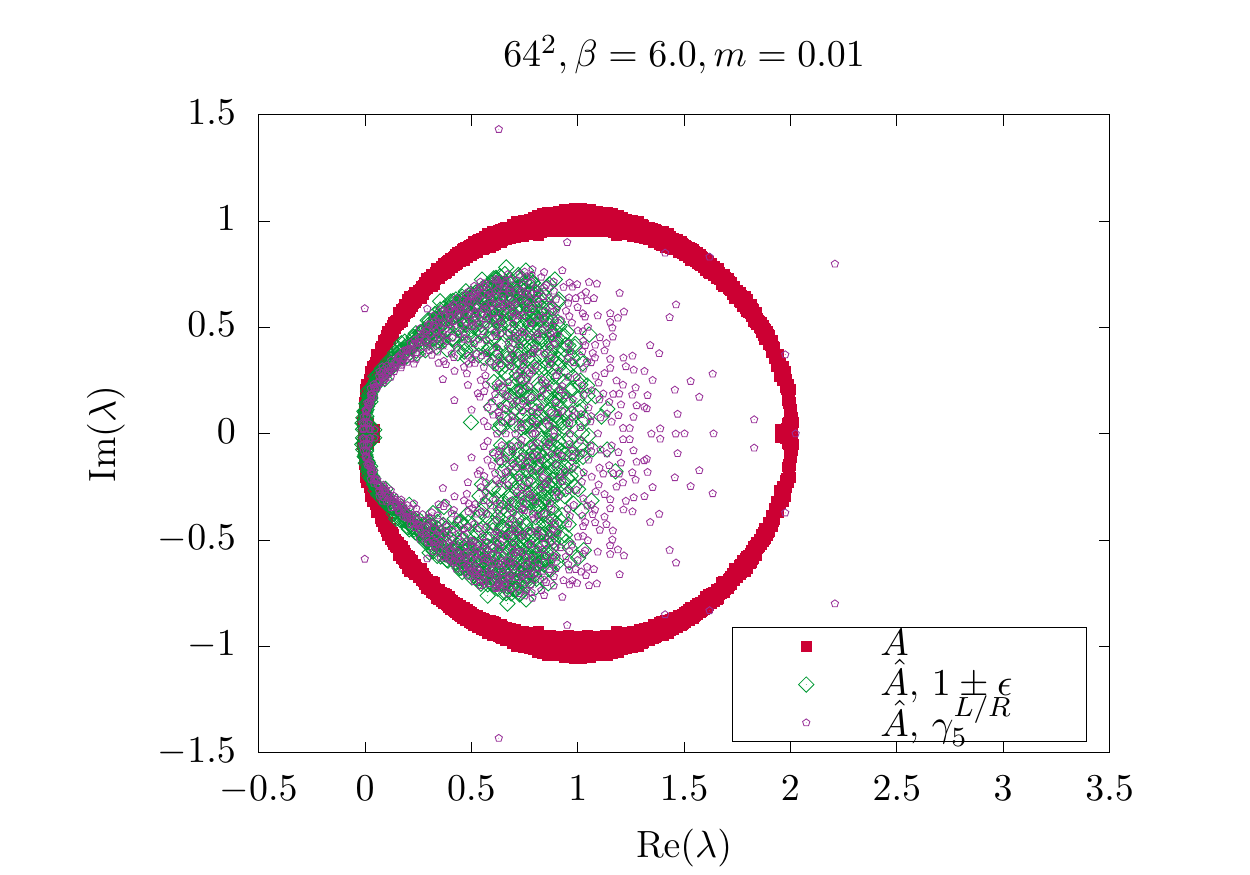}~ \includegraphics[width=0.47\linewidth]{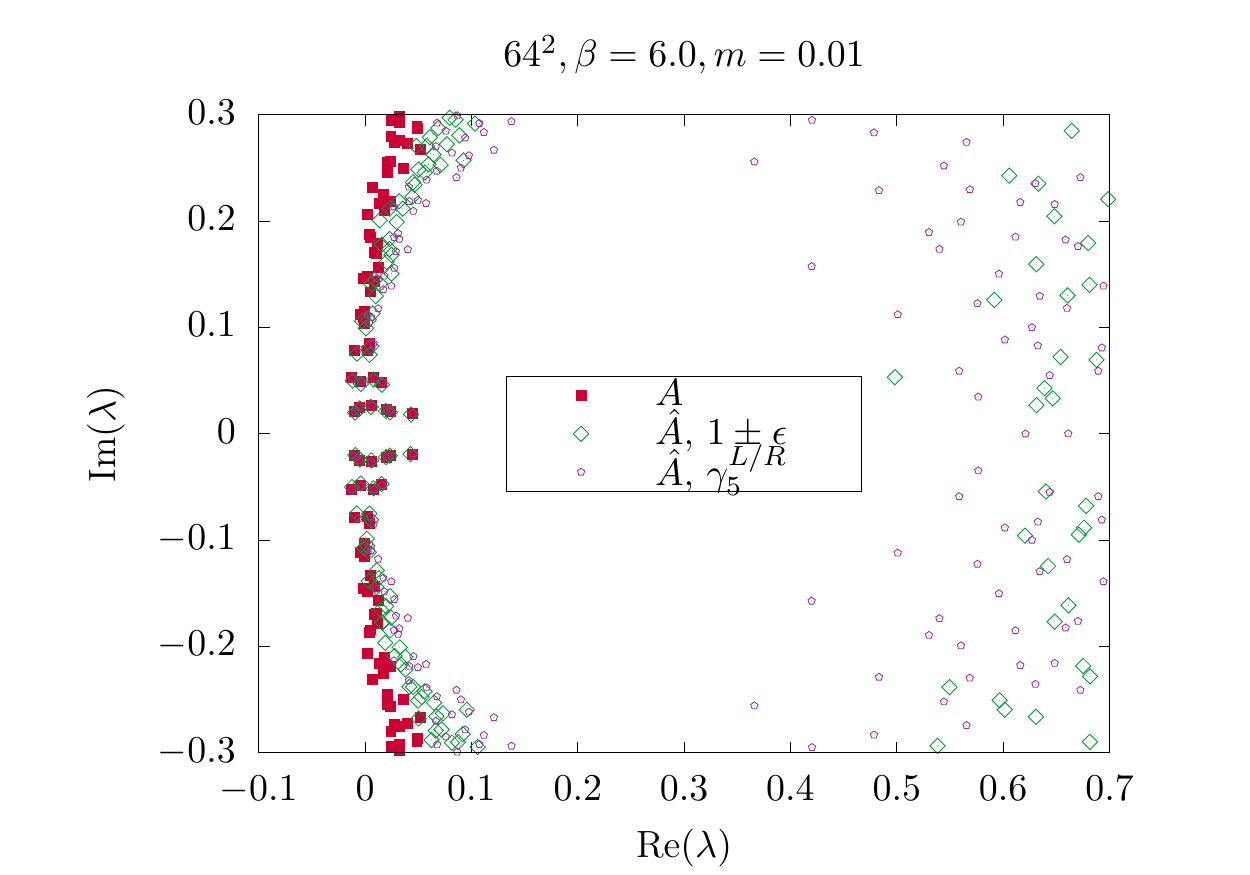}
  \caption{On the left, the spectrum of the Dirac-K\"ahler preconditioned operator from an interacting gauge field, overlaid with the spectrum of the operator coarsened via na\"ive chiral doubling and complex eigenpair doubling. On the right, a zoom in on the low portion spectrum which shows eigenvalues in the left-half plane and further emphasized complex eigenpairing.}
\label{fig:zoomcircleherm}
\end{figure}

\begin{figure}[t] \centering
{\large{$64^2, \beta = 6.0, m = 0.01$}}\\
  \includegraphics[width=.47\linewidth]{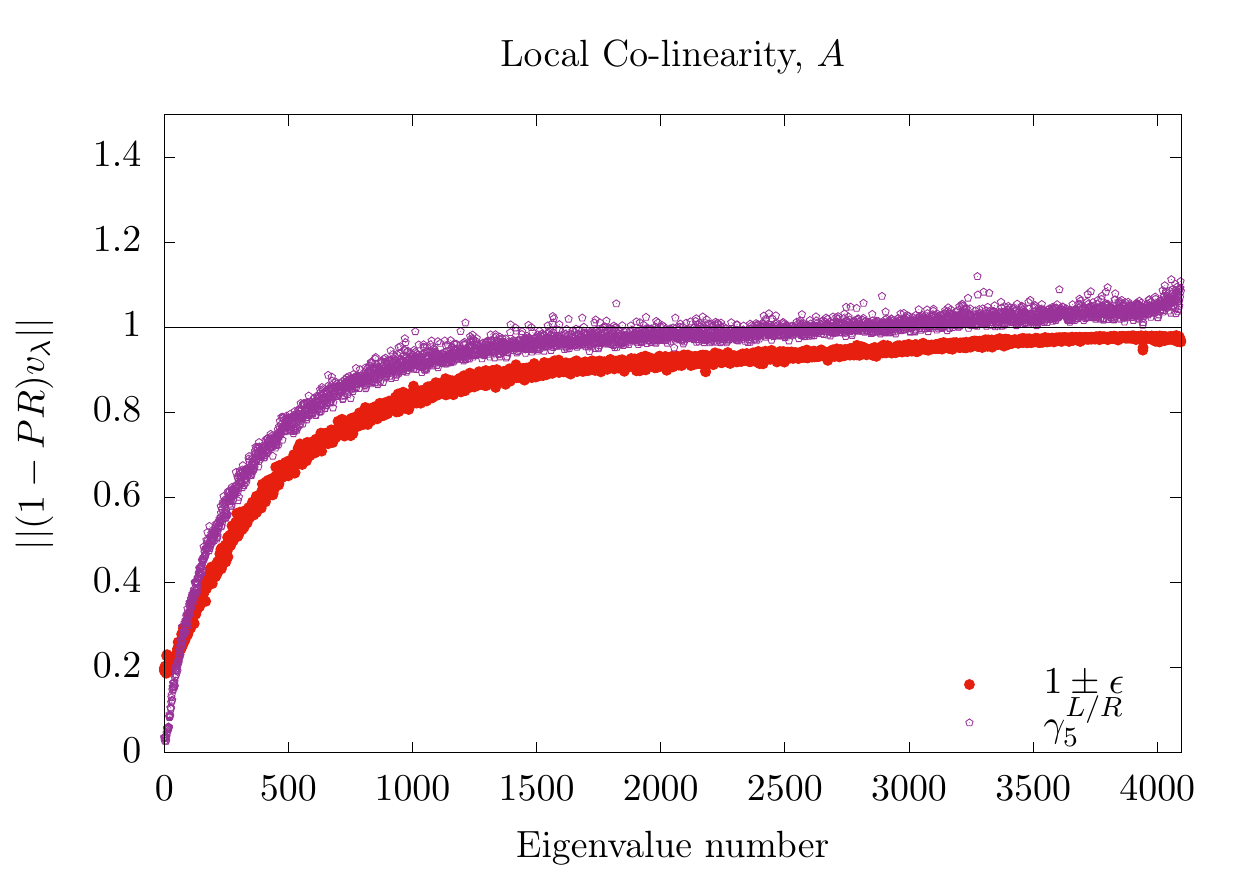}
~\includegraphics[width=0.47\linewidth]{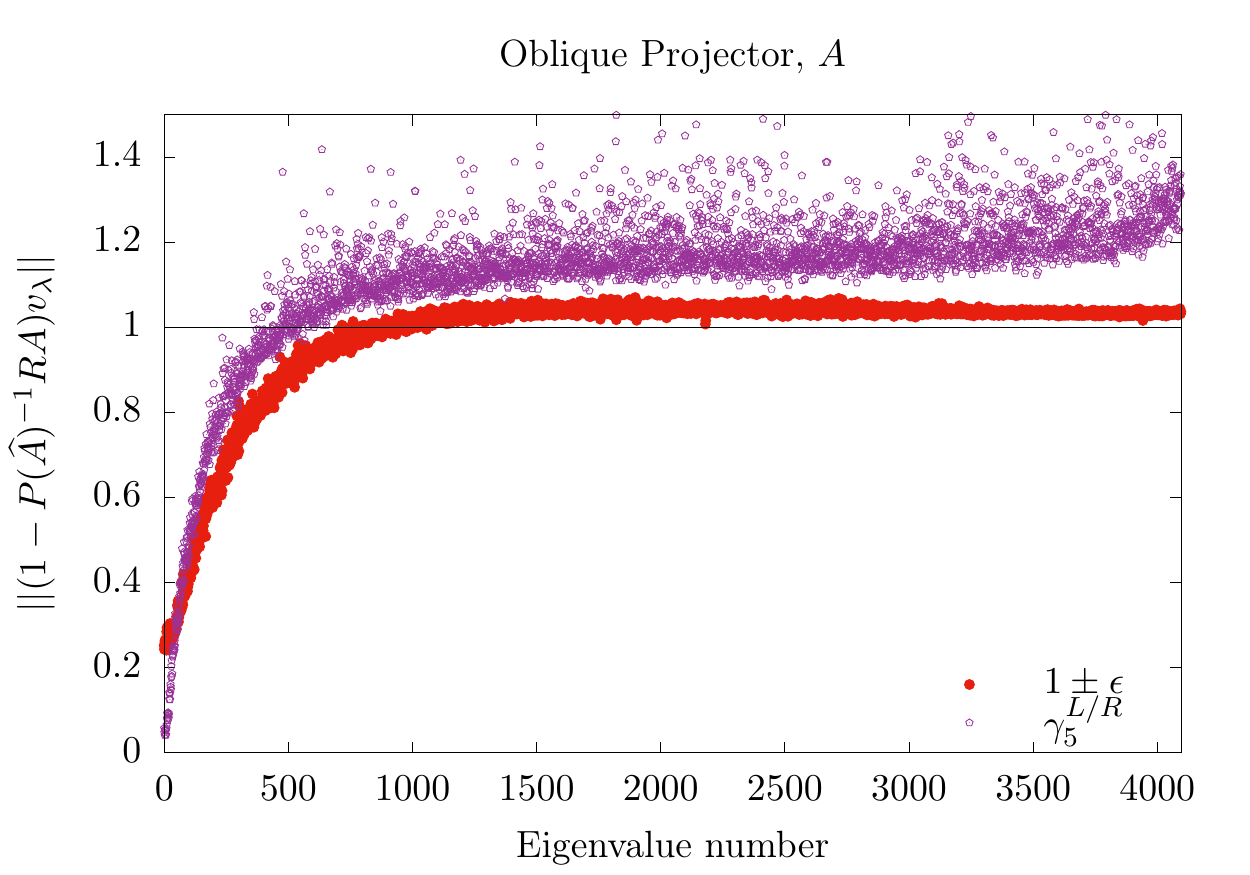}\\\includegraphics[width=.47\linewidth]{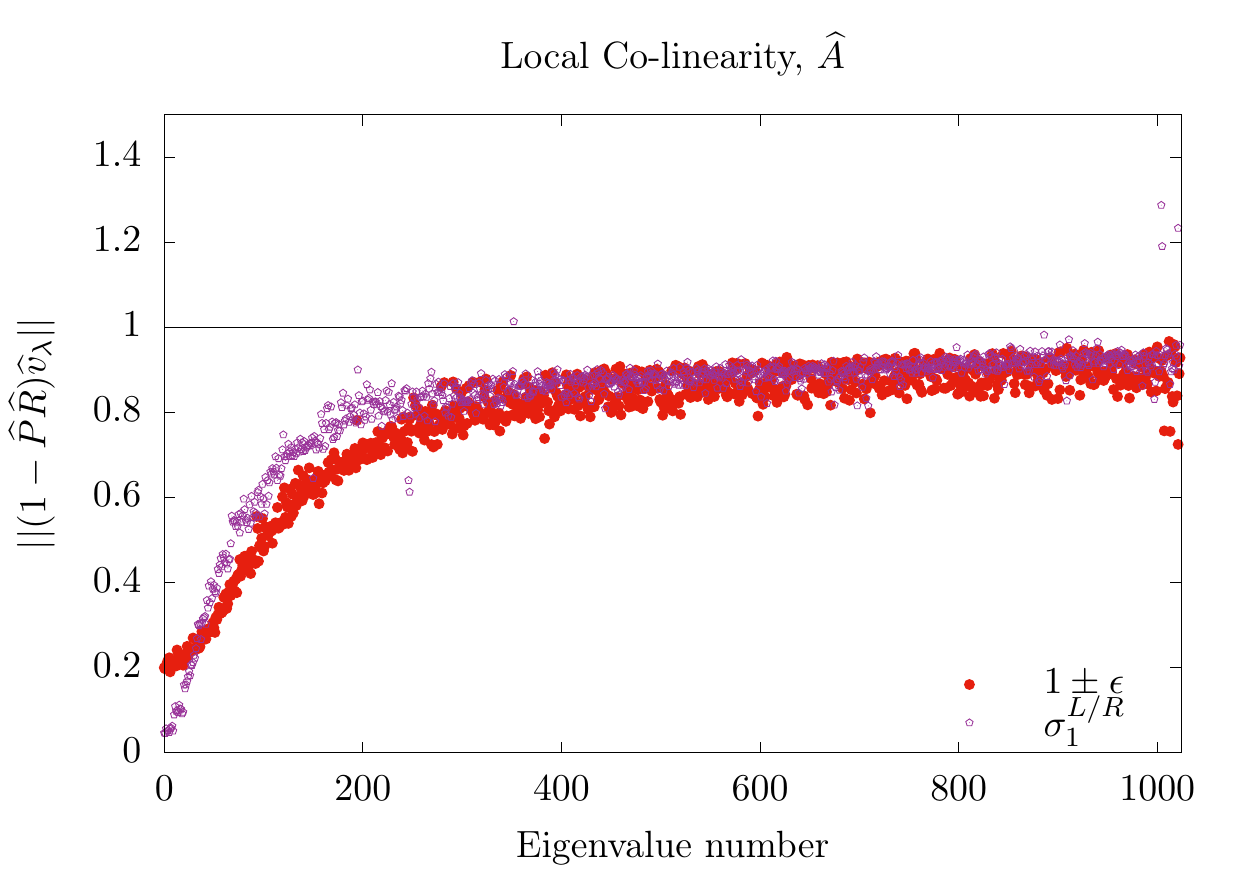}
~\includegraphics[width=0.47\linewidth]{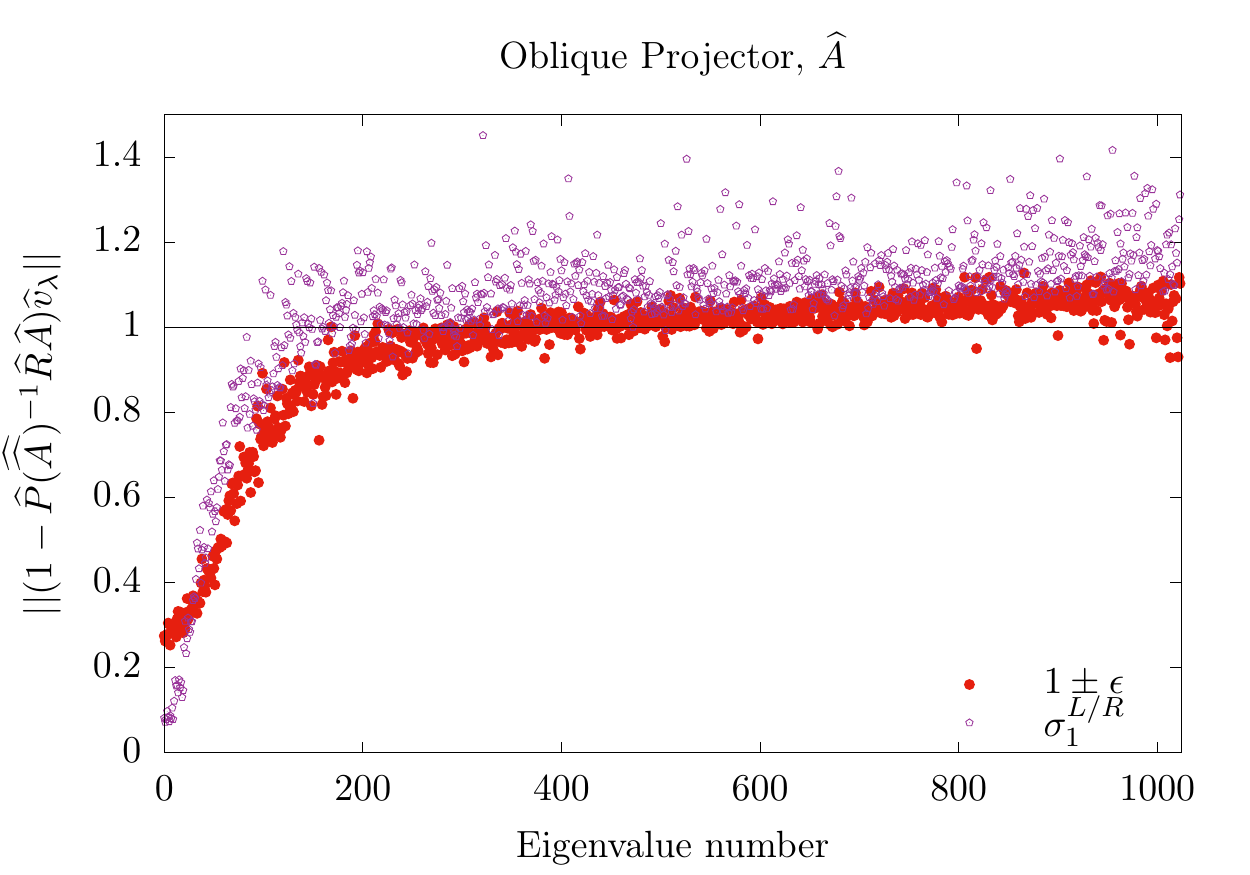}
  \caption{On the left, a measurement of local co-linearity: how well low near-null vectors can reconstruct higher eigenvectors. On the right, the effect of the oblique projector on eigenvectors. The top row considers a representative Dirac-K\"ahler preconditioned operator; the bottom row considers its coarsening. Both panels are sorted by increasing magnitude of the eigenvalues.}
\label{fig:projectorsspectrumcircleherm}
\end{figure}

In Sec.~\ref{sec:hermpres}, we developed a formalism to exactly preserve complex conjugate eigenpairs for a coarsened K\"ahler-Dirac preconditioned operator. This used an {\emph{asymmetric}} coarsening which gave a $\sigma_1^{L/R}$ on the coarse level. This formulation is largely successful, however, it can suffer from anomalously large real eigenvalues in the negative half plane, destabilizing the MG preconditioned solve, in cases where the {\emph{symmetric}} coarsening proceeded without issues. If these stability issues can be addressed, it may lead to a better algorithm in two dimensions and four dimensions. As appropriate, this will be the topic of a future publication. 

This appendix will follow the structure of Sec.~\ref{sec:projkdp}, where we study the spectrum, local co-linearity, and oblique projector of the asymmetrically coarsened operator {\emph{in the case where a recursive algorithm is successful.}} We will then scan the iteration counts as a function of mass, similar to in Sec.~\ref{sec:elim}, and identify cases where the algorithm breaks down. Last, we will investigate one of these cases. 

In Sec.~\ref{sec:projkdp} we considered a representative spectrum of the K\"ahler-Dirac preconditioned operator and a symmetric coarsening. In the case of {\emph{asymmetric}} coarsening, we again expect the low modes to be preserved well, but additionally come in complex conjugate pairs. This is exactly the case in Fig.~\ref{fig:zoomcircleherm}, where we overlay the spectrum of the asymmetric coarse operator. We also see a ``feature'' of $\sigma_1^{L/R}$ Hermiticity: there are pairs of purely real eigenvalues.

In the case of the Wilson or overlap operator, pairs of purely real eigenvalues have a significant physical interpretation. The smaller real eigenvalue corresponds to a physical chiral mode via the lattice index theorem~\cite{Hernandez:1997bd}, which thus needs to be well captured by a MG algorithm. The paired large real eigenvalue is merely a quirk of being on a finite lattice, and thus lives as an isolated large eigenvalue near the cutoff. On the other hand, the pairs of real eigenvalues for the coarsened K\"ahler-Dirac operator do not have an obvious physical intuition, just as the na\"ive staggered fermion operator does not trivially correspond to an index theorem~\cite{Adams:2009eb}. These purely real eigenvalues are a symptom of unstable solves at larger volumes.

Returning to stable solves, we consider the local co-linearity and the oblique projector under an asymmetric coarsening. These are overlaid on the data for a symmetric coarsening in Fig.~\ref{fig:projectorsspectrumcircleherm}. An asymmetric coarsening is roughly comparable to quality to a symmetric coarsening, indicative of a successful MG algorithm.\footnote{In general, the local-colinearity is not bounded by 1 when $R \neq P^\dagger$. This is because $\left(1 - P R\right)$ is {\emph{not}} a normal operator. Thus, for a normalized vector $v$, $v^\dagger \left(1 - R^\dagger P^\dagger\right)\left(1 - P R\right) v$ isn't bounded by 1. This can be realized by the bi-orthonormal basis $p_1 = (1/2,1/2,1/2,1/2),~p_2 = (1/2,1/2,1/2,-3/2),~r_1 = (1/2,1/2,1/2,1/2),~r_2 = (1/2,-1/2,1/2,-1/2)$.}

\begin{figure}[t] \centering
  \includegraphics[width=0.47\linewidth]{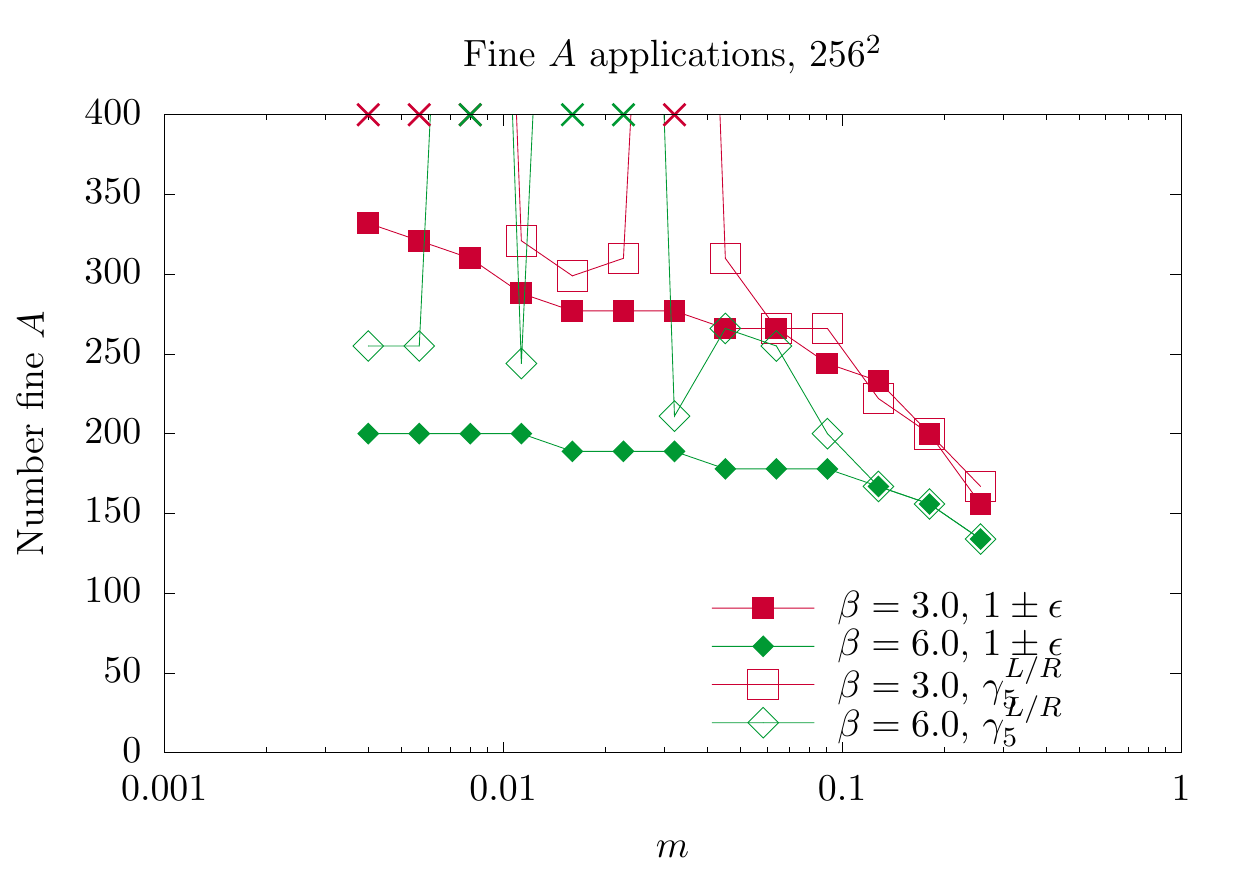}~\includegraphics[width=0.47\linewidth]{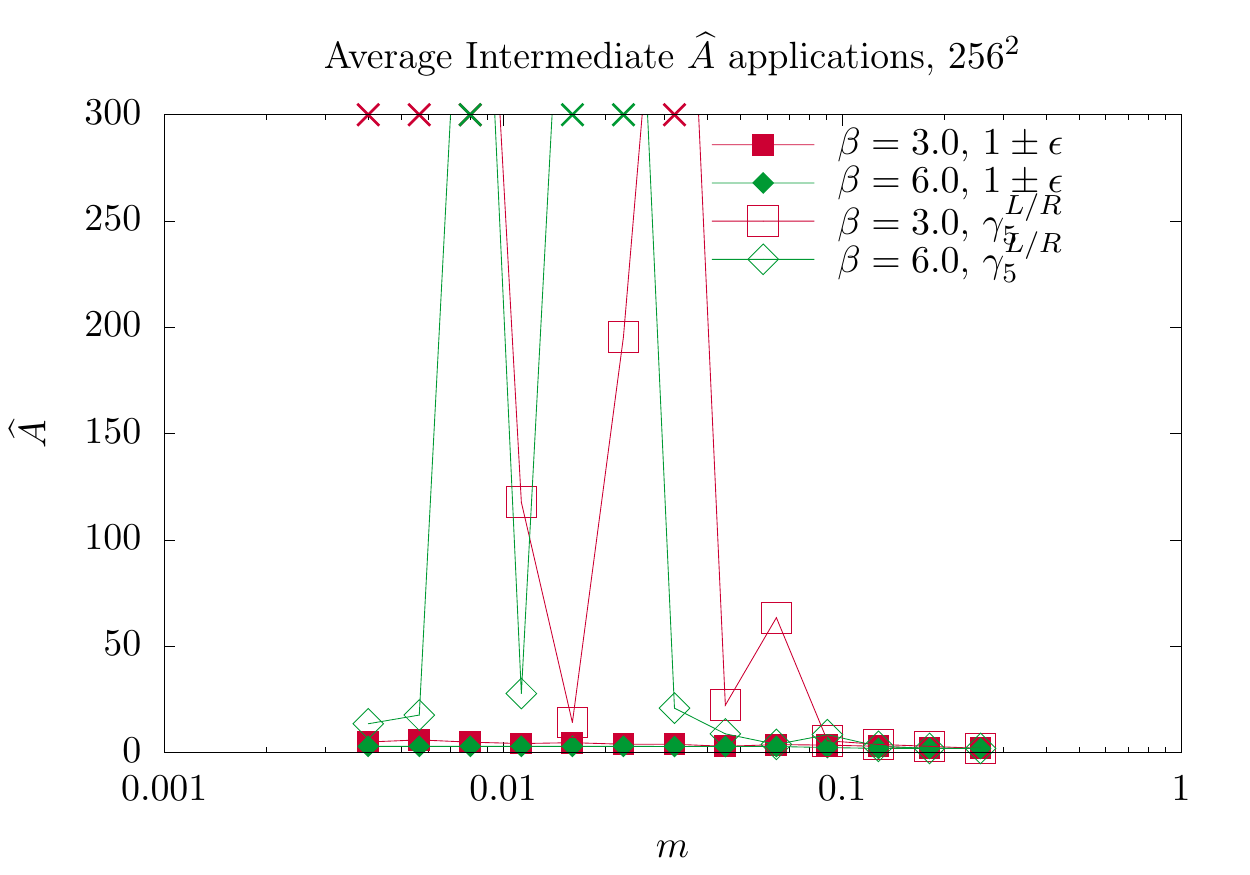}
  \caption{On the left, the number of outer iterations, and on the right, the average number of iterations in the K cycle on the intermediate level, as a function of mass at fixed $\beta$. Values marked with ``$\times$'' indicate a failure to converge when an asymmetric coarsening is used. All data points are from a single configuration per $\beta$ but are representative of a more general behavior.}
\label{fig:iterationscompare}
\end{figure}

As a next task, we consider MG preconditioned solves with the asymmetric coarsened operator. We will only present a subset of the cases considered in Sec.~\ref{sec:elim} and instead focus on the cases where the solve is unstable: large volumes. The number of fine operator applications and average intermediate applications are presented in Fig.~\ref{fig:iterationscompare}. In the cases where a data point is marked by a ``$\times$'', the solve failed. The failures are largely confined to smaller masses, but not with a discernable pattern; indeed, for $\beta = 6.0$, the lowest masses had stable solves!

\begin{figure}[t] \centering
   \includegraphics[width=0.47\linewidth]{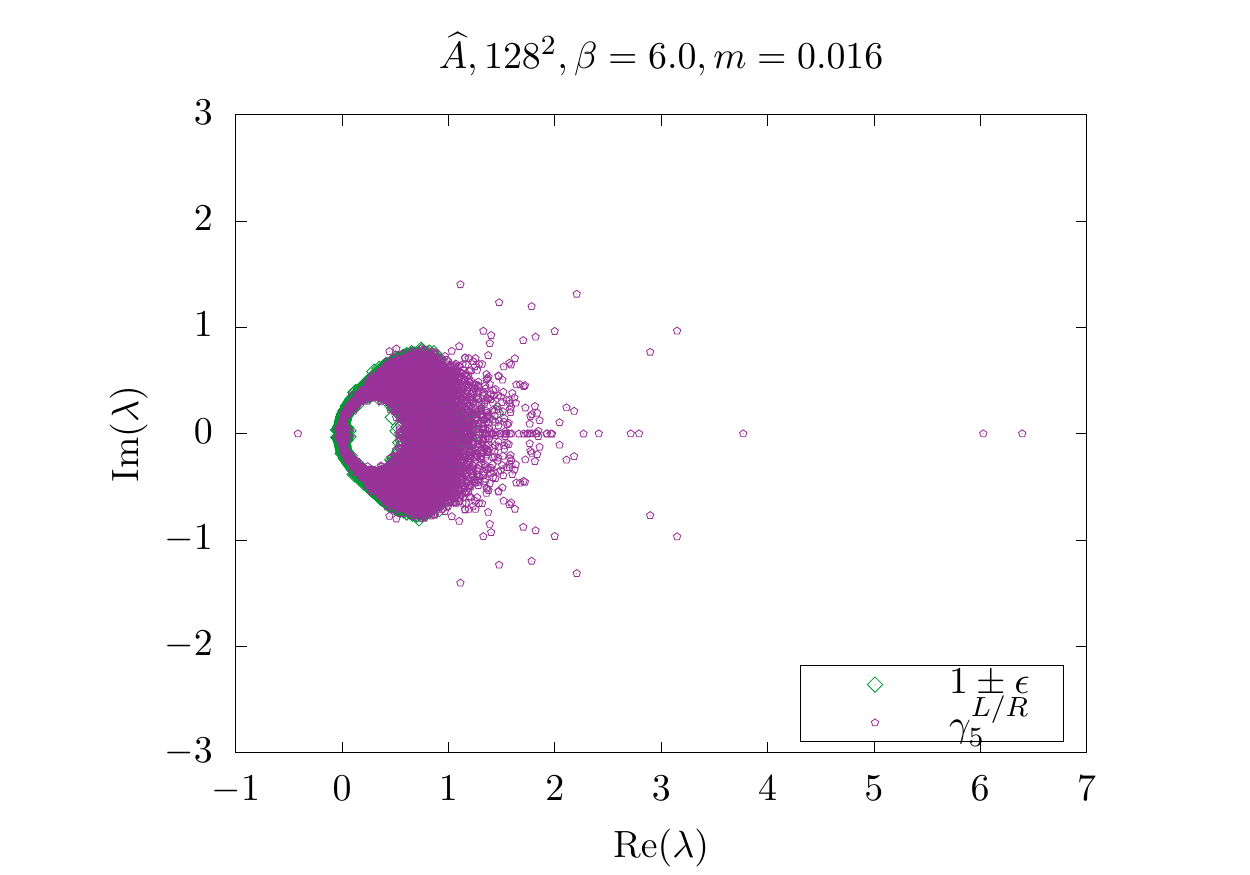}~   \includegraphics[width=0.47\linewidth]{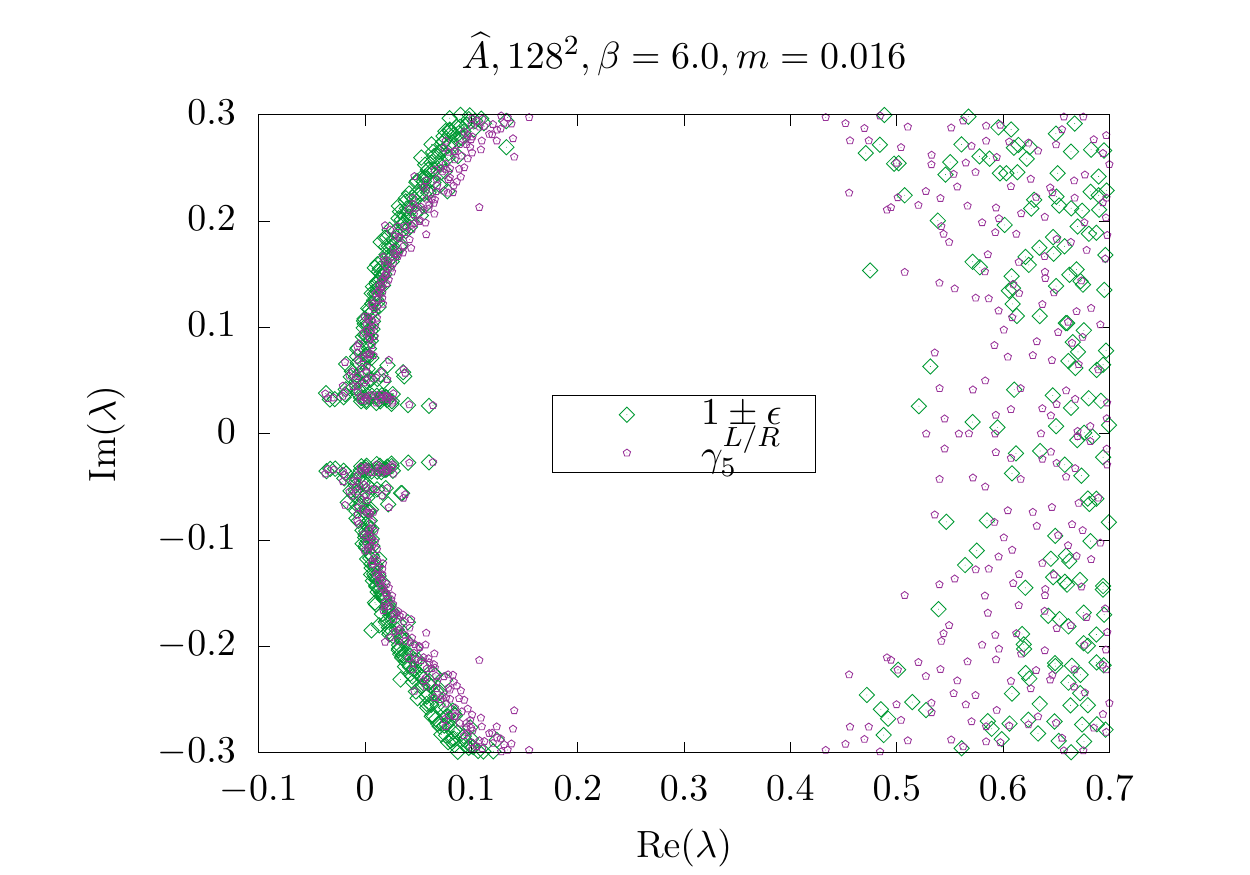}
  \caption{The spectrum of a representative coarse operator from a symmetric coarsening with the asymmetric coarsening overlaid. There are large purely real eigenvalues. {\textbf{Not included is a large negative eigenvalue $\lambda = -26.75$.}} Computing the fine operator spectrum was prohibitatively expensive.}
\label{fig:zoomcircleohno}
\end{figure}

We present the spectrum of the asymmetric coarsened operator, where an MG solve with an asymmetric coarsened operator fails, in the left panel of Fig.~\ref{fig:zoomcircleohno}, where we see there are now large, real eigenvalues far in the right plane and also in the left plane. There is also a large negative real eigenvalue at approximately -26.75. These pathological real eigenvalues are not part of the low subspace and are therefore not well captured by our MG algorithm. However, in the right panel, we see that the low spectrum is still well behaved. It is a point of future research to see if these anomalously large, real eigenvalues can be addressed.

\end{document}